\documentclass[%
reprint,
 amsmath,amssymb,
prb,
]{revtex4-1}
\usepackage{amssymb}
\usepackage{multirow}
\usepackage{graphicx}
\usepackage{dcolumn}
\usepackage{bm}
\usepackage{esint}
\usepackage{color}

\begin{document}


\title{Topological quantum phase transitions in the spin-singlet
superconductor with Rashba and Dresselhaus (110) spin-orbit
couplings}

\author{Jiabin You$^{1,}$}
\email{jiabinyou@gmail.com}

\author{A. H. Chan$^{2,}$}
\email{phycahp@nus.edu.sg}

\author{C. H. Oh$^{1,2,}$}
\email{phyohch@nus.edu.sg}

\author{Vlatko Vedral$^{1,2,3,}$}
\email{phyvv@nus.edu.sg}

\affiliation{$^1$Centre for Quantum Technologies, National
University of Singapore, 117543, Singapore\\$^2$Department of
Physics, National University of Singapore, 117542,
Singapore\\$^3$Department of Physics, University of Oxford,
Clarendon Laboratory, Oxford, OX1 3PU, United Kingdom}

\date{\today}

\begin{abstract}

We examine the topological properties of a spin-singlet
superconductor with Rashba and Dresselhaus (110) spin-orbit
couplings. We demonstrate that there are several topological
invariants in the Bogoliubov-de Gennes (BdG) Hamiltonian by symmetry
analysis. We use the Pfaffian invariant $\mathcal{P}$ for the
particle-hole symmetry to demonstrate all the possible phase
diagrams of the BdG Hamiltonian. We find that the edge spectrum is
either Dirac cone or flat band which supports the emergence of the
Majorana fermion in this system. For the Majorana flat bands, a
higher symmetric BdG Hamiltonian is needed to make them
topologically stable. The Pfaffian invariant $\mathcal{P}(k_{y})$
and the winding number $\mathcal{W}(k_{y})$ are used in determining
the location of the Majorana flat bands.

\end{abstract}

\pacs{Valid PACS appear here}
\maketitle


\section{introduction}

A topological superconductor has topologically protected gapless
edge states, some of which can host Majorana fermions
\cite{PhysRevB.61.10267,PhysRevLett.100.096407,PhysRevB.82.134521,PhysRevB.81.125318,PhysRevLett.105.177002,PhysRevLett.108.107005,PhysRevLett.109.150408,PhysRevB.87.054501,PhysRevLett.104.040502,PhysRevLett.105.217001}.
There are two kinds of edge states in the topological
superconductor. One is a Dirac cone, the other is a flat band,
namely, dispersionless zero-energy state
\cite{PhysRevB.84.060504,PhysRevB.86.161108,PhysRevB.83.224511,PhysRevB.84.020501,Chris2012}.
The Dirac cone can be found in the fully gapped topological
superconductors when the Chern number of the occupied bands is
nonzero. However, the flat band can appear in the fully gapped
topological superconductors as well as the gapless topological
superconductors which, apart from the particle-hole symmetry, have
some extra symmetries in the Hamiltonian. Such flat bands are known
to occur at the zigzag and bearded edge in graphene
\cite{PhysRevB.54.17954}, in the noncentrosymmetric superconductor
\cite{PhysRevB.84.060504,PhysRevB.84.020501,PhysRevB.83.224511,PhysRevB.87.054501}
and in other systems with topologically stable Dirac points
\cite{JETP.93.66}. The flat dispersion implies a peak in the density
of states which has a visible experimental signature in the
tunneling conductance measurements \cite{PhysRevLett.105.097002}.
The zero-bias conductance peak has been observed in recent
experiments on the InSb nanowire \cite{Mourik2012,Das2012} and
$\text{Cu}_{x}\text{Bi}_{2}\text{Se}_{3}$
\cite{PhysRevLett.107.217001,PhysRevLett.108.057001} and might be
due to the flat bands.

In this paper, we shall investigate the spin-singlet superconductor
with the Rashba and Dresselhaus (110) spin-orbit couplings.
Especially, we focus on the Hamiltonian with spin-orbit coupling of
Dresselhaus (110) type which is a gapless topological system
containing two kinds of edge states mentioned above. We also apply a
magnetic field to the superconductor and consider the Zeeman effect.
The Bogoliubov-de Gennes (BdG) Hamiltonian of the superconductor is
generically particle-hole symmetric so that we can associate a
Pfaffian invariant $\mathcal{P}$ with it as a topological invariant
of the system. The Pfaffian invariant $\mathcal{P}$ can be used in
distinguishing the topologically nontrivial phase from the trivial
one. The nontrivial topological phase in this BdG Hamiltonian is a
Majorana type which can be exploited for implementing fault-tolerant
topological quantum computing schemes that are inherently
decoherence-free \cite{Kitaev2001,RevModPhys.80.1083}. All the
possible phase diagrams in the BdG Hamiltonian are demonstrated in
our study. Furthermore, we find that the BdG Hamiltonian can have
partial particle-hole symmetry and chiral symmetry which can be used
to define the one dimensional Pfaffian invariant
$\mathcal{P}(k_{y})$ and the winding number $\mathcal{W}(k_{y})$.
The Pfaffian invariant $\mathcal{P}(k_{y})$ and the winding number
$\mathcal{W}(k_{y})$ can be used in determining the location of the
zero-energy Majorana flat band. We show that only when the system
has these extra symmetries, the Majorana flat band will emerge so
that a zero-bias conductance peak in the tunneling conductance
measurements can be observed
\cite{Mourik2012,Das2012,PhysRevLett.107.217001,PhysRevLett.108.057001}.

The paper is organized as follows. The BdG Hamiltonian for the
spin-singlet superconductor with the Rashba and Dresselhaus (110)
spin-orbit couplings is given in Sec. \ref{model}. In Sec.
\ref{symofBdG}, we discuss the symmetries of the BdG Hamiltonian and
the topological invariants associated with these symmetries are
given in Sec. \ref{tiofBdG}. All the possible phase diagrams of the
BdG Hamiltonian are discussed in Sec. \ref{pdofBdG} whilst the edge
spectra of the BdG Hamiltonian are demonstrated in Sec.
\ref{MBSofBdG}. Finally, we give a brief summary in Sec.
\ref{summary}.

\section{model}\label{model}

We model the spin-singlet superconductor on a square lattice. The
kinetic energy is
\begin{equation}
\label{Hkin}
\begin{split}
H_{\text{kin}}=-t\sum\limits_{is}\sum\limits_{\hat{\nu}=\hat{x},\hat{y}}(c_{i+\hat{\nu}s}^{\dag}c_{is}+c_{i-\hat{\nu}s}^{\dag}c_{is})-\mu\sum\limits_{is}c_{is}^{\dag}c_{is},
\end{split}
\end{equation}
where $c_{is}^{\dag}(c_{is})$ is the creation (annihilation)
operator of the electron with spin $s=(\uparrow,\downarrow)$ at site
$i=(i_{x},i_{y})$, $\hat{x}$ $(\hat{y})$ is the unit vector in the
$x$ ($y$) direction, and $t$ is the hopping amplitude and $\mu$ is
the chemical potential. For the spin-singlet superconductor, we
study the $s$-wave and $d$-wave pairings in this work. The $s$-wave
superconducting term in the square lattice is
\begin{equation}
\label{Hs}
\begin{split}
H_{s}=\sum\limits_{i}[(\Delta_{s_1}+i\Delta_{s_2})c_{i\uparrow}^{\dag}c_{i\downarrow}^{\dag}+\text{H.c.}].
\end{split}
\end{equation}
Similarly, the $d$-wave superconducting term is
\begin{equation}
\label{Hd}
\begin{split}
H_{d}=&\sum\limits_{i}[\frac{\Delta_{d_1}}{2}(c_{i-\hat{y}\uparrow}^{\dag}c_{i\downarrow}^{\dag}+c_{i+\hat{y}\uparrow}^{\dag}c_{i\downarrow}^{\dag}\\
&-c_{i-\hat{x}\uparrow}^{\dag}c_{i\downarrow}^{\dag}-c_{i+\hat{x}\uparrow}^{\dag}c_{i\downarrow}^{\dag})\\
&+i\frac{\Delta_{d_2}}{4}(c_{i-\hat{x}+\hat{y}\uparrow}^{\dag}c_{i\downarrow}^{\dag}+c_{i+\hat{x}-\hat{y}\uparrow}^{\dag}c_{i\downarrow}^{\dag}\\
&-c_{i+\hat{x}+\hat{y}\uparrow}^{\dag}c_{i\downarrow}^{\dag}-c_{i-\hat{x}-\hat{y}\uparrow}^{\dag}c_{i\downarrow}^{\dag})+\text{H.c.}].\\
\end{split}
\end{equation}
We assume that all the superconducting parameters $\Delta_{s_1}$,
$\Delta_{s_2}$, $\Delta_{d_1}$ and $\Delta_{d_2}$ are uniform in the
whole superconductor. The spin-orbit couplings can arise from
structure inversion asymmetry of a confinement potential (e.g.,
external electric field) or bulk inversion asymmetry of the
underlying crystal (e.g., the zinc blende structure) \cite{Winkler}.
These two kinds of asymmetries lead to the well-known Rashba and
Dresselhaus spin-orbit couplings. The Rashba spin-orbit coupling in
the square lattice is of the form
\begin{equation}
\label{HRashba}
\begin{split}
H_{\text{R}}=&-\frac{\alpha}{2}\sum\limits_{i}[(c_{i-\hat{x}\downarrow}^{\dag}c_{i\uparrow}-c_{i+\hat{x}\downarrow}^{\dag}c_{i\uparrow})\\
&+i(c_{i-\hat{y}\downarrow}^{\dag}c_{i\uparrow}-c_{i+\hat{y}\downarrow}^{\dag}c_{i\uparrow})+\text{H.c.}],\\
\end{split}
\end{equation}
where $\alpha$ is the coupling strength of the Rashba spin-orbit
coupling. The Dresselhaus (110) spin-orbit coupling is formulated as
\begin{equation}
\label{HDresselhaus}
\begin{split}
H_{\text{D}}^{110}=&-i\frac{\beta}{2}\sum\limits_{iss'}(\tau_{z})_{ss'}(c_{i-\hat{x}s}^{\dag}c_{is'}-c_{i+\hat{x}s}^{\dag}c_{is'}),\\
\end{split}
\end{equation}
where $\beta$ is the coupling strength for the Dresselhaus (110)
spin-orbit coupling. (110) are the common-used Miller index. We also
apply an arbitrary magnetic field to the superconductor. By
neglecting the orbital effect of the magnetic field, we consider the
Zeeman effect as
\begin{equation}
\label{HZeeman}
\begin{split}
H_{\text{Z}}=\sum\limits_{iss'}(\mathbf{V}\cdot\mathbf{\tau})_{ss'}c_{is}^{\dag}c_{is'},
\end{split}
\end{equation}
where
$\mathbf{V}=\frac{g\mu_{B}}{2}(B_{x},B_{y},B_{z})\equiv(V_{x},V_{y},V_{z})$
and $\mathbf{\tau}=(\tau_{x},\tau_{y},\tau_{z})$ are Pauli matrices
operating on spin space. Therefore, the spin-singlet superconductor
with the Rashba and Dresselhaus (110) spin-orbit couplings in an
arbitrary magnetic field is dictated by the Hamiltonian
$H=H_{\text{kin}}+H_{s}+H_{d}+H_{R}+H_{\text{D}}^{110}+H_{\text{Z}}$.
In the momentum space, the Hamiltonian is recast into
$H=\frac{1}{2}\sum_{\mathbf{k}}\psi_{\mathbf{k}}^{\dag}\mathcal{H(\mathbf{k})}\psi_{\mathbf{k}}$
with
$\psi_{\mathbf{k}}^{\dag}=(c_{\mathbf{k}\uparrow}^{\dag},c_{\mathbf{k}\downarrow}^{\dag},c_{\mathbf{-k}\uparrow},c_{\mathbf{-k}\downarrow})$,
where
$c_{\mathbf{k}s}^{\dag}=(1/\sqrt{N})\sum_{\mathbf{l}}e^{i\mathbf{k}\cdot\mathbf{l}}c_{\mathbf{l}s}^{\dag}$.
Finally, the Bogoliubov-de Gennes Hamiltonian
$\mathcal{H}(\mathbf{k})$ for the superconductor is
\begin{equation}
\label{HBdG}
\begin{split}
\left[\begin{array}{*{20}c}
{\xi(\mathbf{k})+(\mathcal{L}(\mathbf{k})+\mathbf{V})\cdot\mathbf{\tau}} & {i\Delta(\mathbf{k})\tau_{y}}\\
{-i\Delta^{*}(\mathbf{k})\tau_{y}} & {-\xi(\mathbf{k})+(\mathcal{L}(\mathbf{k})-\mathbf{V})\cdot\mathbf{\tau^{*}}}\\
\end{array}\right]
\end{split}
\end{equation}
where $\xi(\mathbf{k})=-2t(\cos{k_{x}}+\cos{k_{y}})-\mu$,
$\Delta(\mathbf{k})=(\Delta_{s_{1}}+i\Delta_{s_{2}})+(\Delta_{d_{1}}(\cos{k_{y}}-\cos{k_{x}})+i\Delta_{d_{2}}\sin{k_{x}}\sin{k_{y}})$
and
$\mathcal{L}(\mathbf{k})=(\alpha\sin{k_{y}},-\alpha\sin{k_{x}},\beta\sin{k_{x}})$.

\section{symmetries of the BdG Hamiltonian}\label{symofBdG}

For the general BdG Hamiltonian Eq. (\ref{HBdG}), it satisfies the
particle-hole symmetry
\begin{equation}
\label{PHS}
\begin{split}
\Xi^{-1}\mathcal{H}(\mathbf{k})\Xi=-\mathcal{H}(-\mathbf{k}),
\end{split}
\end{equation}
where $\Xi=\Lambda K$, $\Lambda=\sigma_{x}\otimes\tau_{0}$ and $K$
is the complex conjugation operator. We find that apart from the
particle-hole symmetry, the BdG Hamiltonian can satisfy some extra
symmetries, namely, partial particle-hole symmetry, chiral symmetry
and partial chiral symmetry when we set some parameters in the
Hamiltonian Eq. (\ref{HBdG}) to $0$. The particle-hole-$k_{x}$ and
particle-hole-$k_{y}$ symmetries are defined as
\begin{equation}
\label{PHkxS}
\begin{split}
\Xi_{k_{x}}^{-1}\mathcal{H}(k_{x},k_{y})\Xi_{k_{x}}=-\mathcal{H}(-k_{x},k_{y})
\end{split}
\end{equation}
and
\begin{equation}
\label{PHkyS}
\begin{split}
\Xi_{k_{y}}^{-1}\mathcal{H}(k_{x},k_{y})\Xi_{k_{y}}=-\mathcal{H}(k_{x},-k_{y}),
\end{split}
\end{equation}
where $\Xi_{k_{x}}$ ($\Xi_{k_{y}}$) takes the $k_{x}$ ($k_{y}$) in
the Hamiltonian to $-k_{x}$ ($-k_{y}$). The chiral symmetry is
\begin{equation}
\label{CS}
\begin{split}
\Sigma^{-1}\mathcal{H}(\mathbf{k})\Sigma=-\mathcal{H}(\mathbf{k}).
\end{split}
\end{equation}
The chiral-$k_{x}$ and chiral-$k_{y}$ symmetries are defined as
\begin{equation}
\label{CkxS}
\begin{split}
\Sigma_{k_{x}}^{-1}\mathcal{H}(k_{x},k_{y})\Sigma_{k_{x}}=-\mathcal{H}(-k_{x},k_{y})
\end{split}
\end{equation}
and
\begin{equation}
\label{CkyS}
\begin{split}
\Sigma_{k_{y}}^{-1}\mathcal{H}(k_{x},k_{y})\Sigma_{k_{y}}=-\mathcal{H}(k_{x},-k_{y}),
\end{split}
\end{equation}
where $\Sigma_{k_{x}}$ ($\Sigma_{k_{y}}$) takes the $k_{x}$
($k_{y}$) in the Hamiltonian to $-k_{x}$ ($-k_{y}$).

We are interested in the BdG Hamiltonian which has one or more extra
symmetries. In the following sections, we would like to consider
these kinds of BdG Hamiltonian as listed in Tab. (\ref{symmandti}).
The spin-singlet superconductor with Rashba spin-orbit coupling has
been extensively investigated in the reference
\cite{PhysRevB.82.134521}. Here we only consider the general
$d_{x^2-y^2}+id_{xy}+s$ pairing case (a) for the superconductor with
Rashba spin-orbit coupling. We shall focus on the superconductor
with Dresselhaus (110) spin-orbit coupling as shown in the case
(b)-(g) in Tab. (\ref{symmandti}).
\begin{table*}[htbp]
\begin{tabular}{|c|c|c|c|c|c|}
\hline
 case & spin-orbit coupling & magnetic field & pairing symmetry & Hamiltonian symmetry & topological invariant \\
\hline
 (a)& $\alpha$ & $V_{z}$ & $\Delta_{s_{1}},\Delta_{d_{1}},\Delta_{d_{2}}$ & $\Xi, \Sigma_{k_{x}}$ & $\mathcal{P}$, $\mathcal{W}$ \\
\hline
 (b)& \multirow{6}{*}{$\beta$} & \multirow{6}{*}{$V_{x},V_{y}$} & $\Delta_{s_{1}}$ & $\Xi, \Xi_{k_{x}}, \Sigma, \Sigma_{k_{y}}$ & $\mathcal{P}$, $\mathcal{P}(k_{y})$, $\mathcal{W}$, $\mathcal{W}(k_{y})$ \\
\cline{1-1}\cline{4-6}
 (c)& & & $\Delta_{s_{1}},\Delta_{s_{2}}$ & $\Xi, \Xi_{k_{x}}$ & $\mathcal{P}$, $\mathcal{P}(k_{y})$ \\
\cline{1-1}\cline{4-6}
 (d)& & & $\Delta_{d_{1}}$ & $\Xi, \Xi_{k_{x}}, \Sigma, \Sigma_{k_{y}}$ & $\mathcal{P}$, $\mathcal{P}(k_{y})$, $\mathcal{W}$, $\mathcal{W}(k_{y})$ \\
\cline{1-1}\cline{4-6}
 (e)& & & $\Delta_{d_{1}},\Delta_{d_{2}}$ & $\Xi, \Sigma_{k_{y}}$ & $\mathcal{P}$, $\mathcal{W}$ \\
\cline{1-1}\cline{4-6}
 (f)& & & $\Delta_{s_{1}},\Delta_{d_{1}}$ & $\Xi, \Xi_{k_{x}}, \Sigma$ & $\mathcal{P}$, $\mathcal{P}(k_{y})$, $\mathcal{W}(k_{y})$ \\
\cline{1-1}\cline{4-6}
 (g)& & & $\Delta_{s_{1}},\Delta_{d_{1}},\Delta_{d_{2}}$ & $\Xi, \Sigma_{k_{y}}$ & $\mathcal{P}$, $\mathcal{W}$ \\
\hline
\end{tabular}
\caption{The BdG Hamiltonian with extra symmetries, namely, the
particle-hole symmetry and the particle-hole-$k_{x}$ symmetry,
$\Xi=\Xi_{k_{x}}=\sigma_{x}K$, the chiral symmetry and the
chiral-$k_{y}$ symmetry, $\Sigma=\Sigma_{k_{y}}=i\sigma_{y}\tau_{x}$
and the chiral-$k_{x}$ symmetry,
$\Sigma_{k_{x}}=i\sigma_{y}\tau_{z}$. The topological invariants for
these extra symmetries are also shown in the last
column.}\label{symmandti}
\end{table*}

\section{topological invariants of the BdG Hamiltonian}\label{tiofBdG}

For the fully gapped Hamiltonian, we can always define Chern number
as a topological invariant of the system given by
\begin{equation}
\begin{split}
\mathcal{C}=\frac{1}{2\pi}\int_{FBZ}d^{2}\mathbf{k}\mathcal{F}^{-}(\mathbf{k}).
\end{split}
\end{equation}
Here
$\mathcal{F}^{-}(\mathbf{k})=\epsilon^{ij}\partial_{k_{i}}A_{j}^{-}(\mathbf{k})$
is the strength of the gauge field
$\mathbf{A}^{-}(\mathbf{k})=i\sum_{n<0}\langle\psi_{n}(\mathbf{k})|\nabla\psi_{n}(\mathbf{k})\rangle$,
where $\psi_{n}(\mathbf{k})$ is the eigenstate of the Hamiltonian.
The integral is carried out in the first Brillouin zone (FBZ) and
the summation is carried out for the occupied states. If the
Hamiltonian has some extra symmetries, more topological invariants
can be introduced into the system.

We first consider the particle-hole symmetry Eq. (\ref{PHS}) which
can be reduced to $\Lambda
\mathcal{H}(\mathbf{k})\Lambda=-\mathcal{H}^{*}(-\mathbf{k})$. We
find that under this symmetry $\mathcal{H}(\mathbf{K})\Lambda$ is an
antisymmetric matrix with
$(\mathcal{H}(\mathbf{K})\Lambda)^{T}=-\mathcal{H}(\mathbf{K})\Lambda$,
where $\mathbf{K}$ is the particle-hole symmetric momenta satisfying
$\mathbf{K}=-\mathbf{K}+\mathbf{G}$ and $\mathbf{G}$ is the
reciprocal lattice vector of the square lattice. With this property,
we can define the Pfaffian invariant for the particle-hole symmetric
Hamiltonian as \cite{PhysRevB.82.184525}
\begin{equation}
\begin{split}
\mathcal{P}=\text{sgn}\left\{\frac{\text{Pf}[\mathcal{H}(\mathbf{K_{1}})\Lambda]\text{Pf}[\mathcal{H}(\mathbf{K_{4}})\Lambda]}{\text{Pf}[\mathcal{H}(\mathbf{K_{2}})\Lambda]\text{Pf}[\mathcal{H}(\mathbf{K_{3}})\Lambda]}\right\},
\end{split}
\end{equation}
where $\mathbf{K_{1}}=(0,0)$, $\mathbf{K_{2}}=(\pi,0)$,
$\mathbf{K_{3}}=(0,\pi)$ and $\mathbf{K_{4}}=(\pi,\pi)$ are the four
particle-hole symmetric momenta in the first Brillouin zone of the
square lattice. Here we shall show that the Pfaffian invariant
$\mathcal{P}$ is the parity of the Chern number $\mathcal{C}$,
$\mathcal{P}=(-1)^{\mathcal{C}}$. For the $2n\times2n$ antisymmetric
matrix $\mathcal{H}(\mathbf{K})\Lambda$, we have
$\text{Pf}[\mathcal{H}(\mathbf{K})\Lambda]^{*}=(-1)^{n}\text{Pf}[\mathcal{H}(\mathbf{K})\Lambda]$.
Therefore,
$(i^{n}\text{Pf}[\mathcal{H}(\mathbf{K})\Lambda])^{*}=i^{n}\text{Pf}[\mathcal{H}(\mathbf{K})\Lambda]$
is real so that we can define a quantity
$S[\mathcal{H}(\mathbf{K})]=\text{sgn}\{i^{n}\text{Pf}[\mathcal{H}(\mathbf{K})\Lambda]\}$
for any particle-hole symmetric Hamiltonian. Suppose
$\mathcal{H}(\mathbf{K})$ is diagonalized by the transformation
$\mathcal{H}(\mathbf{K})=U(\mathbf{K})D(\mathbf{K})U^{\dag}(\mathbf{K})$,
where $D(\mathbf{K})$ is a diagonal matrix of eigenvalues
$\text{diag}\{E_{n}(\mathbf{K}),\cdot\cdot\cdot,E_{1}(\mathbf{K}),-E_{1}(\mathbf{K}),\cdot\cdot\cdot,-E_{n}(\mathbf{K})\}$
and the columns of the unitary matrix $U(\mathbf{K})$ are the
eigenvectors of $\mathcal{H}(\mathbf{K})$. The eigenvectors for
positive eigenvalues in $U(\mathbf{K})$ are chosen to be related to
the eigenvectors for negative eigenvalues by particle-hole symmetry.
With this convention, we find that $U^{\dag}\Lambda=\Gamma U^{T}$,
where $\Gamma=\sigma_{x}\tau_{x}$. Therefore,
$S[\mathcal{H}(\mathbf{K})]$ can be further reduced to
\begin{equation}
\begin{split}
S[\mathcal{H}(\mathbf{K})]&=\text{sgn}\{i^{n}\text{Pf}[\mathcal{H}(\mathbf{K})\Lambda]\},\\
&=\text{sgn}\{i^{n}\text{Pf}[U(\mathbf{K})D(\mathbf{K})U^{\dag}(\mathbf{K})\Lambda]\},\\
&=\text{sgn}\{i^{n}\text{Pf}[U(\mathbf{K})D(\mathbf{K})\Gamma U^{T}(\mathbf{K})]\},\\
&=\text{sgn}\{i^{n}\det{U(\mathbf{K})}\text{Pf}[D(\mathbf{K})\Gamma]\}.\\
\end{split}
\end{equation}
Since
$\text{Pf}[D(\mathbf{K})\Gamma]=\prod_{n>0}E_{n}(\mathbf{K})>0$ and
$|\det{U(\mathbf{K})}|=1$, we arrive at
\begin{equation}
\begin{split}
S[\mathcal{H}(\mathbf{K})]=i^{n}\det{U(\mathbf{K})}.
\end{split}
\end{equation}
Note that
$\mathbf{A}(\mathbf{k})=i\sum_{n}\langle\psi_{n}(\mathbf{k})|\nabla\psi_{n}(\mathbf{k})\rangle$
is a total derivative \cite{PhysRevB.82.134521}
$\mathbf{A}(\mathbf{k})=i\nabla\ln[\det U(\mathbf{k})]$. Therefore,
consider a pair of particle-hole symmetric momenta $\mathbf{K_{1}}$
and $\mathbf{K_{2}}$, we find that
\begin{equation}
\begin{split}
\frac{\det U(\mathbf{K_{2}})}{\det U(\mathbf{K_{1}})}=e^{-iS_{1,2}},
\end{split}
\end{equation}
where
$S_{1,2}=\int_{\mathbf{K_{1}}}^{\mathbf{K_{2}}}\mathbf{A}(\mathbf{k})\cdot
d\mathbf{k}$ and the line integral runs from $\mathbf{K_{1}}$ to
$\mathbf{K_{2}}$. Since
$\mathbf{A}^{+}(\mathbf{k})=i\sum_{n>0}\langle\psi_{n}(\mathbf{k})|\nabla\psi_{n}(\mathbf{k})\rangle=\mathbf{A}^{-}(\mathbf{-k})$,
we find that
$S_{1,2}=\int_{\gamma_{1}}\mathbf{A}^{-}(\mathbf{k})\cdot
d\mathbf{k}$, where $\gamma_{1}$ is the line from $(-\pi,0)$ to
$(\pi,0)$. Similarly,
\begin{equation}
\begin{split}
\frac{\det U(\mathbf{K_{4}})}{\det U(\mathbf{K_{3}})}=e^{-iS_{3,4}},
\end{split}
\end{equation}
where $S_{3,4}=\int_{\gamma_{2}}\mathbf{A}^{-}(\mathbf{k})\cdot
d\mathbf{k}$ and $\gamma_{2}$ is the line from $(-\pi,\pi)$ to
$(\pi,\pi)$. Therefore,
\begin{equation}
\begin{split}
\frac{\det U(\mathbf{K_{1}})\det U(\mathbf{K_{4}})}{\det
U(\mathbf{K_{2}})\det U(\mathbf{K_{3}})}=e^{iS_{\gamma}},
\end{split}
\end{equation}
where
$S_{\gamma}=\ointctrclockwise_{\gamma}\mathbf{A}^{-}(\mathbf{k})\cdot
d\mathbf{k}$ and $\gamma$ is the directed line surrounding the upper
half Brillouin zone (UHBZ) in the counter clockwise direction. Since
$\mathcal{F}^{-}(\mathbf{k})=\mathcal{F}^{-}(\mathbf{-k})$, we have
\begin{equation}
\begin{split}
S_{\gamma}&=\ointctrclockwise_{\gamma}\mathbf{A}^{-}(\mathbf{k})\cdot d\mathbf{k},\\
&=\int_{UHBZ}d^{2}\mathbf{k}\mathcal{F}^{-}(\mathbf{k}),\\
&=\frac{1}{2}\int_{FBZ}d^{2}\mathbf{k}\mathcal{F}^{-}(\mathbf{k}),\\
&=\pi\mathcal{C}.\\
\end{split}
\end{equation}
Finally, we obtain that
\begin{equation}
\label{defP}
\begin{split}
\mathcal{P}&=\text{sgn}\left\{\frac{\text{Pf}[\mathcal{H}(\mathbf{K_{1}})\Lambda]\text{Pf}[\mathcal{H}(\mathbf{K_{4}})\Lambda]}{\text{Pf}[\mathcal{H}(\mathbf{K_{2}})\Lambda]\text{Pf}[\mathcal{H}(\mathbf{K_{3}})\Lambda]}\right\},\\
&=\frac{\det{U(\mathbf{K_{1}})}\det{U(\mathbf{K_{4}})}}{\det{U(\mathbf{K_{2}})}\det{U(\mathbf{K_{3}})}},\\
&=(-1)^{\mathcal{C}}.\\
\end{split}
\end{equation}
Therefore, the Pfaffian invariant $\mathcal{P}$ is the parity of the
Chern number.

Similarly, if the Hamiltonian has partial particle-hole symmetry,
for example, the particle-hole-$k_{x}$ symmetry Eq. (\ref{PHkxS}),
then we can treat $k_{y}$ as a parameter and define the Pfaffian
invariant $\mathcal{P}(k_{y})$ to identify the location of the edge
states in the edge Brillouin zone
\cite{PhysRevLett.109.150408,Chris2012},
\begin{equation}
\label{Pky}
\begin{split}
\mathcal{P}(k_{y})=\text{sgn}\left\{\frac{\text{Pf}[\mathcal{H}(\pi,k_{y})\Lambda]}{\text{Pf}[\mathcal{H}(0,k_{y})\Lambda]}\right\},
\end{split}
\end{equation}
where $k_{x}=0$ and $k_{x}=\pi$ are the two particle-hole symmetric
momenta in the edge Brillouin zone. Similar to the Eq. (\ref{defP}),
we can get an expression of $\mathcal{P}(k_{y})$ in terms of the
line integral of the vector potential $A_{x}^{-}(\mathbf{k})$ as
\begin{equation}
\label{PkySky}
\begin{split}
\mathcal{P}(k_{y})=(-1)^{\frac{1}{\pi}\int_{-\pi}^{\pi}dk_{x}A_{x}^{-}(\mathbf{k})}.
\end{split}
\end{equation}

If the Hamiltonian has chiral symmetry Eq. (\ref{CS}), then the
winding number can be introduced as a topological invariant of the
system. The $\Sigma$ appearing in this paper can be diagonalized as
$\Sigma=TDT^{\dag}$ with $TT^{\dag}=1$ and
$D=\text{diag}\{i,i,-i,-i\}$. The Hamiltonian
$\mathcal{H}(\mathbf{k})$ is then simultaneously off-diagonalized as
$\mathcal{H}(\mathbf{k})=TQ(\mathbf{k})T^{\dag}$, where
$Q(\mathbf{k})$ is of the form $\left(\begin{array}{*{20}c}
{0} & {q(\mathbf{k})}\\
{q^{\dag}(\mathbf{k})} & {0}\\
\end{array}\right)$. We can thus define the winding number as
\begin{equation}
\begin{split}
\mathcal{W}(k_{y})&=-\frac{1}{4\pi}\int_{-\pi}^{\pi}dk_{x}\text{tr}[\Sigma \mathcal{H}^{-1}(\mathbf{k})\partial_{k_{x}}\mathcal{H}(\mathbf{k})],\\
&=-\frac{1}{4\pi}\int_{-\pi}^{\pi}dk_{x}\text{tr}[D Q^{-1}(\mathbf{k})\partial_{k_{x}}Q(\mathbf{k})],\\
&=\frac{i}{4\pi}\int_{-\pi}^{\pi}dk_{x}\text{tr}[q^{-1}(\mathbf{k})\partial_{k_{x}}q(\mathbf{k})-q^{\dag-1}(\mathbf{k})\partial_{k_{x}}q^{\dag}(\mathbf{k})],\\
&=-\frac{1}{2\pi}\text{Im}\int_{-\pi}^{\pi}dk_{x}\text{tr}[q^{-1}(\mathbf{k})\partial_{k_{x}}q(\mathbf{k})].\\
\end{split}
\end{equation}
Here we shall show that
$\int_{-\pi}^{\pi}dk_{x}\text{tr}[q^{-1}(\mathbf{k})\partial_{k_{x}}q(\mathbf{k})]$
is pure imaginary. It is easy to see
$\text{tr}[q^{-1}\partial_{k_{x}}q]^{*}=-\text{tr}[q^{\dag}\partial_{k_{x}}q^{\dag-1}]$.
From the eigen equation of $Q(\mathbf{k})$, we find that
$qq^{\dag}|\psi_{n}\rangle=E_{n}^{2}|\psi_{n}\rangle$ which leads to
the identity $qq^{\dag}\Psi=\Psi\Pi$, where
$\Pi=\text{diag}\{E_{1}^{2},E_{2}^{2}\}$ and the unitary matrix
$\Psi=(|\psi_{1}\rangle,|\psi_{2}\rangle)$. Therefore, we obtain
$\text{tr}[q^{\dag}\partial_{k_{x}}q^{\dag-1}]=\text{tr}[q^{-1}\partial_{k_{x}}q]+\text{tr}[\Pi\partial_{k_{x}}\Pi^{-1}]$
and accordingly,
\begin{equation}
\begin{split}
\int_{-\pi}^{\pi}dk_{x}\text{tr}[q^{-1}\partial_{k_{x}}q]^{*}=&-\int_{-\pi}^{\pi}dk_{x}\text{tr}[q^{-1}\partial_{k_{x}}q]\\
&-\int_{-\pi}^{\pi}dk_{x}\text{tr}[\Pi\partial_{k_{x}}\Pi^{-1}].\\
\end{split}
\end{equation}
Due to the periodic boundary condition, we have
$E_{n}(k_{x}=-\pi,k_{y})=E_{n}(k_{x}=\pi,k_{y})$ so that
\begin{equation}
\begin{split}
\int_{-\pi}^{\pi}dk_{x}\text{tr}[\Pi\partial_{k_{x}}\Pi^{-1}]=-2\sum_{n=1}^{2}\int_{-\pi}^{\pi}dk_{x}\partial_{k_{x}}\ln
E_{n}(\mathbf{k})=0.
\end{split}
\end{equation}
Thus
$\int_{-\pi}^{\pi}dk_{x}\text{tr}[q^{-1}\partial_{k_{x}}q]^{*}=-\int_{-\pi}^{\pi}dk_{x}\text{tr}[q^{-1}\partial_{k_{x}}q]$
is pure imaginary. Finally, the winding number for the chiral
symmetry Eq. (\ref{CS}) is obtained,
\begin{equation}
\label{Wky}
\begin{split}
\mathcal{W}(k_{y})=-\frac{1}{2\pi
i}\int_{-\pi}^{\pi}dk_{x}\text{tr}[q^{-1}(\mathbf{k})\partial_{k_{x}}q(\mathbf{k})].
\end{split}
\end{equation}
When the Hamiltonian has partial particle-hole symmetry and chiral
symmetry simultaneously, we can find a relation between the Pfaffian
invariant $\mathcal{P}(k_{y})$ and the winding number
$\mathcal{W}({k_{y}})$. According to the reference
\cite{PhysRevB.82.134521},
$\frac{1}{\pi}\int_{-\pi}^{\pi}A_{x}^{-}(\mathbf{k})=\frac{1}{2\pi
i}\int_{-\pi}^{\pi}\text{tr}[q(\mathbf{k})^{-1}\partial_{k_{x}}q(\mathbf{k})]+2N$,
where $N$ is an integer. Substitute this into  Eq. (\ref{PkySky}),
we get that $\mathcal{P}(k_{y})=(-1)^{\mathcal{W}(k_{y})}$.
Therefore, the Pfaffian invariant $\mathcal{P}(k_{y})$ is the parity
of the winding number $\mathcal{W}(k_{y})$.

If the Hamiltonian has partial chiral symmetry, for example, the
chiral-$k_{y}$ symmetry Eq. (\ref{CkyS}), then we can only define
the winding number $\mathcal{W}(k_{y})$ at $k_{y}=0$ and
$k_{y}=\pi$. Consequently, we can associate a topological invariant
$\mathcal{W}$ with the chiral-$k_{y}$ symmetry as
\cite{PhysRevB.82.134521}
\begin{equation}
\begin{split}
\mathcal{W}=(-1)^{\mathcal{W}(0)-\mathcal{W}(\pi)}.
\end{split}
\end{equation}
The topological invariant $\mathcal{W}$ is also the parity of the
Chern number, $\mathcal{W}=(-1)^{\mathcal{C}}$. Therefore, the
Pfaffian invariant $\mathcal{P}$ for the particle-hole symmetry is
equivalent to the topological invariant $\mathcal{W}$ for the
partial chiral symmetry.

\section{phase diagrams of the BdG Hamiltonian}\label{pdofBdG}

\begin{figure}
\begin{tabular}{cc}
\includegraphics[width=4cm]{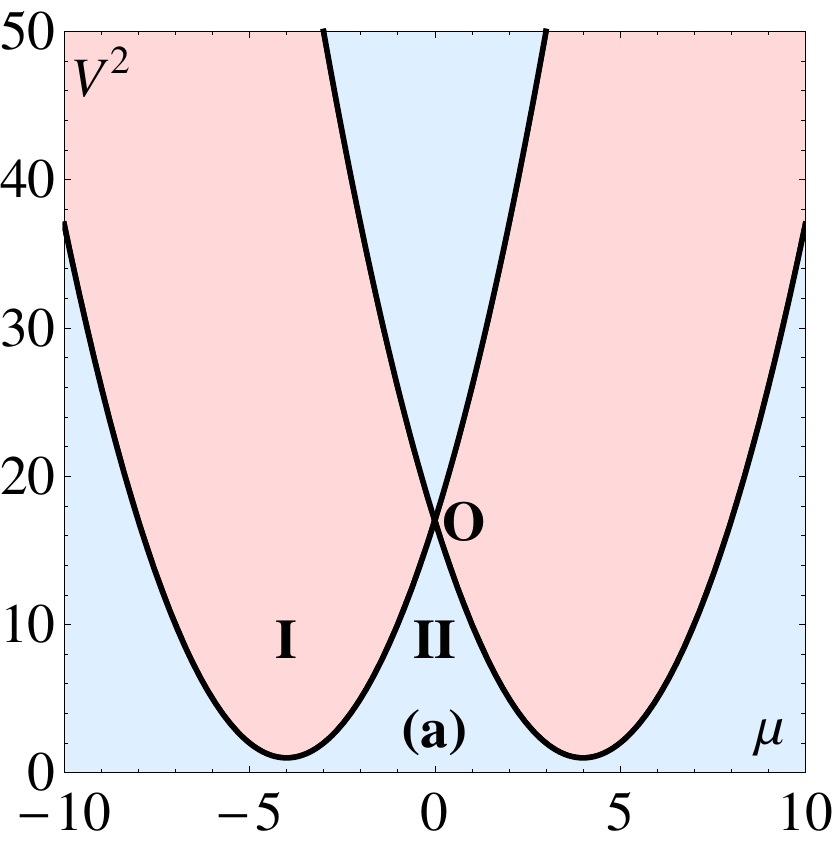} &
\includegraphics[width=4cm]{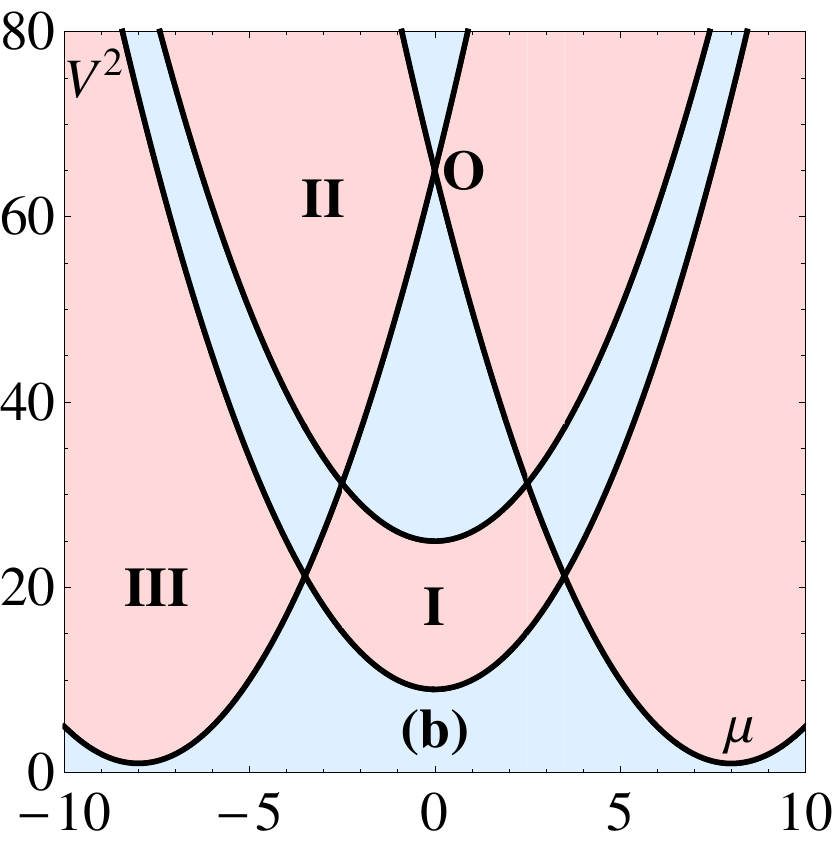} \\
\includegraphics[width=4cm]{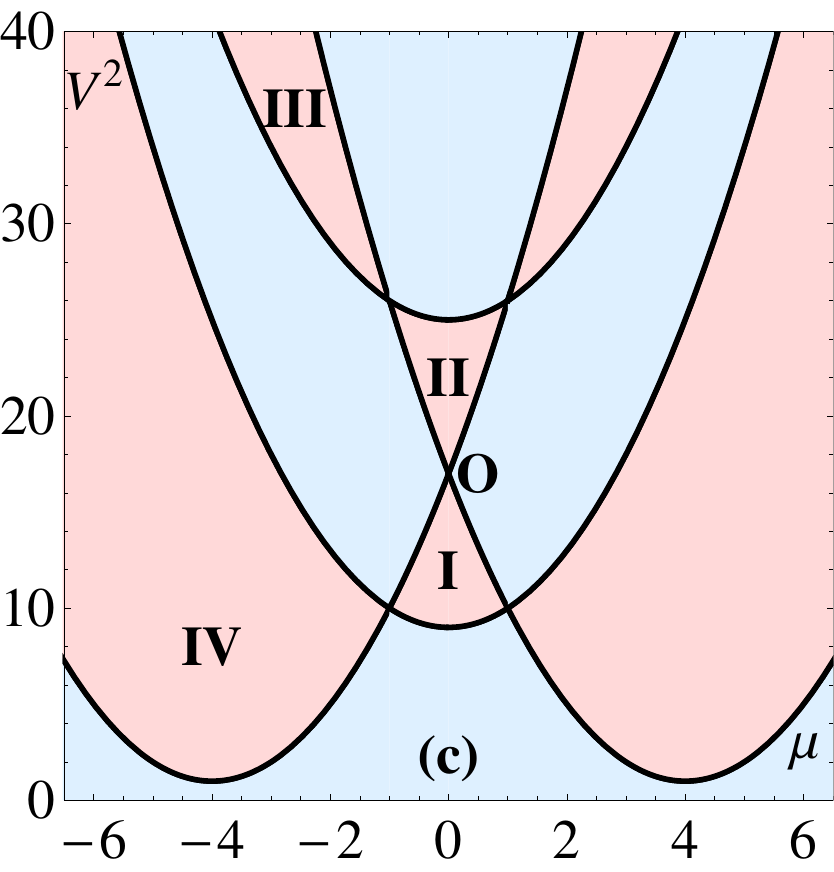} &
\includegraphics[width=4cm]{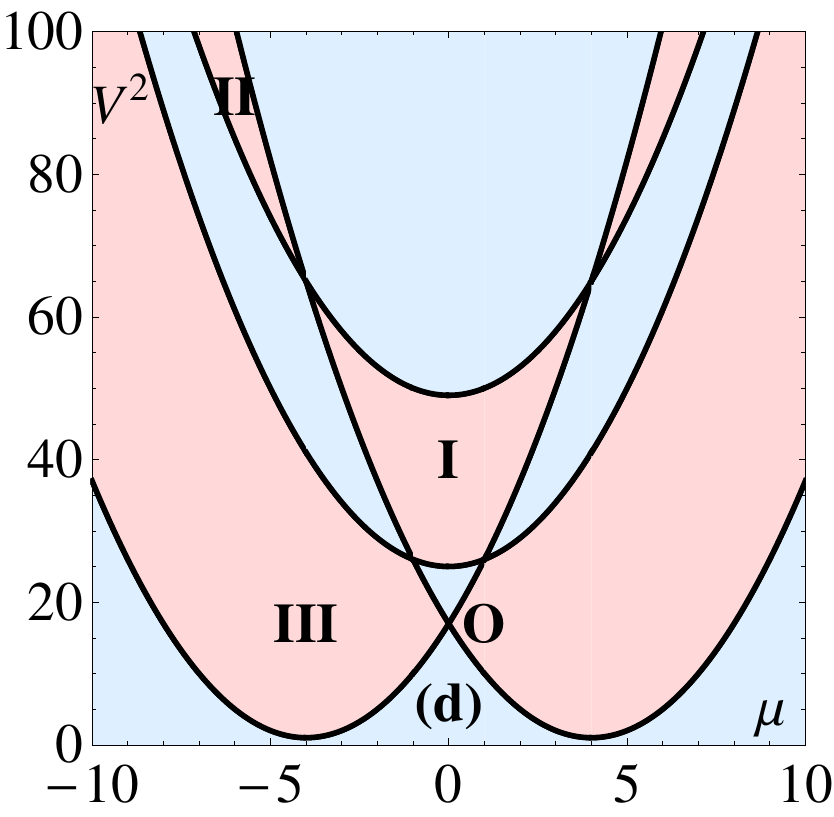} \\
\end{tabular}
\caption{(color online). The possible phase diagrams of the
spin-singlet superconductor with Rashba and Dresselhaus (110)
spin-orbit couplings. (a) is the phase diagram for the pure $s$-wave
or $d$-wave superconductor. (b), (c) and (d) are the phase diagrams
for the mixed $s$-wave and $d$-wave superconductor.}\label{phd}
\end{figure}

In contrast to the even number of Majorana bound states in the
trivial topological phase, the number of Majorana bound states is
odd in the nontrivial topological phase. The Pfaffian invariant
$\mathcal{P}$ is in fact the parity of the number of Majorana bound
states. Therefore, we can use the Pfaffian invariant $\mathcal{P}$
to investigate the topological quantum phase transitions in the BdG
Hamiltonian Eq. (\ref{HBdG}). The phase diagrams are shown in Fig.
(\ref{phd}). Our interest is in the red region where the Pfaffian
invariant $\mathcal{P}=-1$ which means that the system has an odd
number of Majorana bound states at the edge and is thus in the
nontrivial topological phase. The explicit expression of the
Pfaffian invariant Eq. (\ref{defP}) for the general case of the BdG
Hamiltonian is
\begin{widetext}
\begin{equation}
\begin{split}
\mathcal{P}=\text{sgn}\left\{\frac{[(\mu+4t)^{2}+(\Delta_{s_{1}}^{2}+\Delta_{s_{2}}^{2})-V^{2}][(\mu-4t)^{2}+(\Delta_{s_{1}}^{2}+\Delta_{s_{2}}^{2})-V^{2}]}{[\mu^{2}+(\Delta_{s_{1}}+2\Delta_{d_{1}})^{2}+\Delta_{s_{2}}^{2}-V^{2}][\mu^{2}+(\Delta_{s_{1}}-2\Delta_{d_{1}})^{2}+\Delta_{s_{2}}^{2}-V^{2}]}\right\}.
\end{split}
\end{equation}
\end{widetext}
Therefore, the phase diagram is divided by the following four
parabolas in the plane of $V^{2}\sim\mu$: (i)
$V^{2}=(\mu+4t)^{2}+(\Delta_{s_{1}}^{2}+\Delta_{s_{2}}^{2})$; (ii)
$V^{2}=(\mu-4t)^{2}+(\Delta_{s_{1}}^{2}+\Delta_{s_{2}}^{2})$; (iii)
$V^{2}=\mu^{2}+(\Delta_{s_{1}}+2\Delta_{d_{1}})^{2}+\Delta_{s_{2}}^{2}$
and (iv)
$V^{2}=\mu^{2}+(\Delta_{s_{1}}-2\Delta_{d_{1}})^{2}+\Delta_{s_{2}}^{2}$,
where $V^{2}=V_{x}^{2}+V_{y}^{2}+V_{z}^{2}$. Notice that the
Pfaffian invariant $\mathcal{P}$ has nothing to do with the
spin-orbit couplings. Thus the topological phases can exist even
without the spin-orbit couplings. However, the spin-orbit couplings
can open a gap to render the Majorana fermion located at the edge of
the system; otherwise the Majorana fermion will spread into the
bulk. Now we turn to discuss all the possible phase diagrams in the
BdG Hamiltonian. When $\Delta_{s_{1}}\Delta_{d_{1}}=0$, the phase
diagram is only divided by the parabolas (i) and (ii) and is shown
in Fig. (\ref{phd}a). When $\Delta_{s_{1}}\Delta_{d_{1}}\ne0$, there
are three topologically different cases in the phase diagrams as
follows. Let us first define the intersection point of the parabolas
(i) and (ii) as $O$, then the phase diagram where the parabolas
(iii) and (iv) are both below $O$ is shown in Fig. (\ref{phd}b); the
phase diagram where the parabolas (iii) and (iv) are on either side
of $O$ is shown in Fig. (\ref{phd}c); the phase diagram where the
parabolas (iii) and (iv) are both above $O$ is shown in Fig.
(\ref{phd}d). Furthermore, if we assume
$\Delta_{s_{1}}\Delta_{d_{1}}>0$, then the phase diagram is as Fig.
(\ref{phd}b) when
$\Delta_{d_{1}}^{2}-\Delta_{s_{1}}\Delta_{d_{1}}<\Delta_{d_{1}}^{2}+\Delta_{s_{1}}\Delta_{d_{1}}<4t^{2}$;
the phase diagram is as Fig. (\ref{phd}c) when
$\Delta_{d_{1}}^{2}-\Delta_{s_{1}}\Delta_{d_{1}}<4t^{2}<\Delta_{d_{1}}^{2}+\Delta_{s_{1}}\Delta_{d_{1}}$;
the phase diagram is as Fig. (\ref{phd}d) when
$4t^{2}<\Delta_{d_{1}}^{2}-\Delta_{s_{1}}\Delta_{d_{1}}<\Delta_{d_{1}}^{2}+\Delta_{s_{1}}\Delta_{d_{1}}$.
Therefore, we have exhibited all the possible phase diagrams in the
BdG Hamiltonian Eq. (\ref{HBdG}) in this paper. For the pure
$s$-wave and $d$-wave superconductors, the phase diagrams are
topologically equivalent to Fig. (\ref{phd}a); for the
superconductors with mixed $s$-wave and $d$-wave pairing symmetries,
the phase diagrams are topologically equivalent to Fig.
(\ref{phd}b), Fig. (\ref{phd}c) and Fig. (\ref{phd}d) depending on
$t$.

\section{Majorana bound states at the edge of the BdG Hamiltonian}\label{MBSofBdG}

In this Section, we shall demonstrate the Majorana bound states at
the edge of the spin-singlet superconductor in the different cases
as listed in Tab. (\ref{symmandti}). By setting the boundary
conditions of the $x$ direction to be open and the $y$ direction to
be periodic, we diagonalize the Hamiltonian Eq. (\ref{HBdG}) with
the cylindrical symmetry and get the edge spectrum of the
Hamiltonian.

\begin{figure}
\begin{tabular}{c}
\includegraphics[width=8cm]{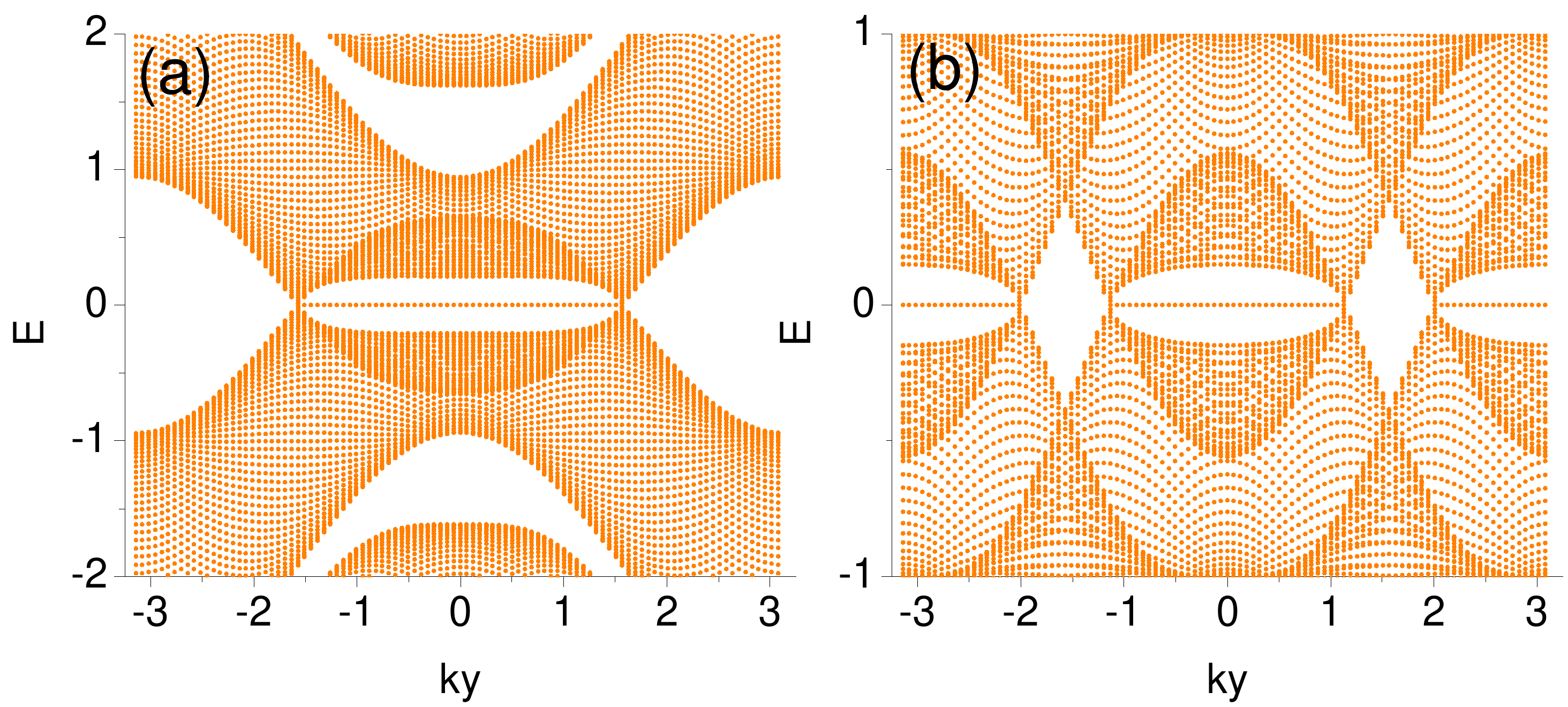}
\end{tabular}
\begin{tabular}{cc}
\includegraphics[width=4cm]{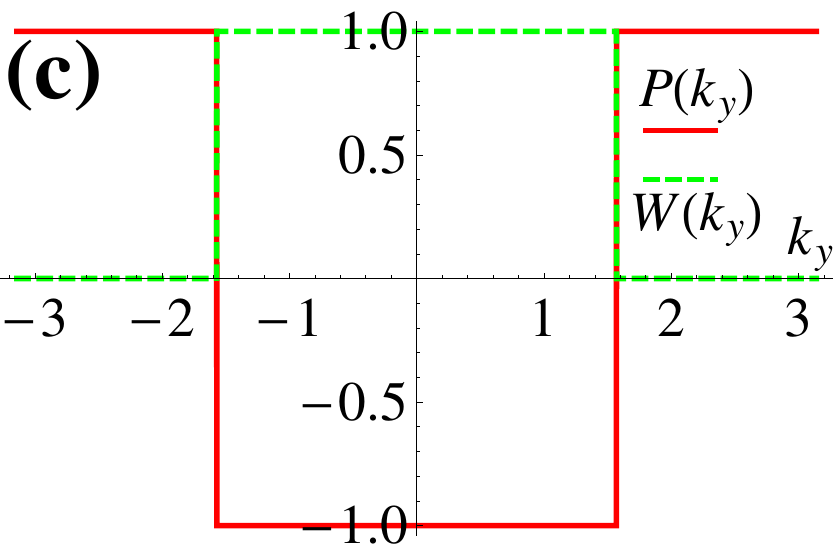} &
\includegraphics[width=4cm]{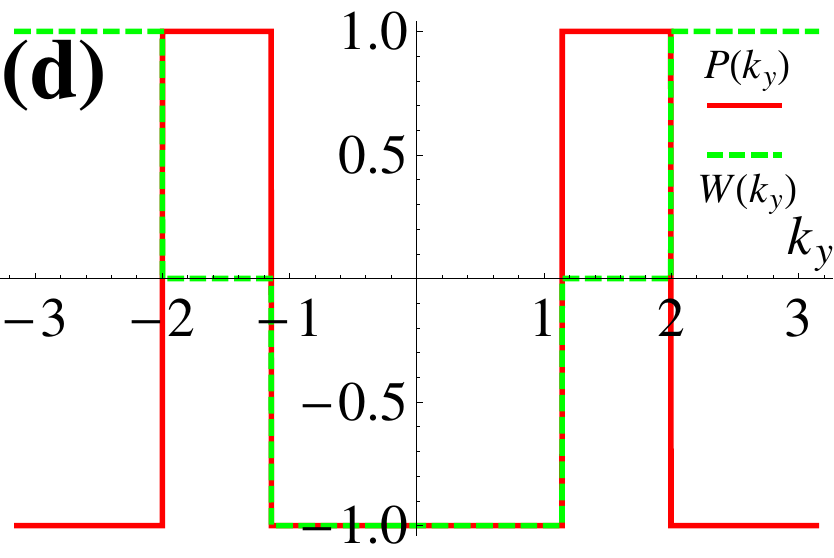} \\
\end{tabular}
\caption{(color online). (a) and (b) are the edge spectra of the
$s$-wave superconductor with Dresselhaus (110) spin-orbit coupling
in case (b) of Tab. (\ref{symmandti}). The open edges are at
$i_{x}=0$ and $i_{x}=50$, $k_{y}$ denotes the momentum in the $y$
direction and $k_{y}\in(-\pi,\pi]$. The parameters are $t=1$,
$\beta=1$, $\Delta_{s_{1}}=1$, $\Delta_{s_{2}}=0$ and (a) $\mu=-4,
V^{2}=5$, (b) $\mu=0, V^{2}=9$, which correspond to regions I and II
in Fig. (\ref{phd}a), respectively. (c) and (d) are the Pfaffian
invariant given by Eq. (\ref{Pky}) and winding number given by Eq.
(\ref{Wky}) for (a) and (b). This figure is cited from the reference
\cite{PhysRevB.87.054501} for completeness.}\label{caseb}
\end{figure}

\begin{figure}
\begin{tabular}{cc}
\includegraphics[width=4cm]{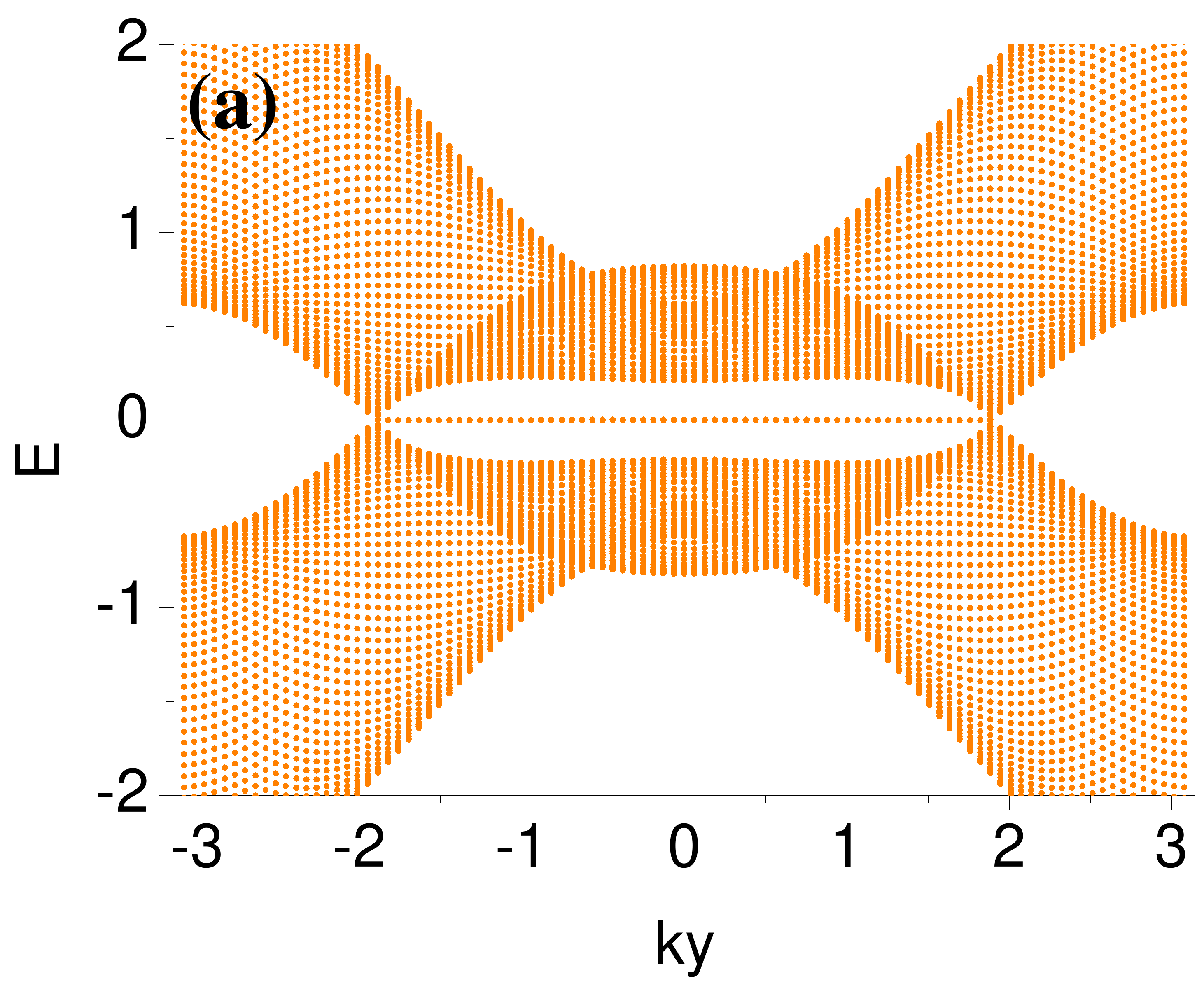} &
\includegraphics[width=4cm]{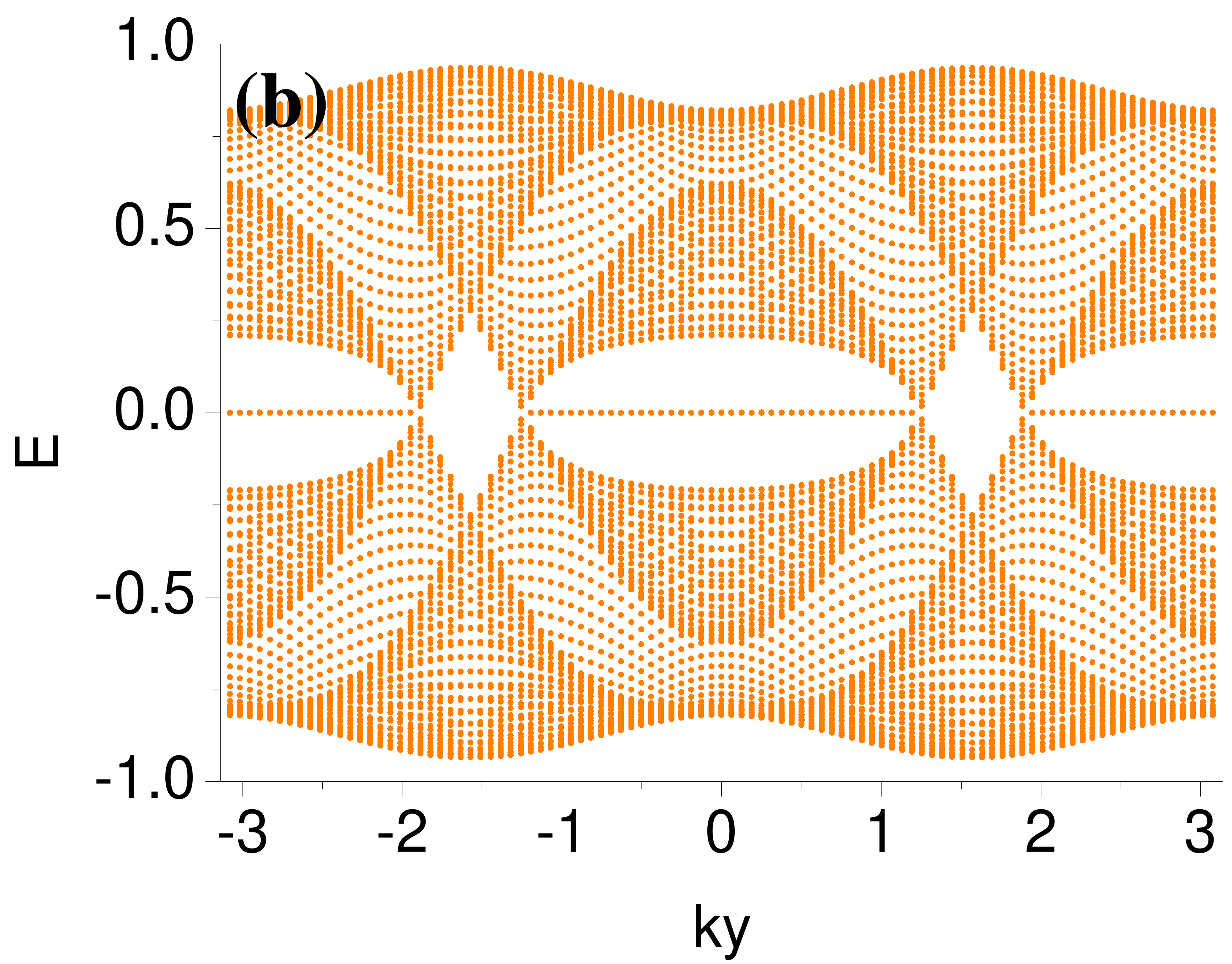} \\
\includegraphics[width=4cm]{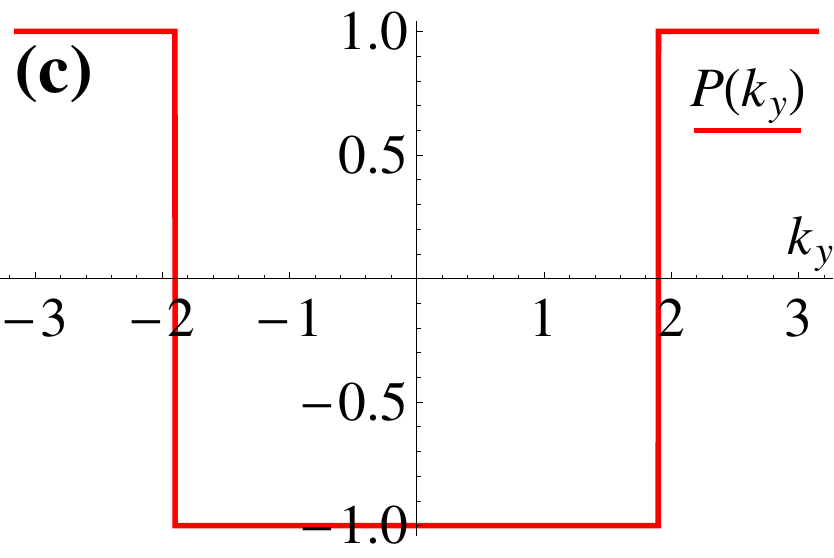} &
\includegraphics[width=4cm]{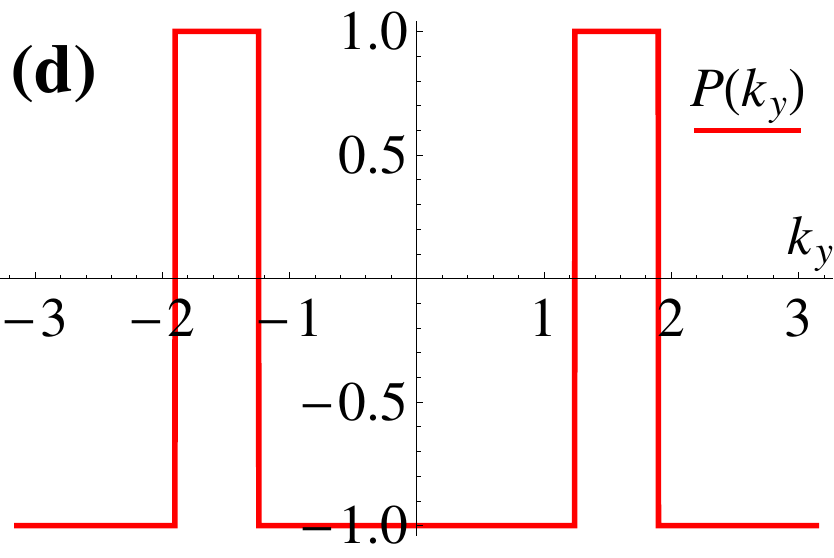} \\
\end{tabular}
\caption{(color online). (a) and (b) are the edge spectra of the
$s$-wave superconductor with Dresselhaus (110) spin-orbit coupling
in case (c) of Tab. (\ref{symmandti}). The parameters are $t=1$,
$\beta=1$, $\Delta_{s_{1}}=1$, $\Delta_{s_{2}}=1$ and (a) $\mu=-4,
V^{2}=9$, (b) $\mu=0, V^{2}=9$, which correspond to regions I and II
in Fig. (\ref{phd}a), respectively. (c) and (d) are the Pfaffian
invariant for (a) and (b).}\label{casec}
\end{figure}

We first discuss the pure $s$-wave superconductor in case (b) and
case (c). The difference between case (b) and case (c) is that the
appearance of the superconducting parameter $\Delta_{s_{2}}$ in case
(c) lowers the symmetry by breaking the chiral symmetry.
Consequently, the number of topological invariants in case (c) is
less than the one in case (b). The phase diagrams of case (b) and
case (c) are topologically equivalent and are shown in Fig.
(\ref{phd}a). The edge spectra of case (b) and case (c) are
exhibited in Fig. (\ref{caseb}) and Fig. (\ref{casec}),
respectively. We observe that the Majorana bound states in these two
cases are flat bands. There are odd number of Majorana flat bands in
the nontrivial topological phase. Although the edge spectra of these
two cases are similar, the symmetries and the topological invariants
protecting them are different as shown in Tab. (\ref{symmandti}). In
case (b), we can use $\mathcal{P}$ and $\mathcal{W}$ to find the
regions in the phase diagram where there exists nontrivial Majorana
flat band and use $\mathcal{P}({k_{y}})$ and $\mathcal{W}({k_{y}})$
to locate the range of flat band in the edge Brillouin zone as shown
in Fig. (\ref{caseb}c) and Fig. (\ref{caseb}d). However, in case
(c), only the Pfaffian invariant $\mathcal{P}$ and
$\mathcal{P}({k_{y}})$ can be used in finding the Majorana flat
bands. We find that when $\mathcal{P}(k_{y})=-1$ or
$\mathcal{W}(k_{y})$ is odd, the topologically nontrivial Majorana
flat bands will emerge in the edge Brillouin zone.

\begin{figure}
\begin{tabular}{cc}
\includegraphics[width=4cm]{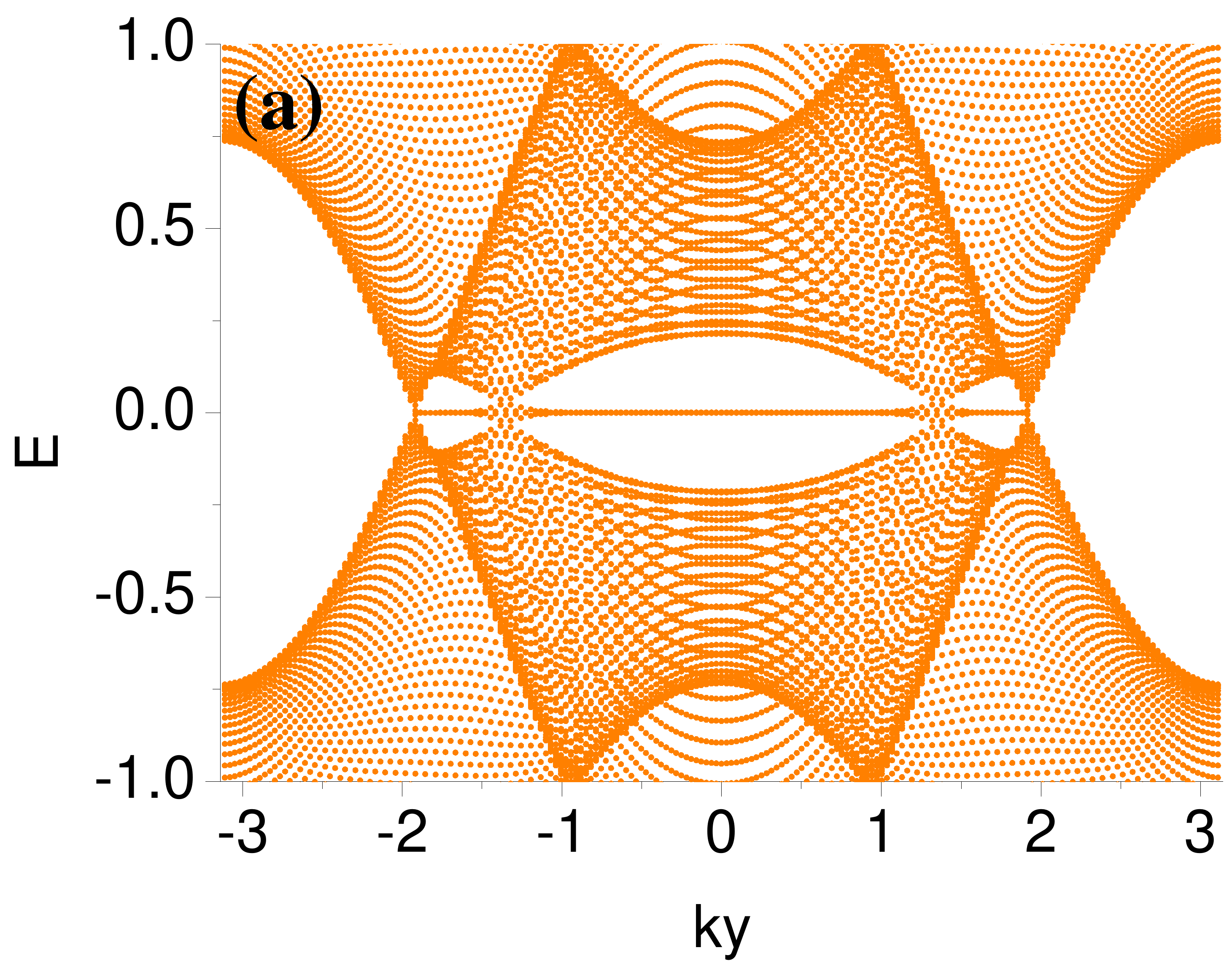} &
\includegraphics[width=4cm]{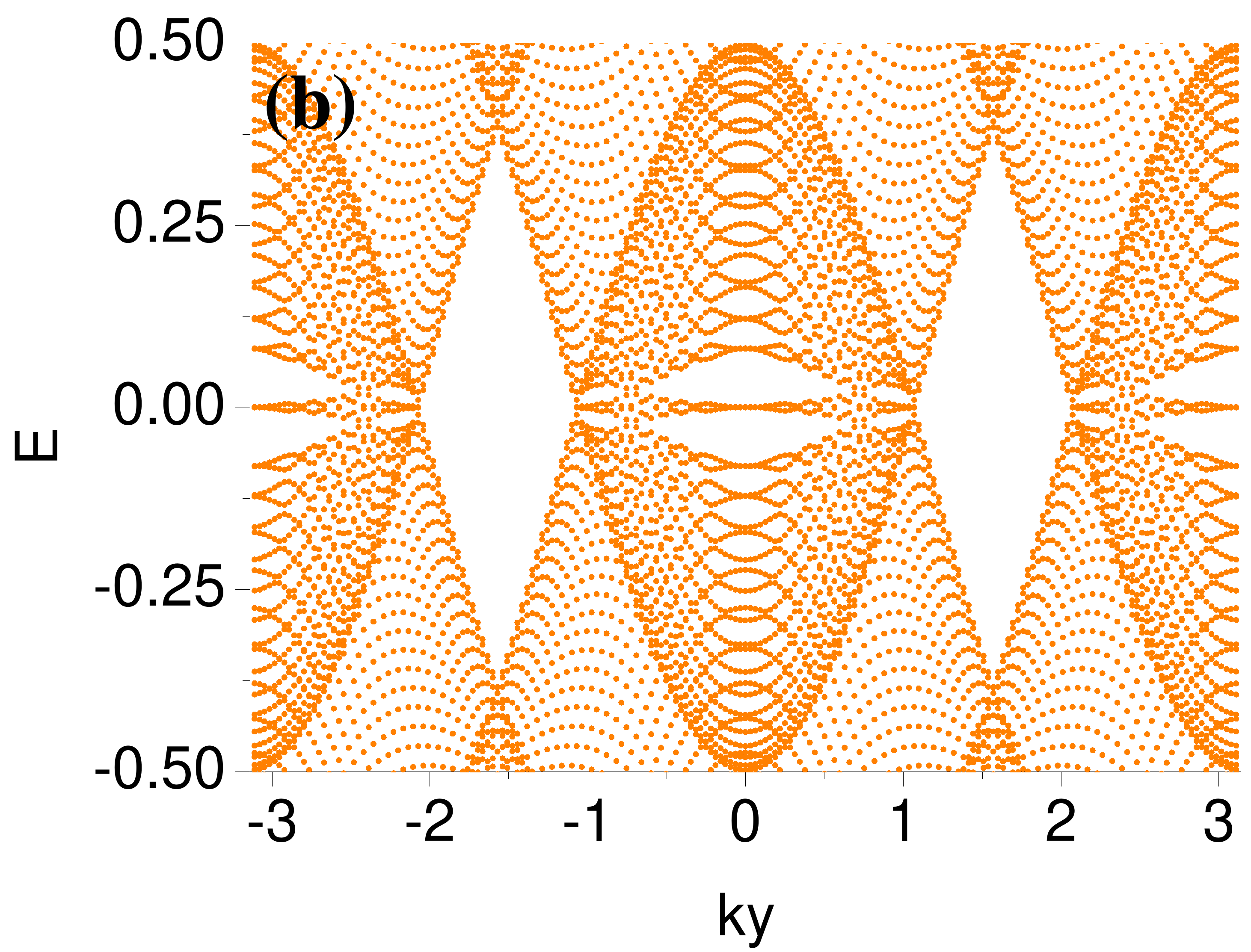} \\
\includegraphics[width=4cm]{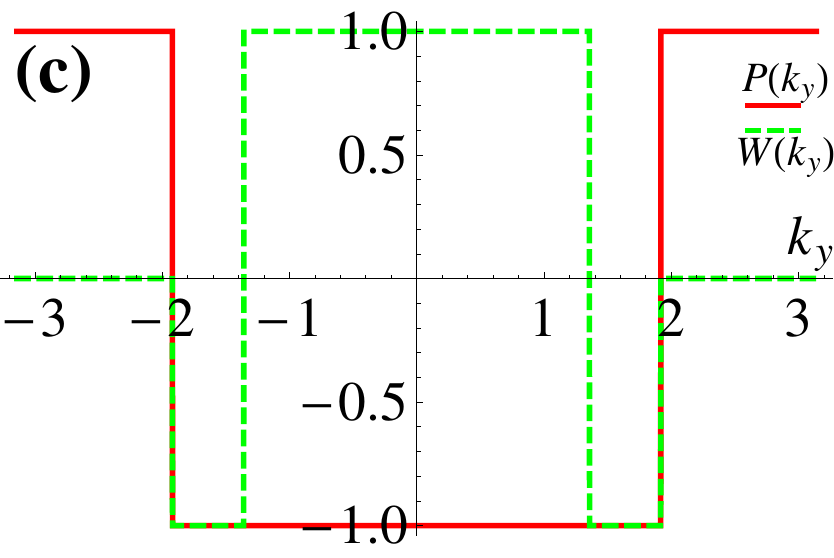} &
\includegraphics[width=4cm]{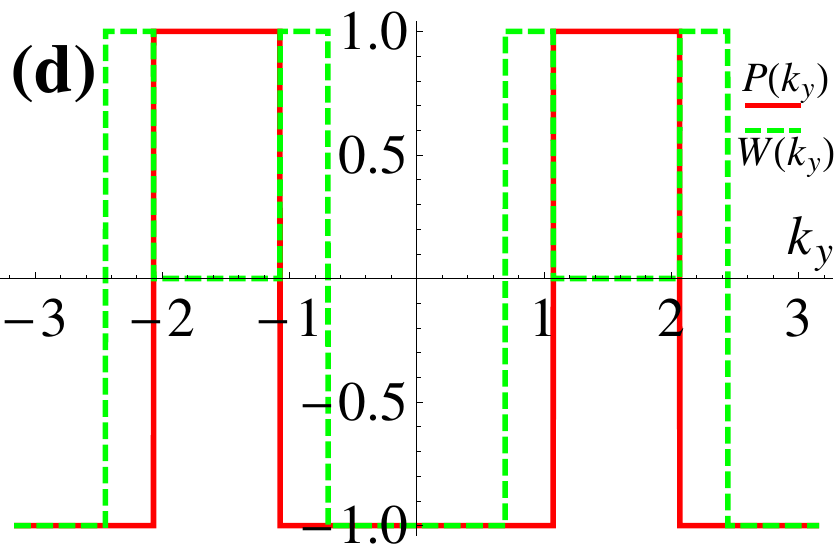} \\
\end{tabular}
\caption{(color online). (a) and (b) are the edge spectra of the
$d_{x^2-y^2}$-wave superconductor with Dresselhaus (110) spin-orbit
coupling in case (d) of Tab. (\ref{symmandti}). The parameters are
$t=1$, $\beta=1$, $\Delta_{d_{1}}=1$, $\Delta_{d_{2}}=0$ and (a)
$\mu=-4, V^{2}=9$, (b) $\mu=0, V^{2}=9$, which correspond to regions
I and II in Fig. (\ref{phd}a), respectively. (c) and (d) are the
Pfaffian invariant and winding number for (a) and (b).}\label{cased}
\end{figure}

\begin{figure}
\begin{tabular}{cc}
\includegraphics[width=4cm]{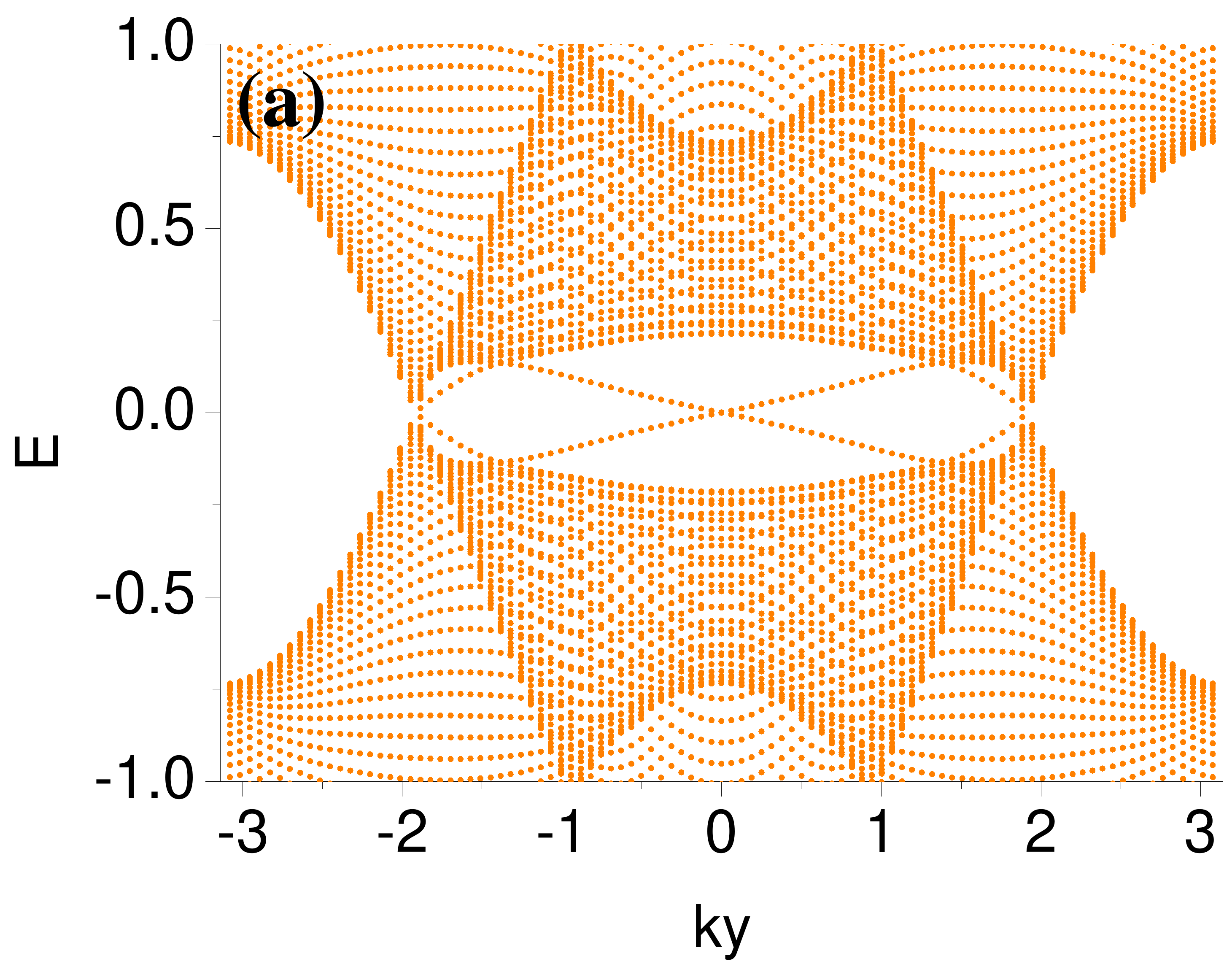} &
\includegraphics[width=4cm]{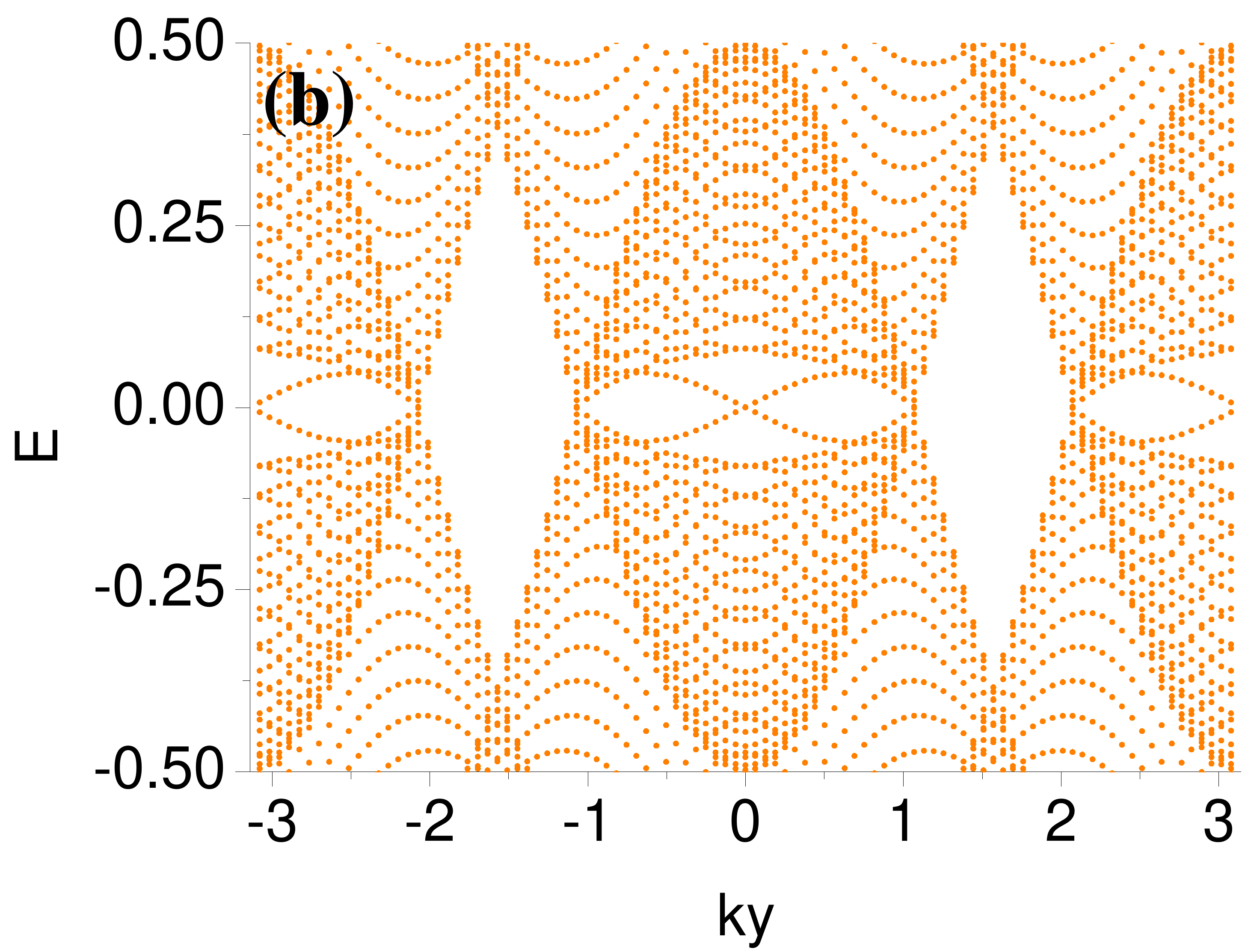} \\
\end{tabular}
\caption{(color online). (a) and (b) are the edge spectra of the
$d_{x^2-y^2}+id_{xy}$-wave superconductor with Dresselhaus (110)
spin-orbit coupling in case (e) of Tab. (\ref{symmandti}). The
parameters are $t=1$, $\beta=1$, $\Delta_{d_{1}}=1$,
$\Delta_{d_{2}}=1$ and (a) $\mu=-4, V^{2}=9$, (b) $\mu=0, V^{2}=9$,
which correspond to regions I and II in Fig. (\ref{phd}a),
respectively.}\label{casee}
\end{figure}

For the pure $d$-wave superconductor in case (d) and (e), the phase
diagram is also shown in Fig. (\ref{phd}a). The edge spectra of the
$d_{x^2-y^2}$-wave superconductor are depicted in Fig.
(\ref{cased}). There are an odd number of Majorana flat bands in the
nontrivial topological phase as shown in Fig. (\ref{cased}a) and the
location of the Majorana flat bands is consistent with the Pfaffian
invariant $\mathcal{P}(k_{y})$ and the winding number
$\mathcal{W}(k_{y})$ as shown in Fig. (\ref{cased}c) and
(\ref{cased}d). Note that the winding number $\mathcal{W}(k_{y})$
can change by some even number in the same phase. However, its
parity, the Pfaffian invariant $\mathcal{P}(k_{y})$ is unchanged in
the same phase since $\mathcal{P}(k_{y})=(-1)^{\mathcal{W}(k_{y})}$.
The edge spectra of the $d_{x^2-y^2}+id_{xy}$-wave superconductor
are displayed in Fig. (\ref{casee}). We find that the Majorana bound
states in this case get dispersed and become Dirac cones. Similarly,
there is an odd number of Dirac cones in the nontrivial topological
phase as shown in Fig. (\ref{casee}a). Comparing case (d) with case
(e), we find that the Majorana flat bands can emerge only when the
Hamiltonian has a chiral or partial particle-hole symmetry since
these symmetries can provide the topological invariants
$\mathcal{P}(k_{y})$ or $\mathcal{W}(k_{y})$ to make the Majorana
flat bands stable in the edge Brillouin zone.

\begin{figure*}
\begin{tabular}{ccccc}
\includegraphics[width=3cm]{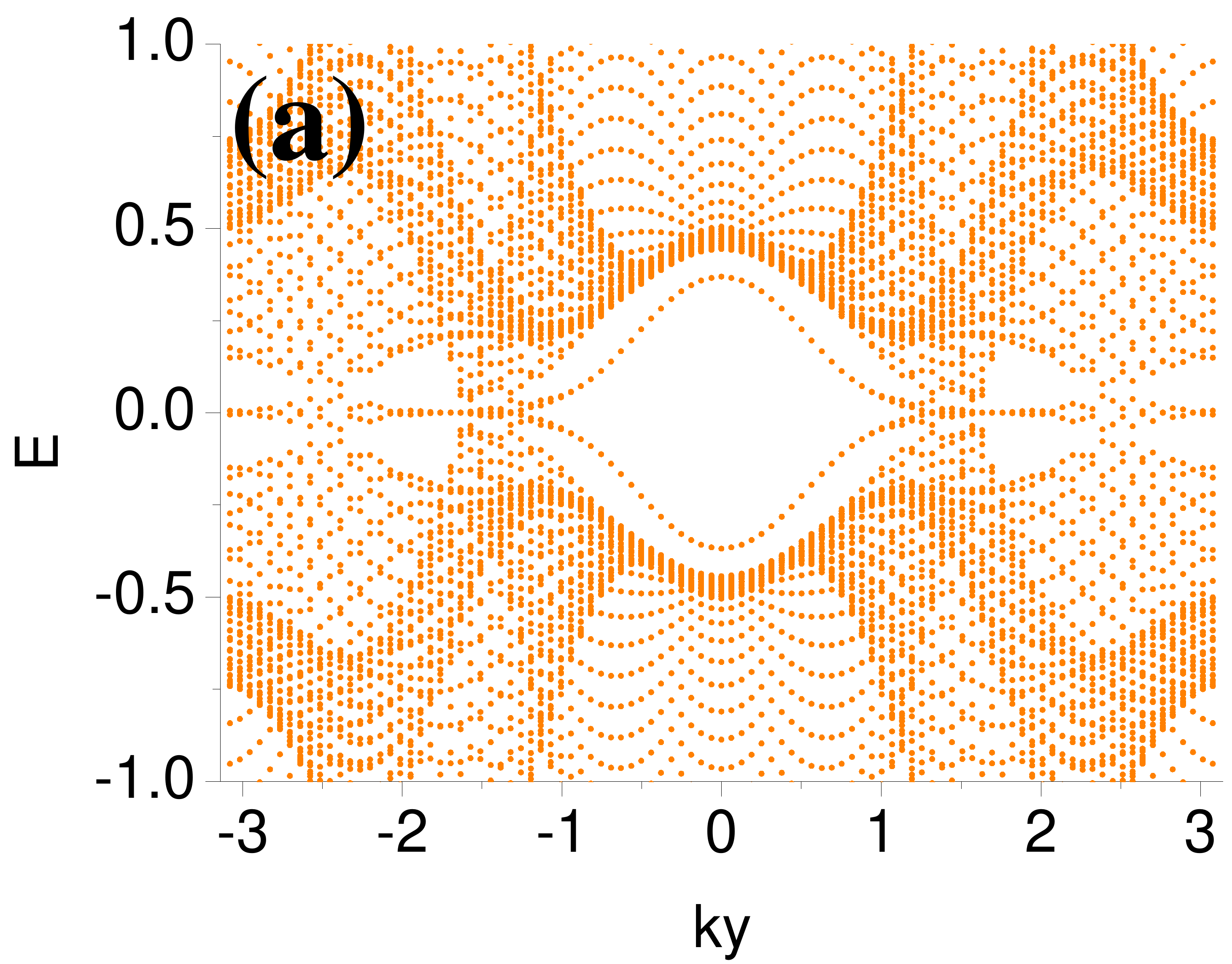} &
\includegraphics[width=3cm]{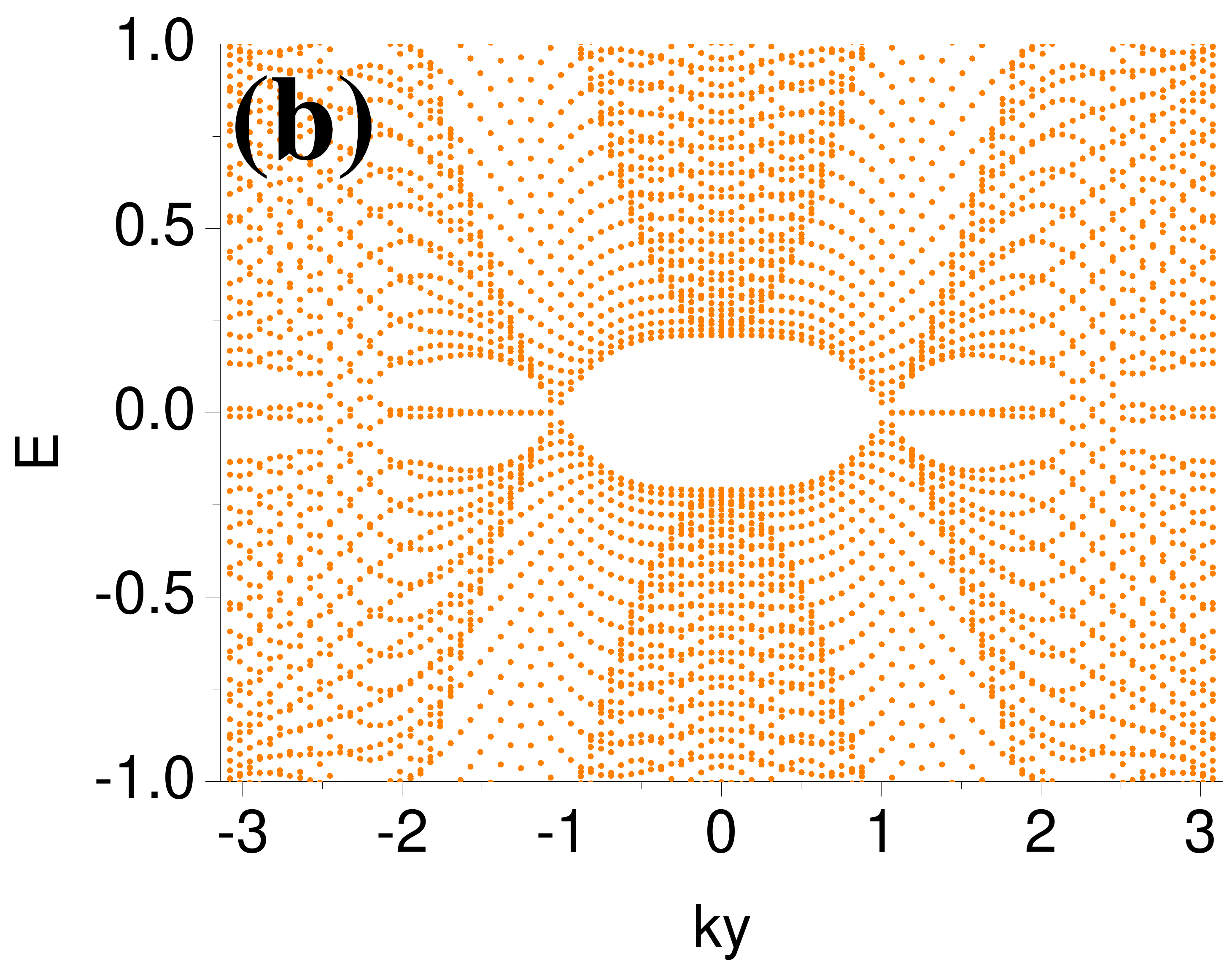} &
\includegraphics[width=3cm]{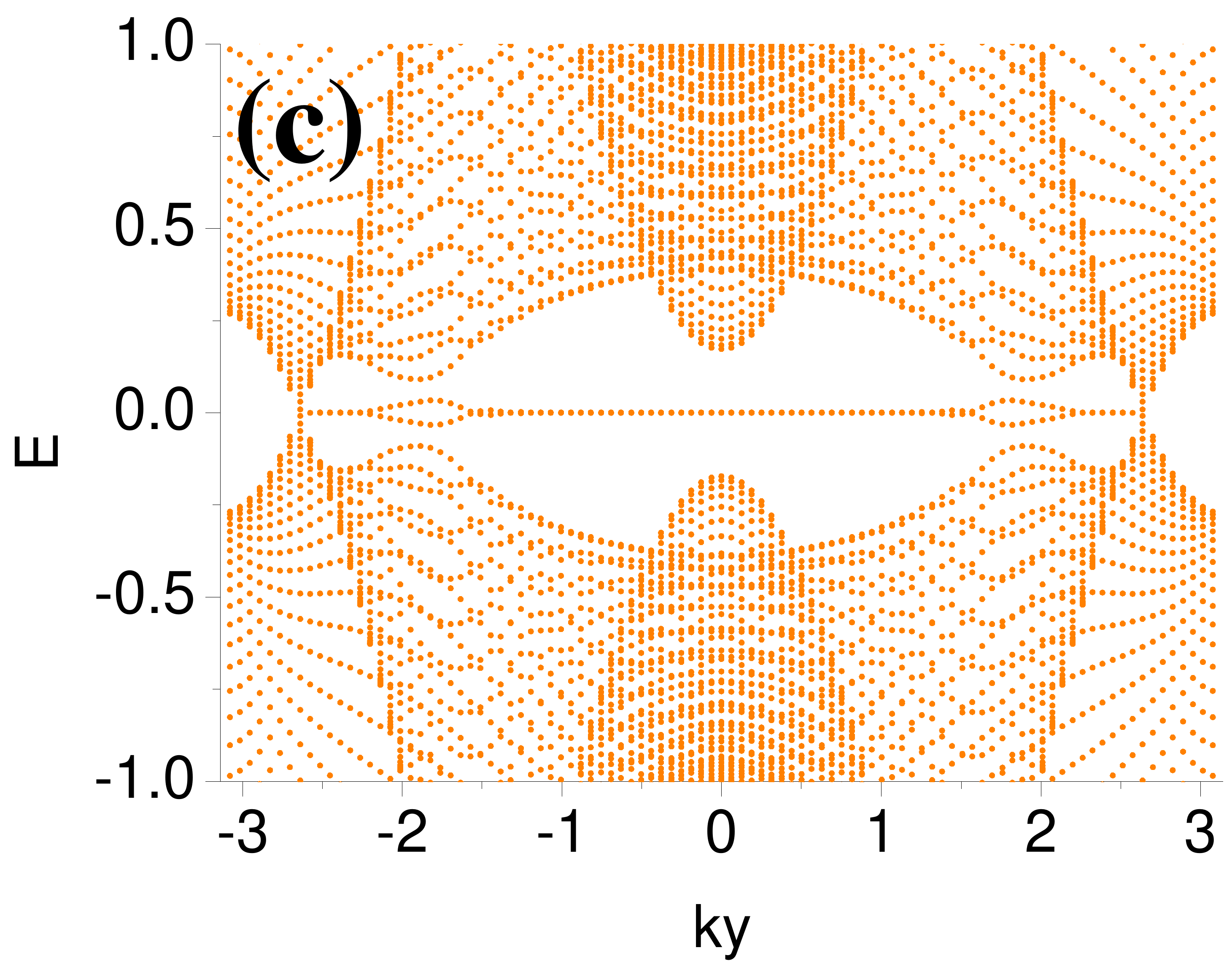} &
\includegraphics[width=3cm]{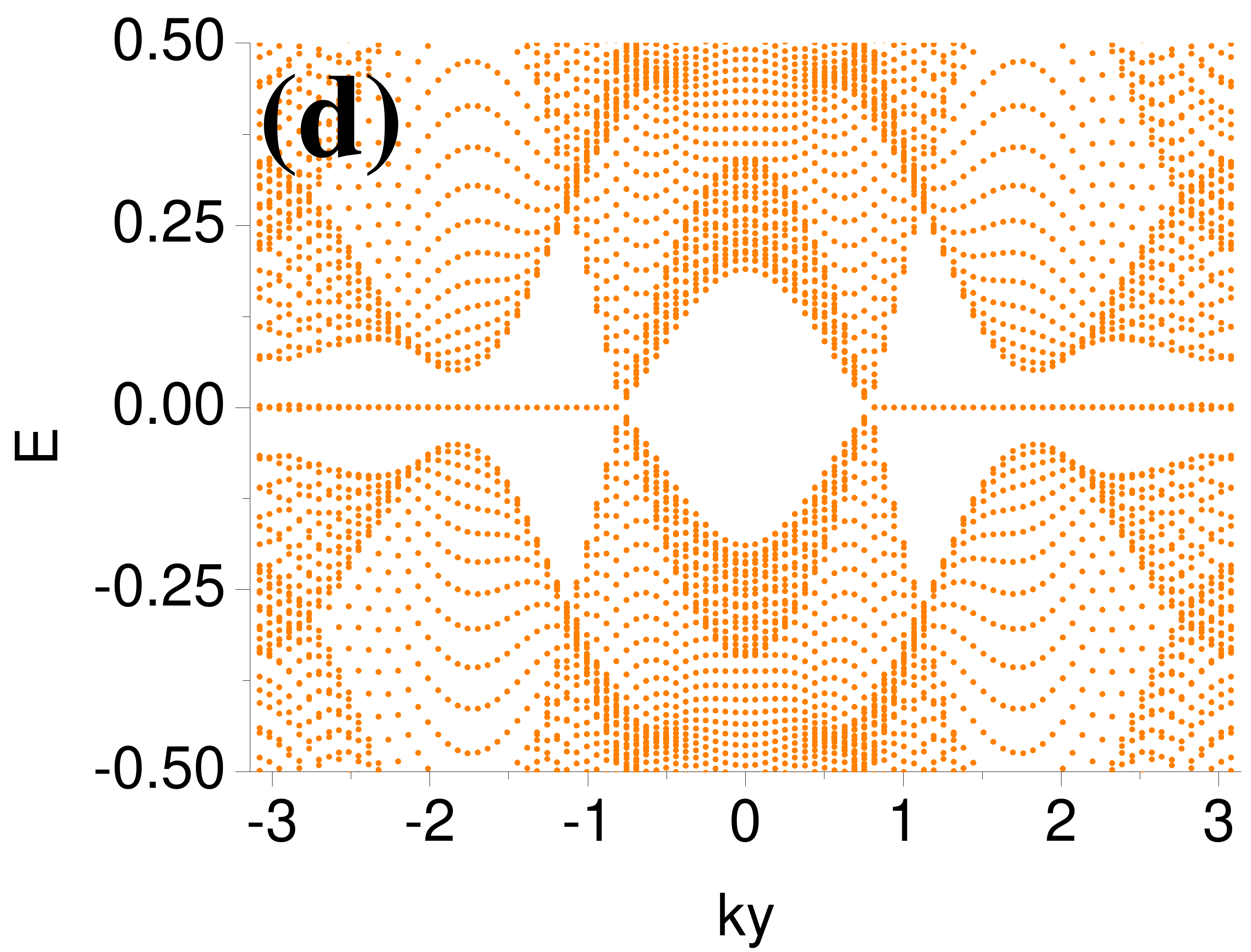} &
\includegraphics[width=3cm]{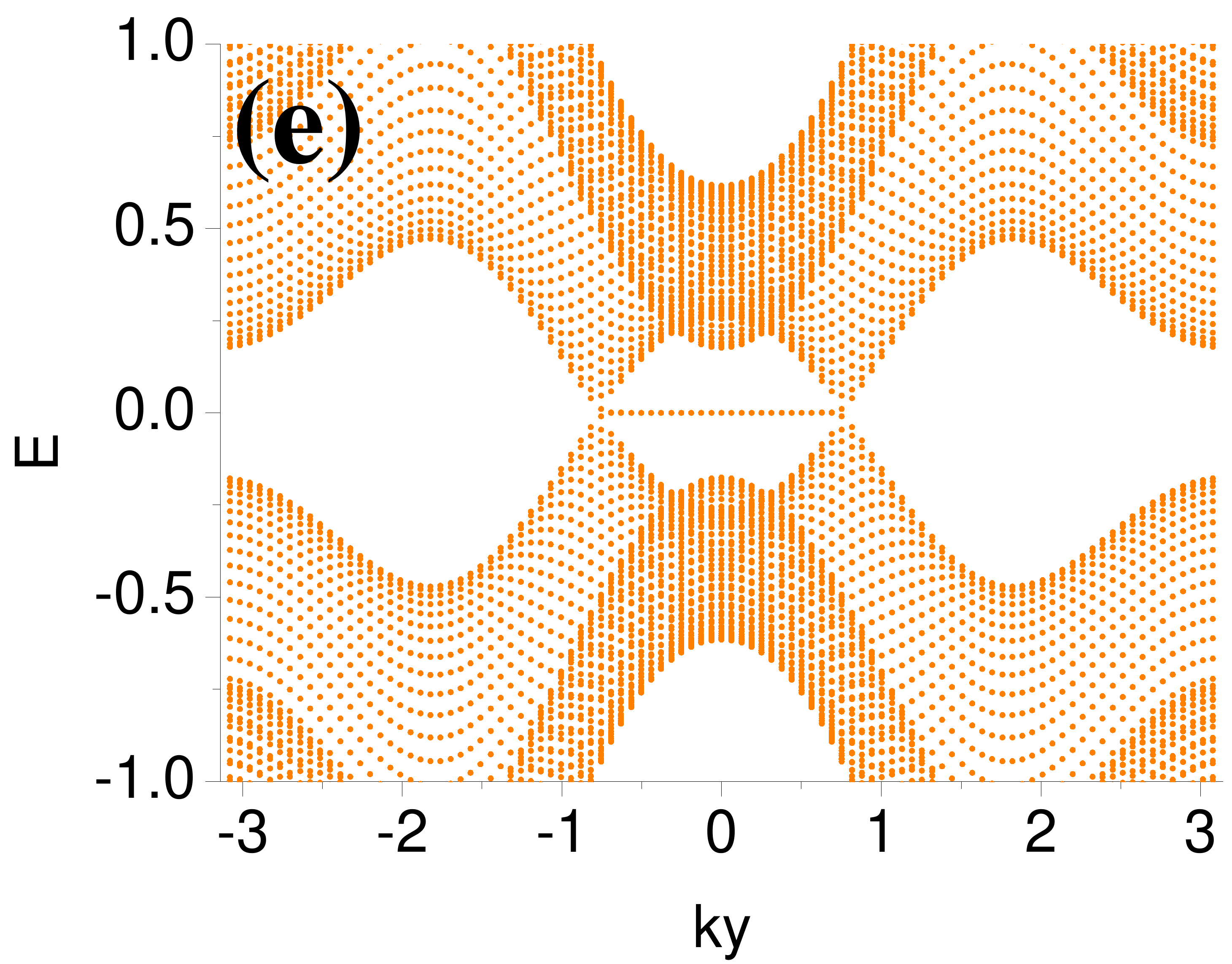} \\
\includegraphics[width=3cm]{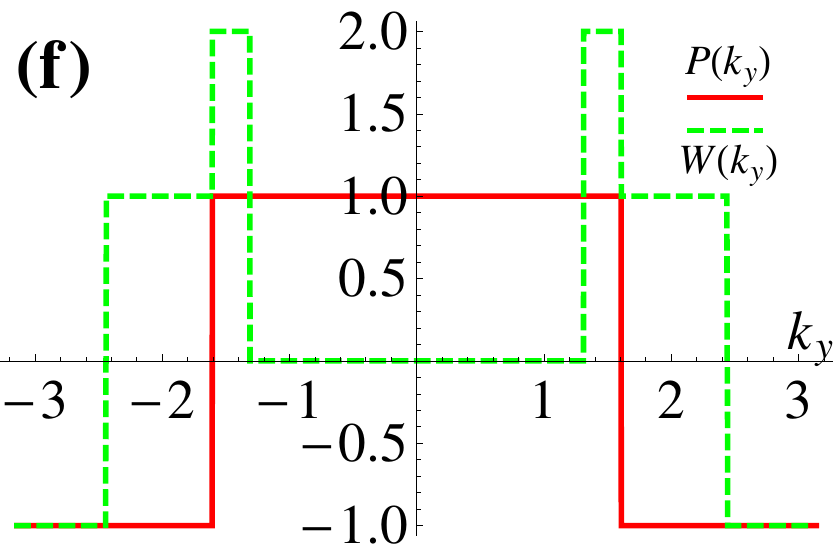} &
\includegraphics[width=3cm]{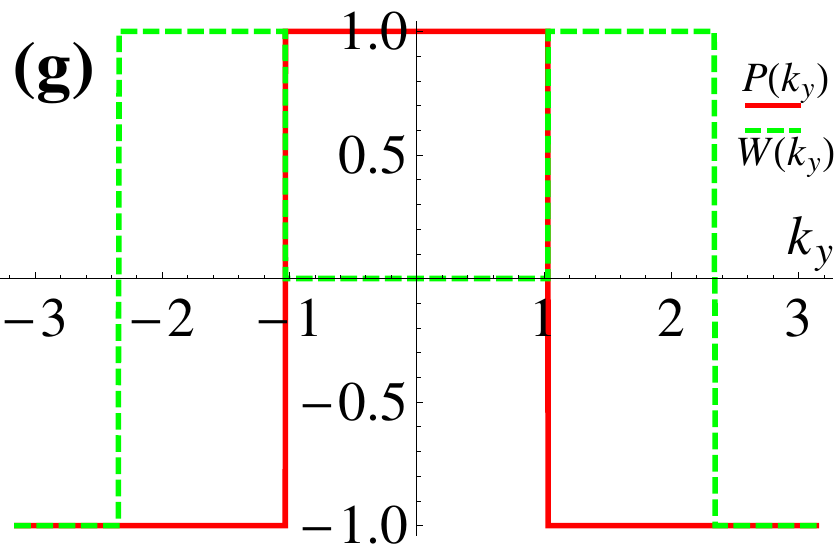} &
\includegraphics[width=3cm]{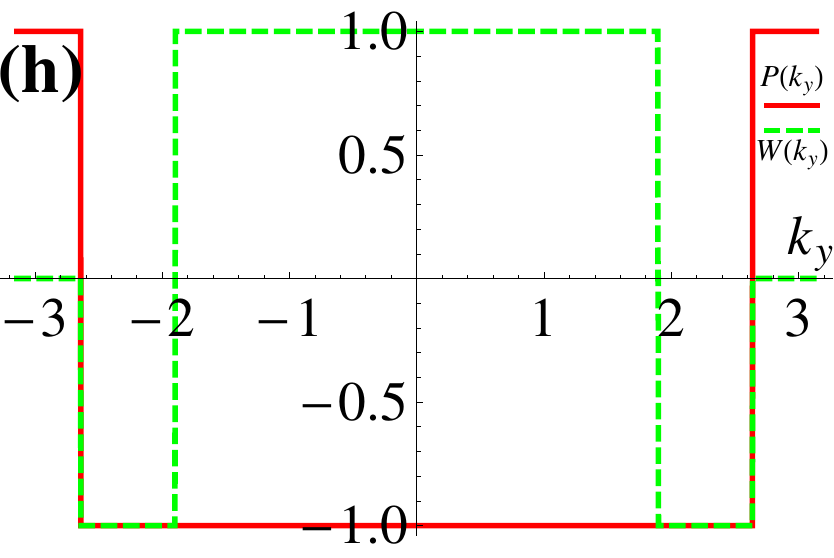} &
\includegraphics[width=3cm]{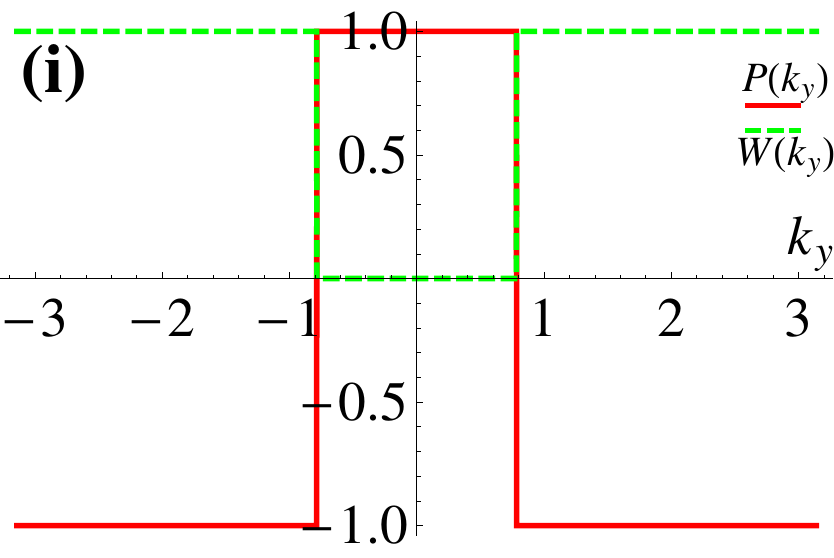} &
\includegraphics[width=3cm]{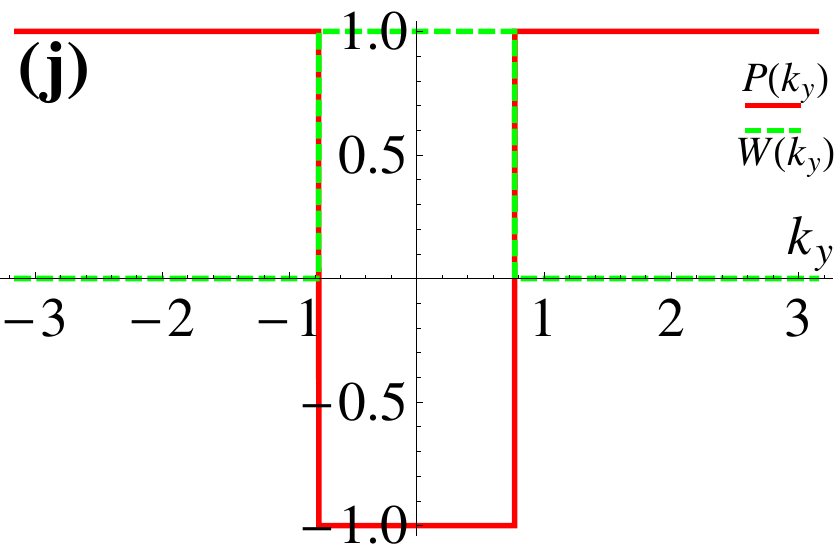} \\
\includegraphics[width=3cm]{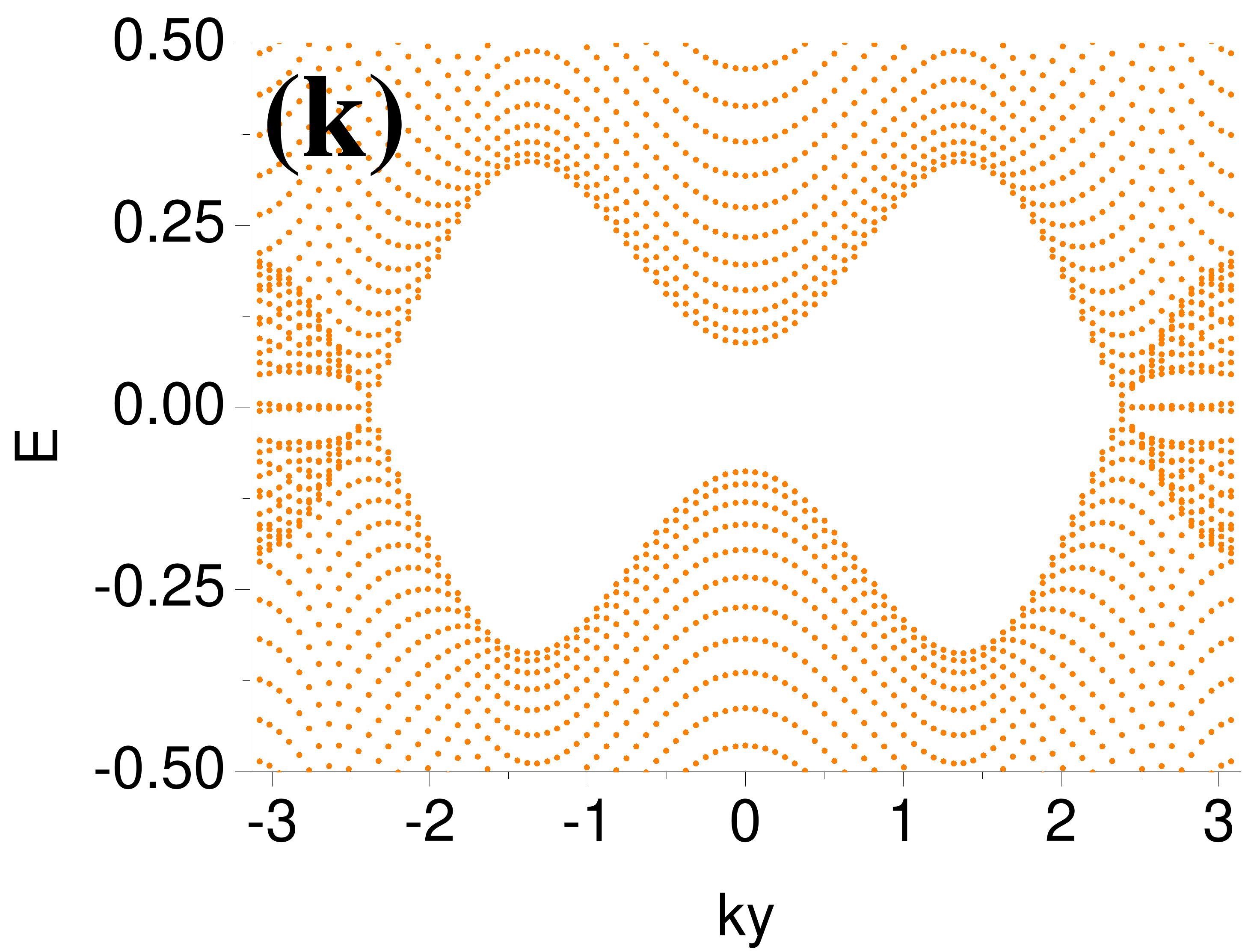} &
\includegraphics[width=3cm]{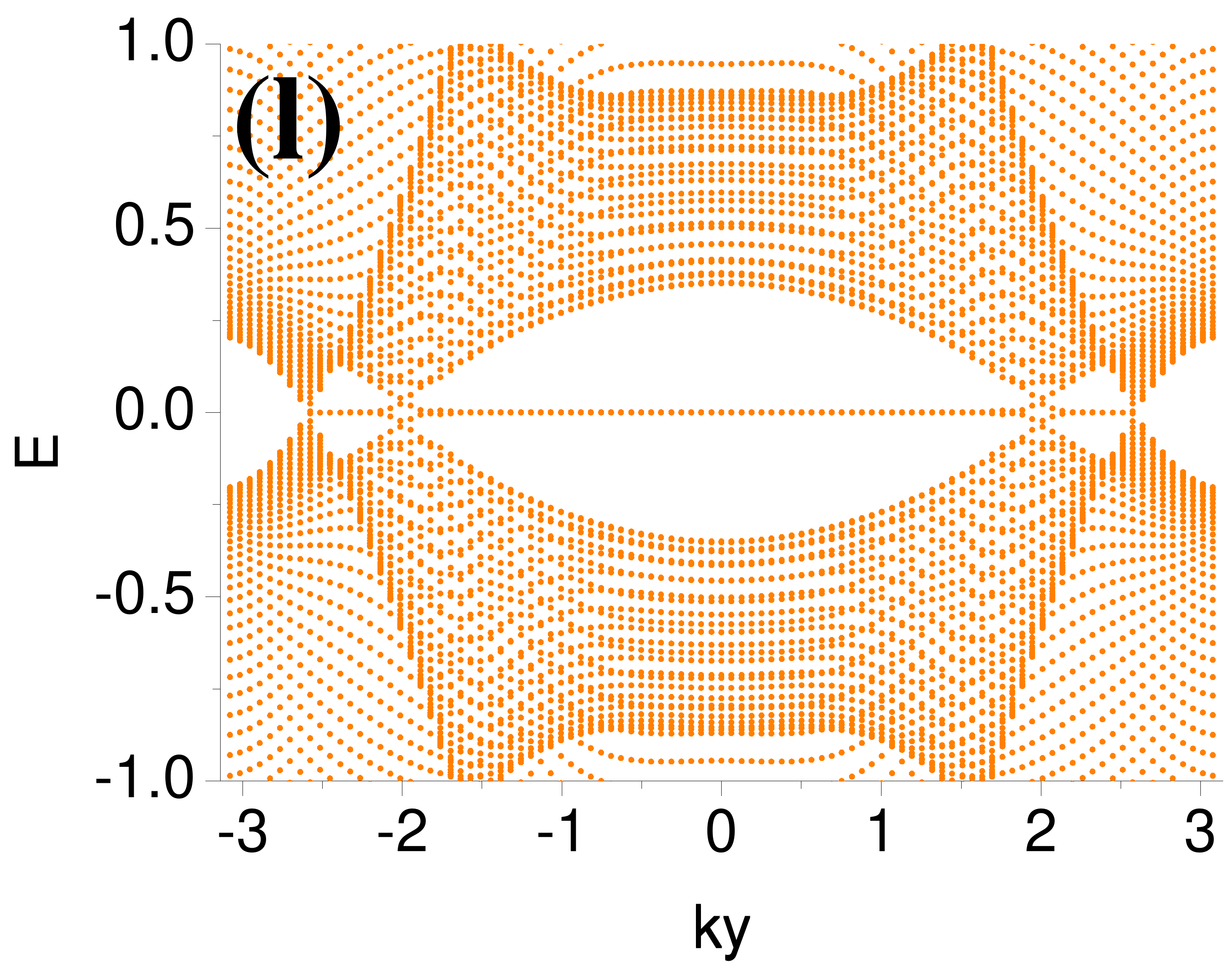} &
\includegraphics[width=3cm]{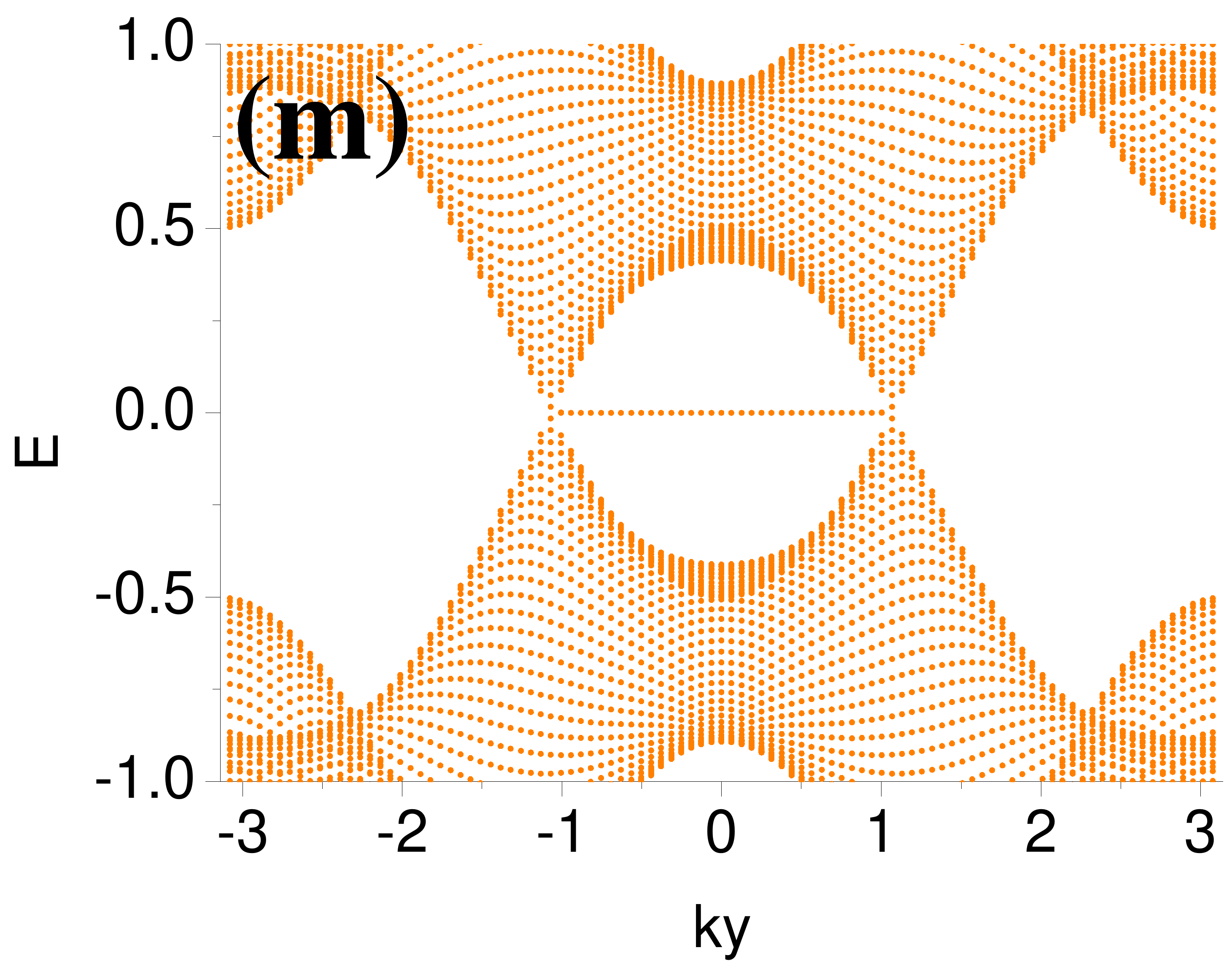} &
\includegraphics[width=3cm]{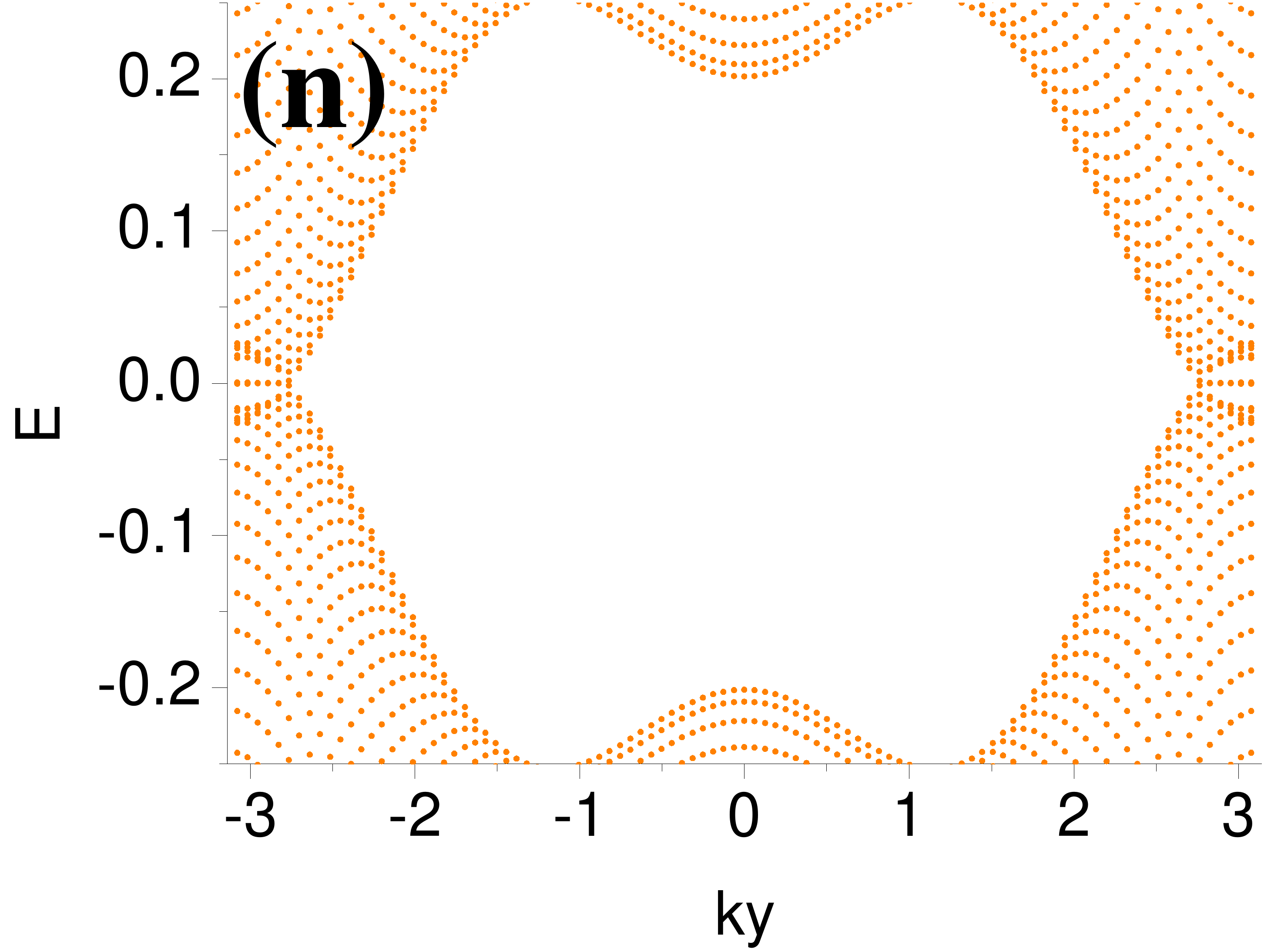} &
\includegraphics[width=3cm]{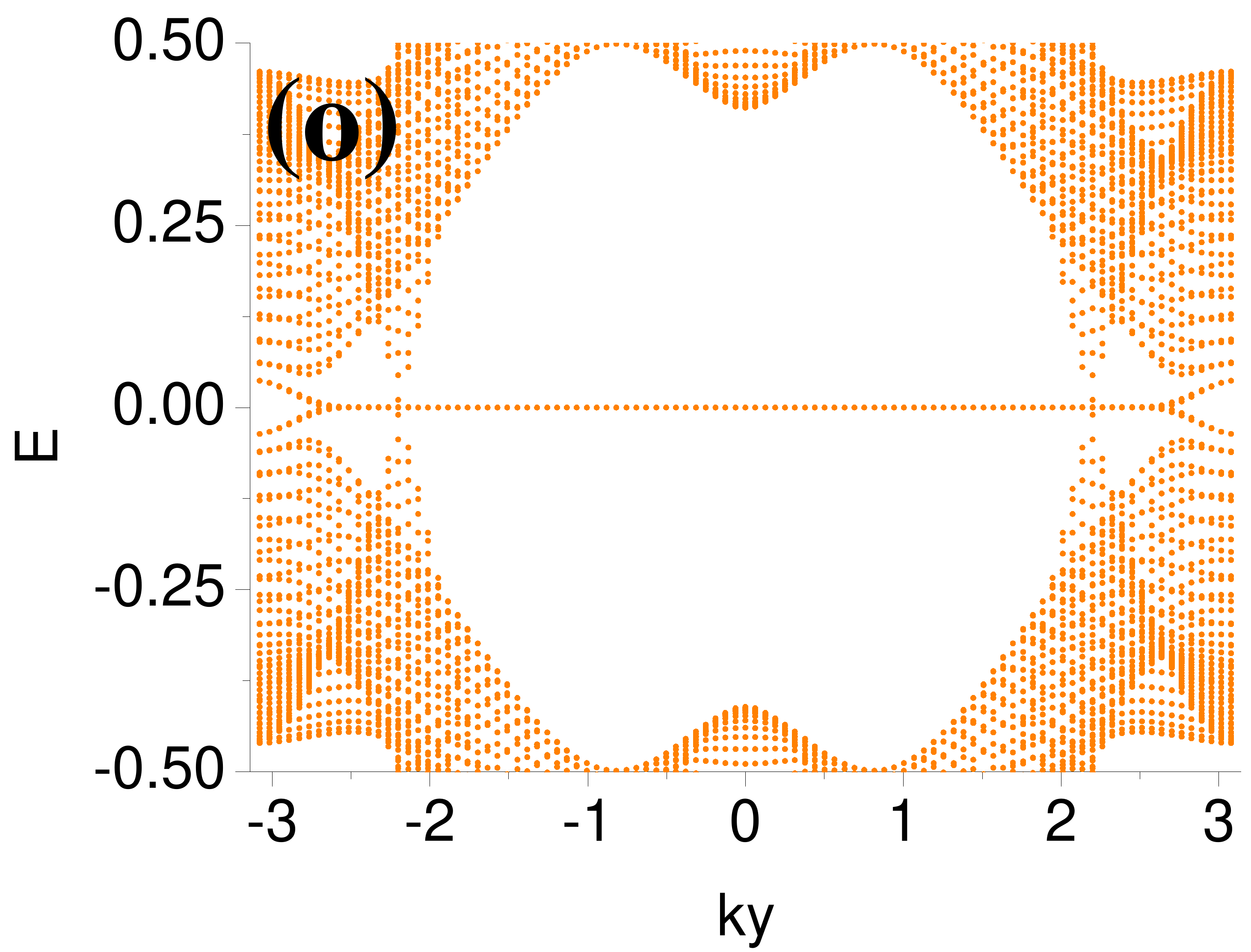} \\
\includegraphics[width=3cm]{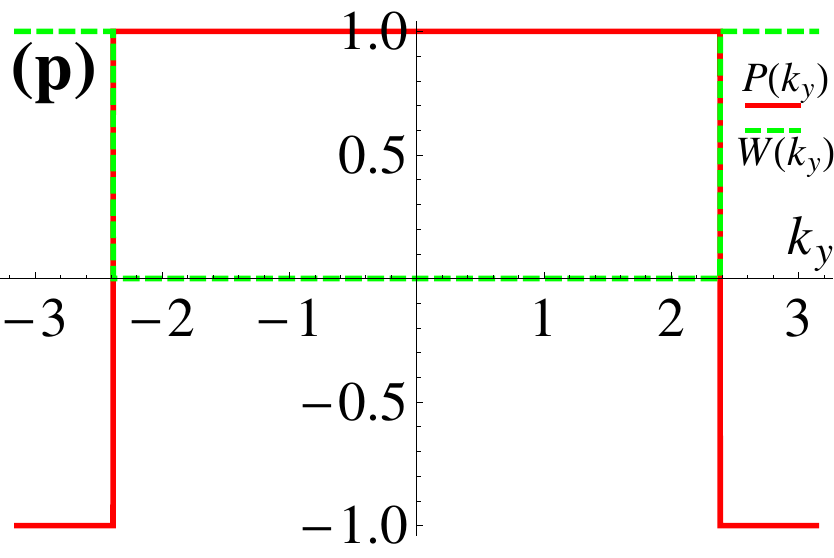} &
\includegraphics[width=3cm]{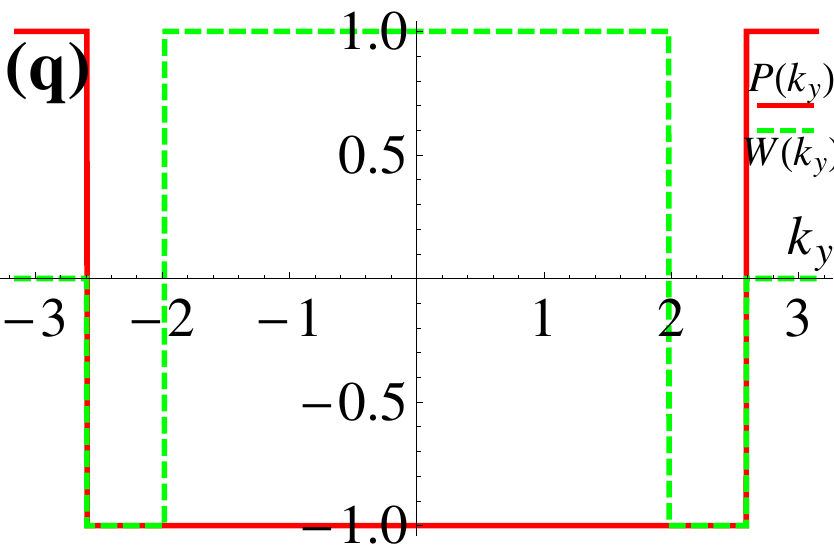} &
\includegraphics[width=3cm]{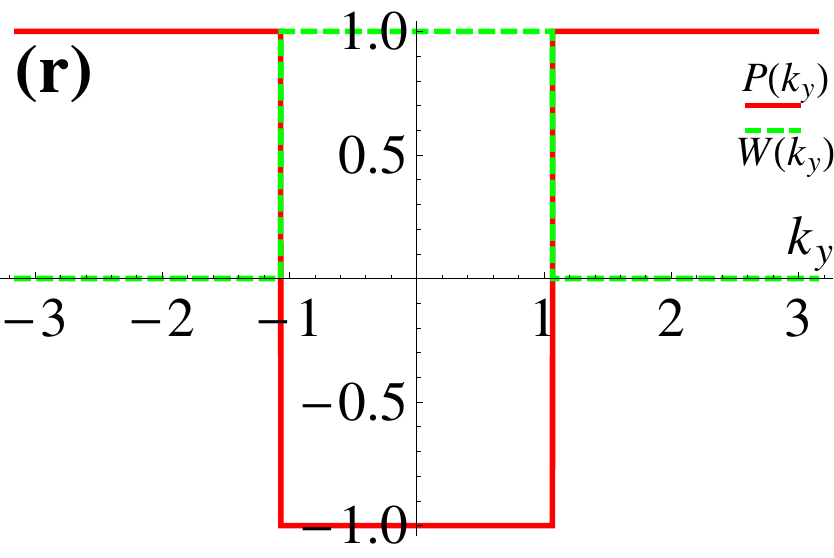} &
\includegraphics[width=3cm]{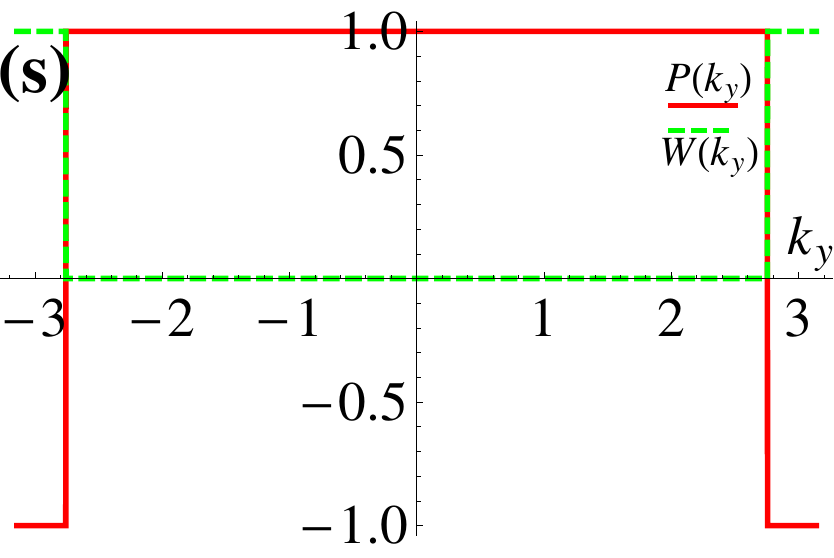} &
\includegraphics[width=3cm]{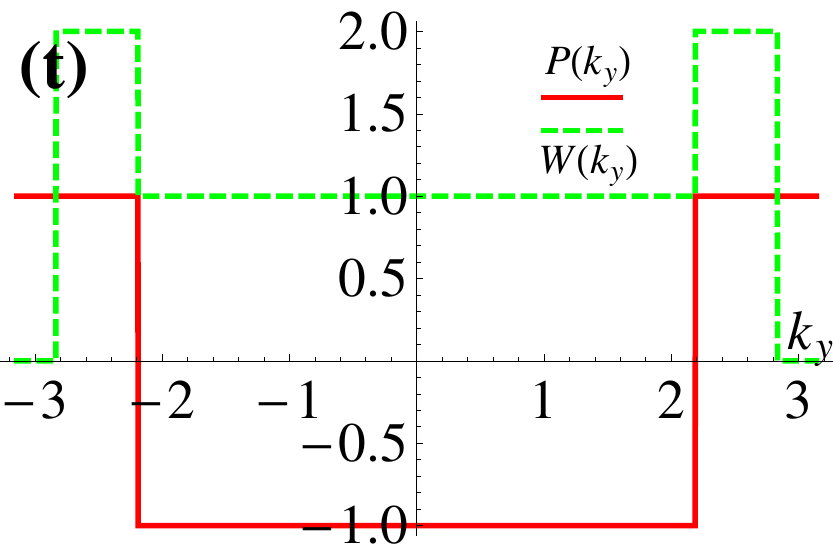} \\
\end{tabular}
\caption{(color online). (a)-(e) and (k)-(o) are the edge spectra of
the $d_{x^2-y^2}+s$-wave superconductor with Dresselhaus (110)
spin-orbit coupling in case (f) of Tab. (\ref{symmandti}). There are
three different kinds of phase diagrams depending on the hopping
amplitude $t$ as shown in Fig. (\ref{phd}b)-(\ref{phd}d). For the
phase diagram of Fig. (\ref{phd}b), the edge spectra are
demonstrated in (a), (b) and (c). The parameters are $t=2$,
$\beta=1$, $\Delta_{s_{1}}=1$, $\Delta_{d_{1}}=2$ and (a) $\mu=0,
V^{2}=16$, (b) $\mu=-2.5, V^{2}=36$, (c) $\mu=-4, V^{2}=20$, which
correspond to regions I, II and III in Fig. (\ref{phd}b),
respectively. (f), (g) and (h) are the Pfaffian invariant and
winding number for (a), (b) and (c). For the phase diagram of Fig.
(\ref{phd}c), the edge spectra are demonstrated in (d), (e), (k) and
(l). The parameters are $t=1$, $\beta=1$, $\Delta_{s_{1}}=1$,
$\Delta_{d_{1}}=2$ and (d) $\mu=0, V^{2}=12$, (e) $\mu=0, V^{2}=20$,
(k) $\mu=-1.8, V^{2}=30$, (l) $\mu=-4.5, V^{2}=25$, which correspond
to regions I, II, III and IV in Fig. (\ref{phd}c), respectively.
(i), (j), (p) and (q) are the Pfaffian invariant and winding number
for (d), (e), (k) and (l). For the phase diagram of Fig.
(\ref{phd}d), the edge spectra are demonstrated in (m), (n) and (o).
The parameters are $t=0.5$, $\beta=1$, $\Delta_{s_{1}}=1$,
$\Delta_{d_{1}}=2$ and (m) $\mu=0, V^{2}=16$, (n) $\mu=-7,
V^{2}=81$, (o) $\mu=-1, V^{2}=5$, which correspond to regions I, II
and III in Fig. (\ref{phd}d), respectively. (r), (s) and (t) are the
Pfaffian invariant and winding number for (m), (n) and
(o).}\label{casef}
\end{figure*}

\begin{figure*}
\begin{tabular}{ccccc}
\includegraphics[width=3cm]{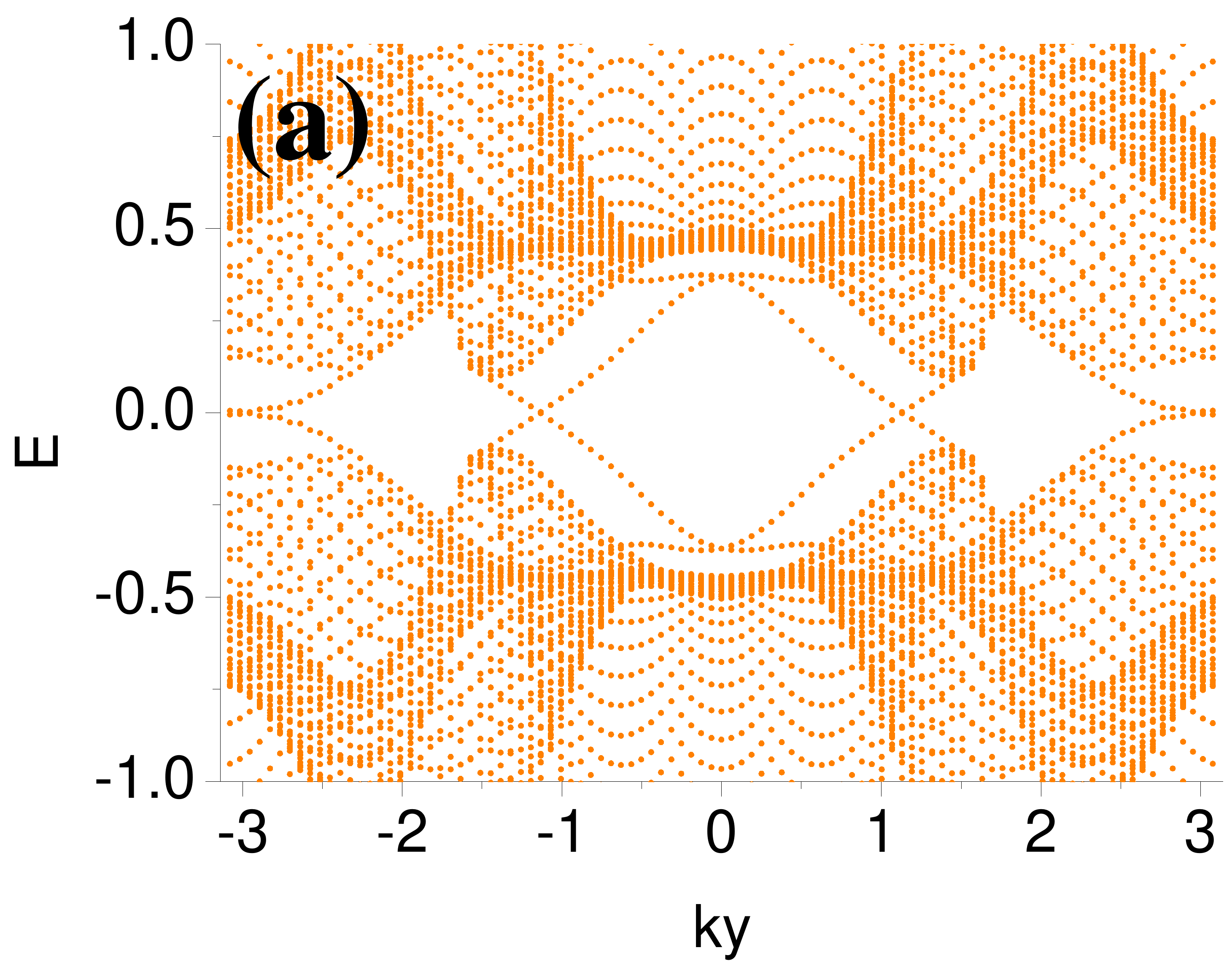} &
\includegraphics[width=3cm]{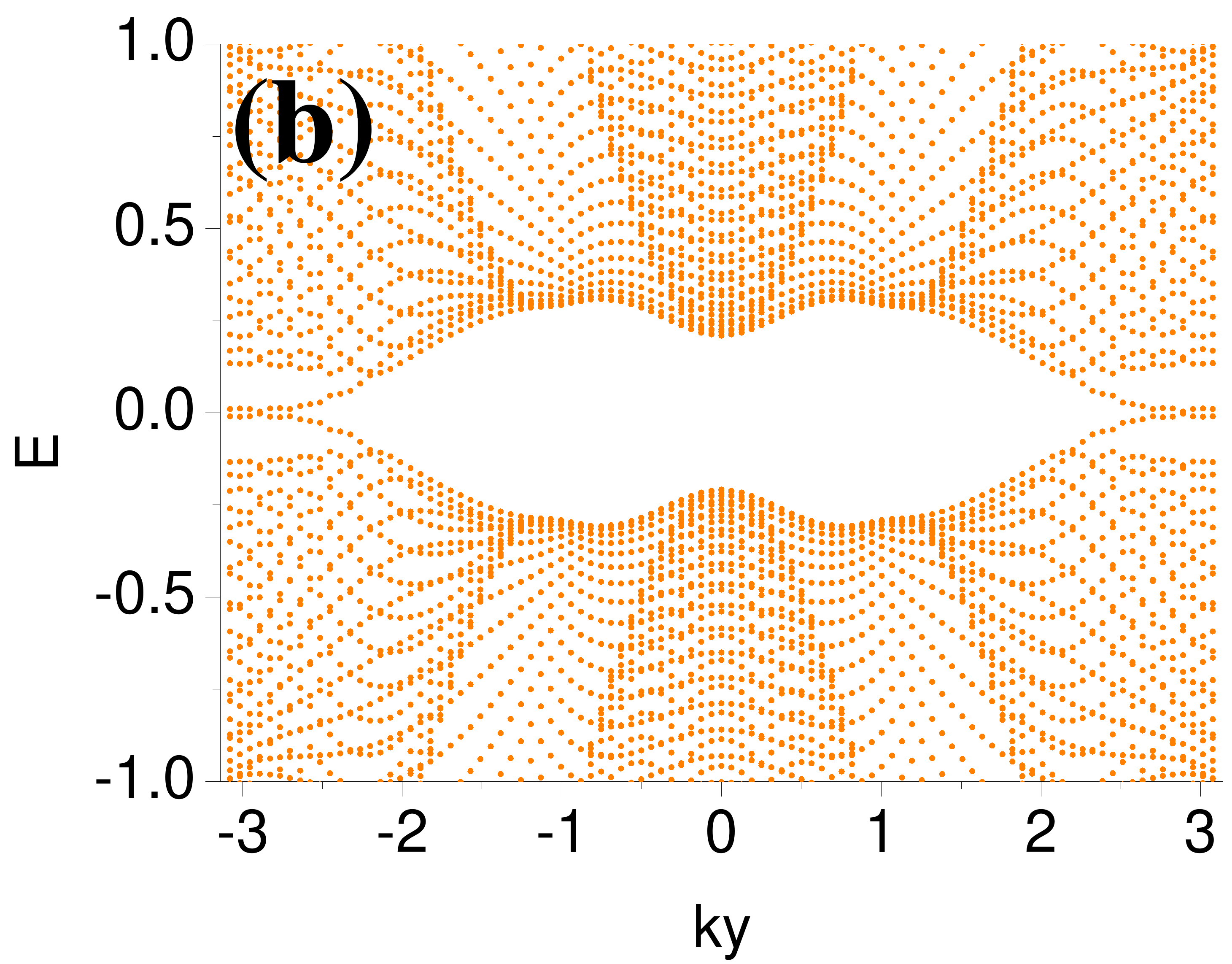} &
\includegraphics[width=3cm]{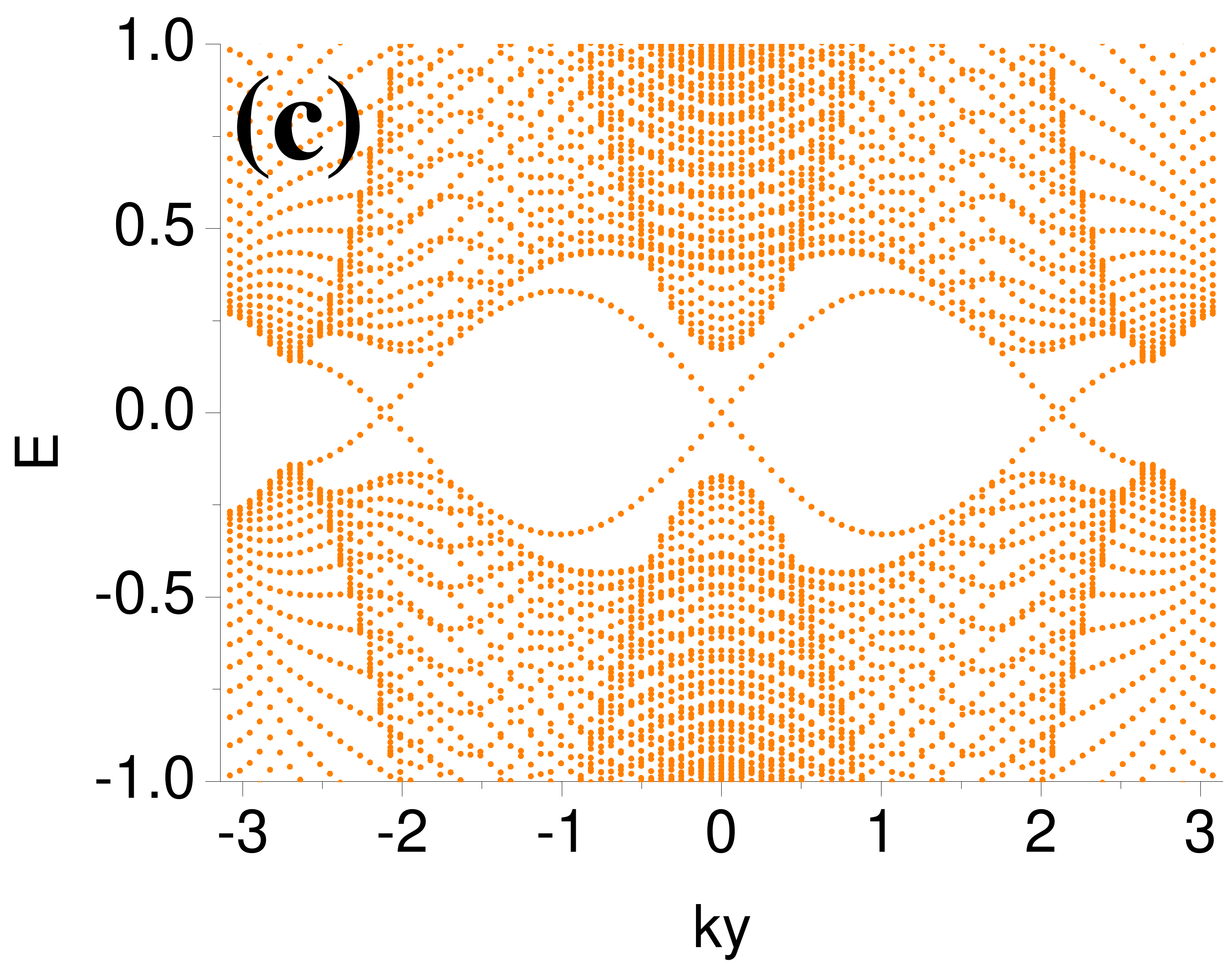} &
\includegraphics[width=3cm]{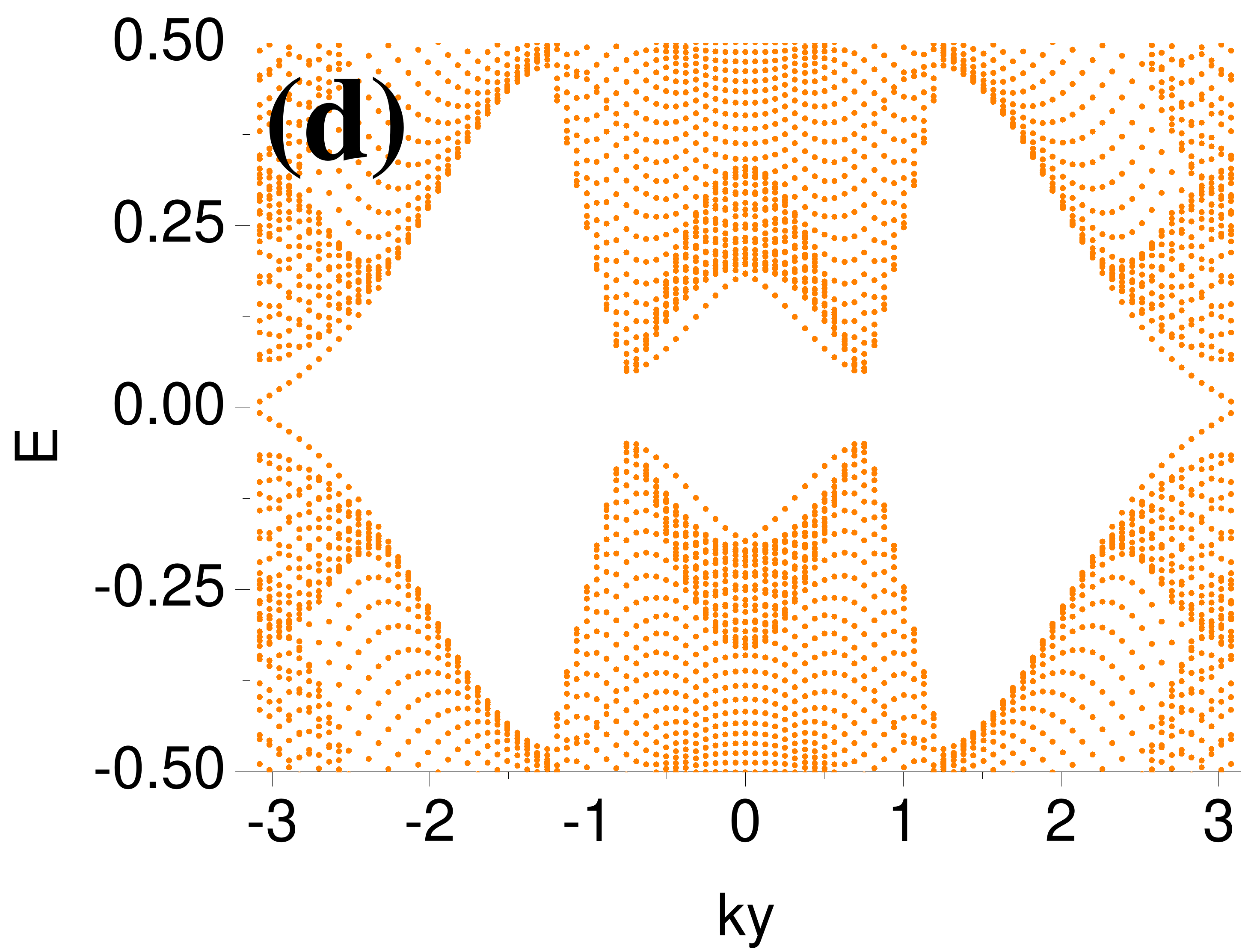} &
\includegraphics[width=3cm]{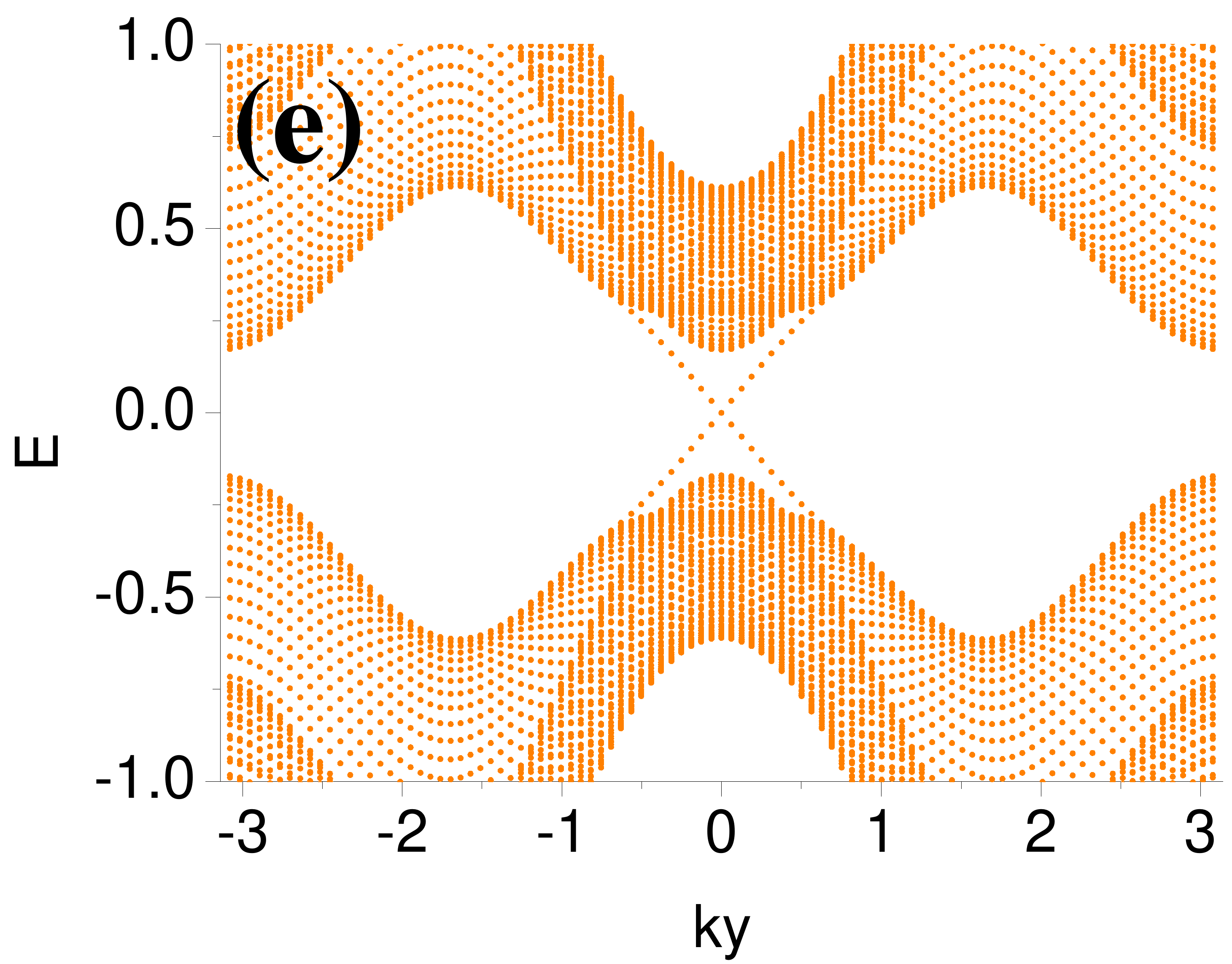} \\
\includegraphics[width=3cm]{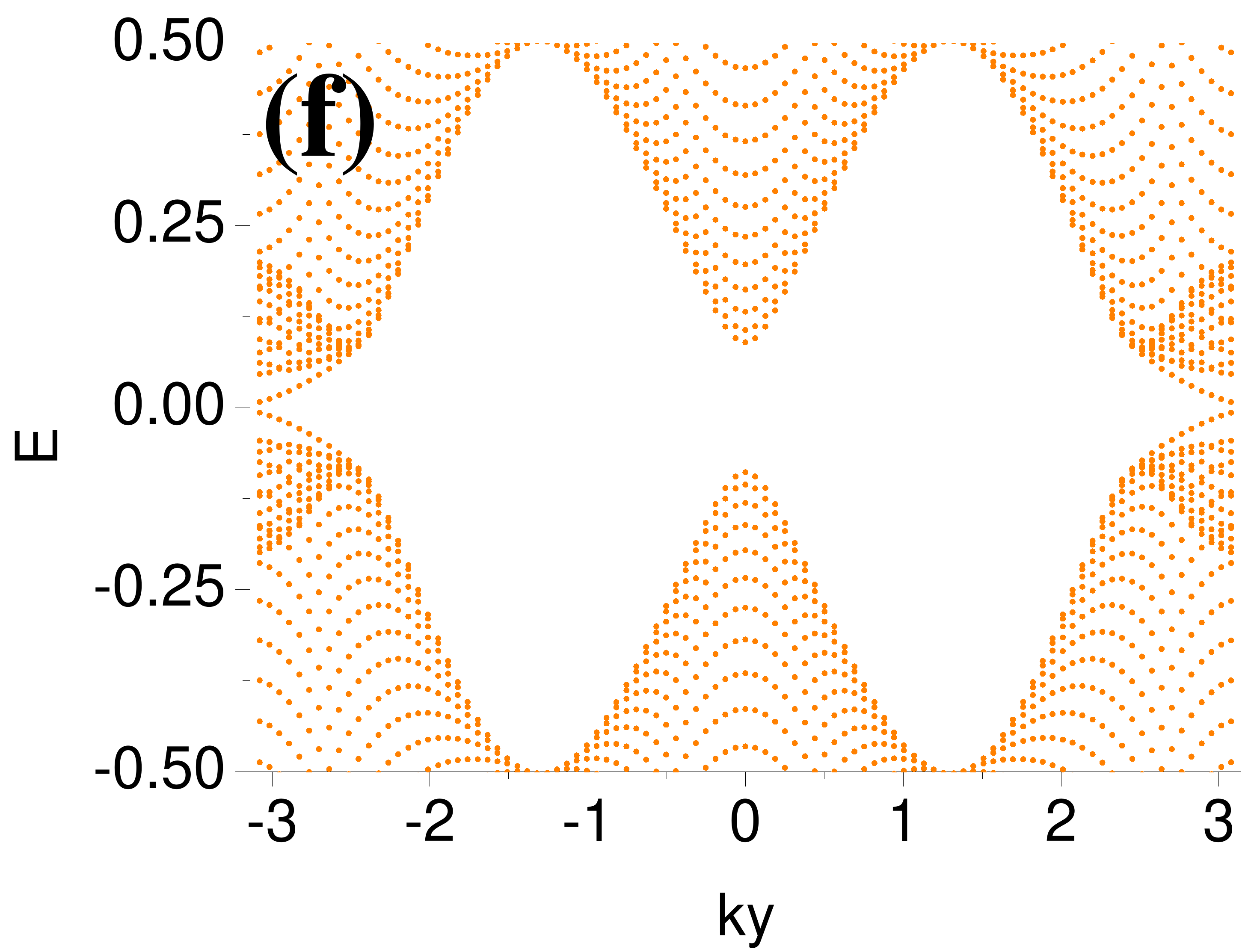} &
\includegraphics[width=3cm]{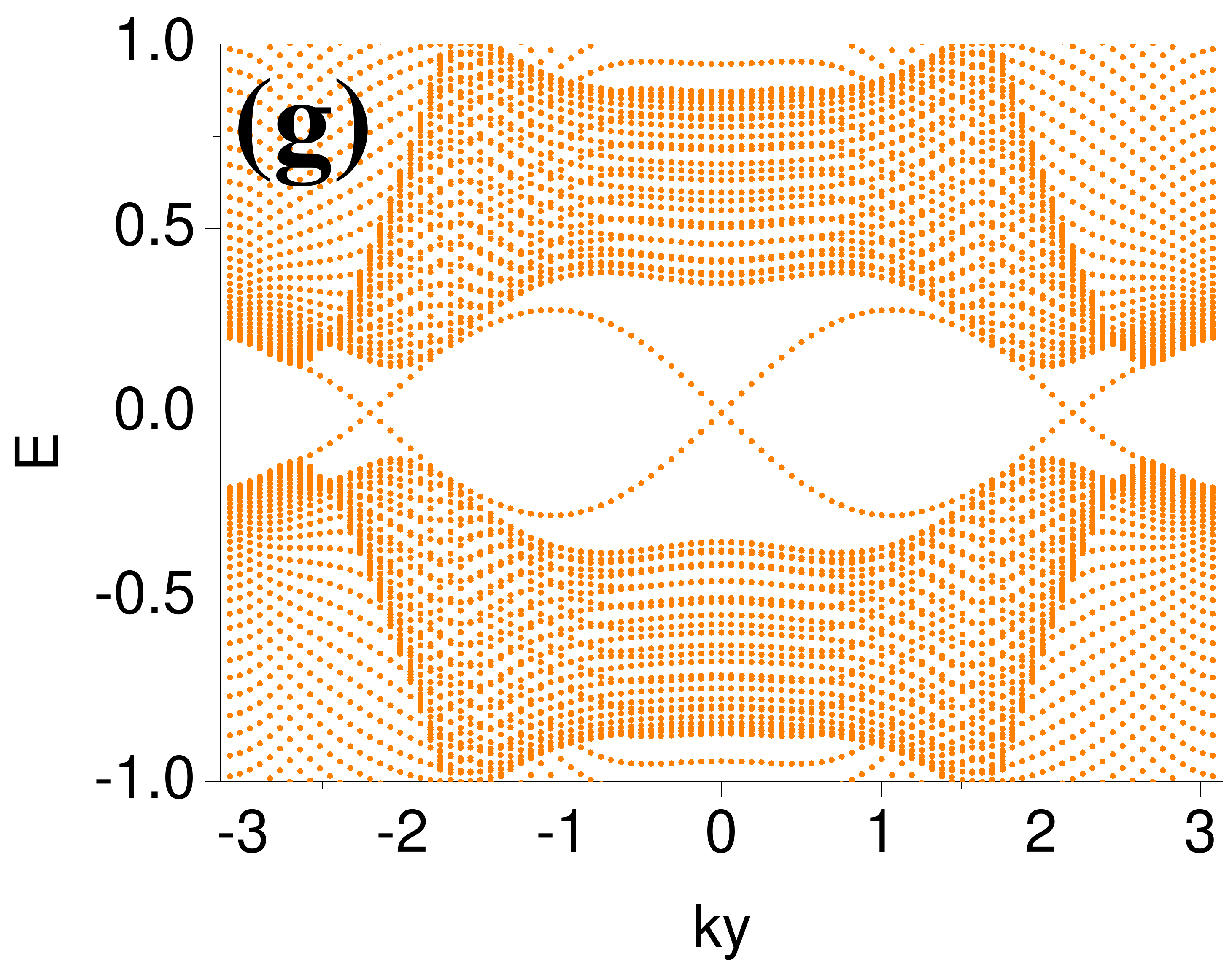} &
\includegraphics[width=3cm]{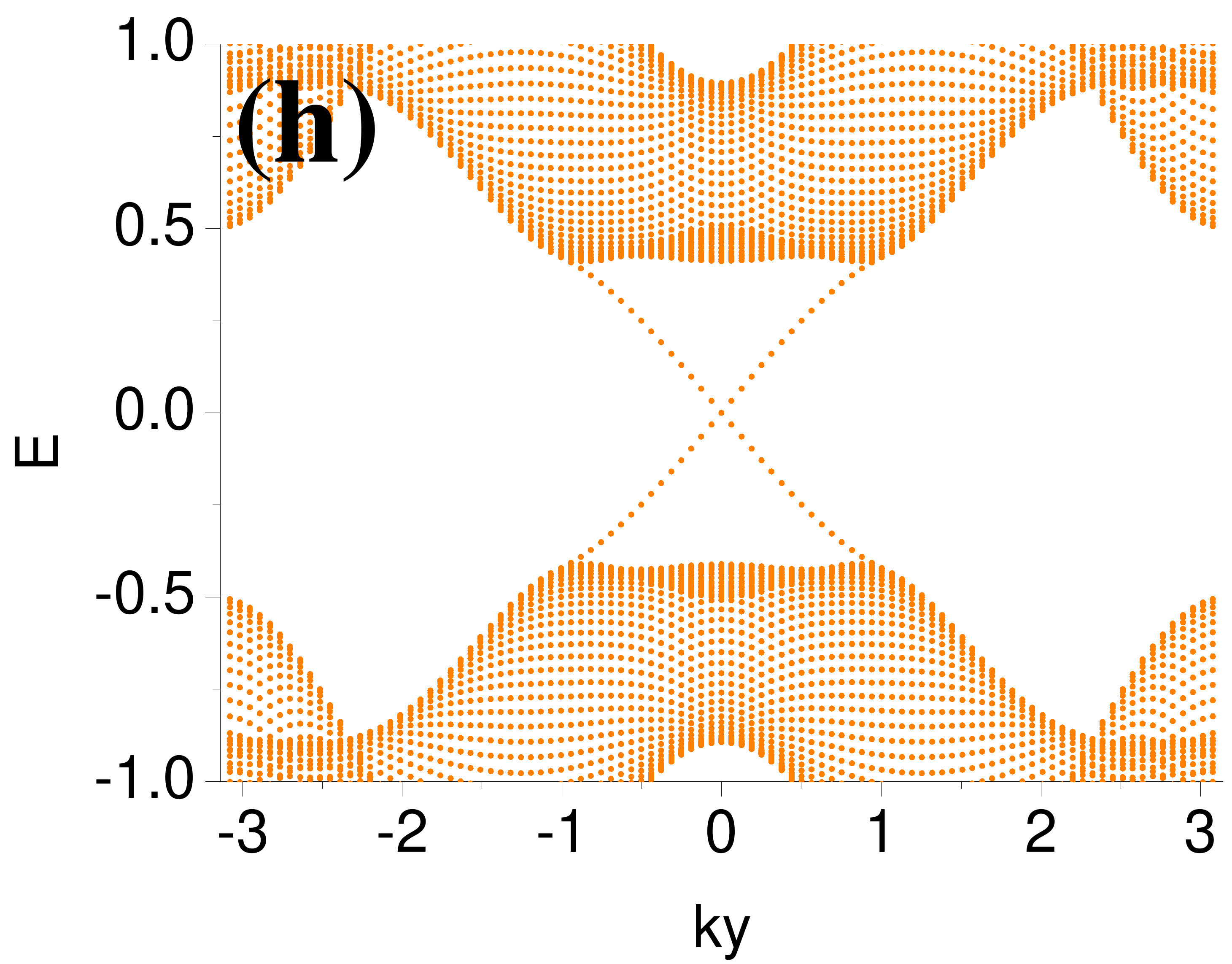} &
\includegraphics[width=3cm]{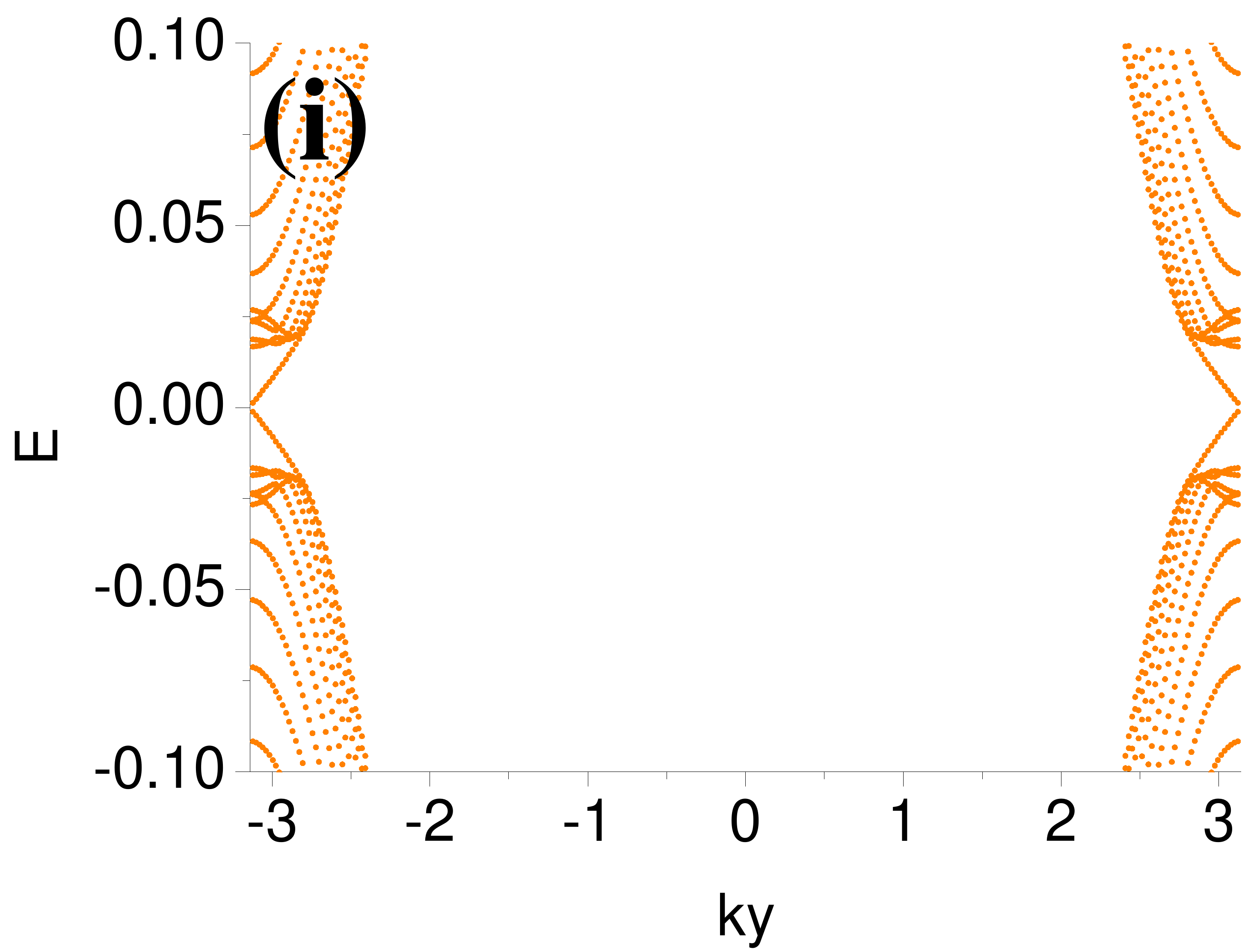} &
\includegraphics[width=3cm]{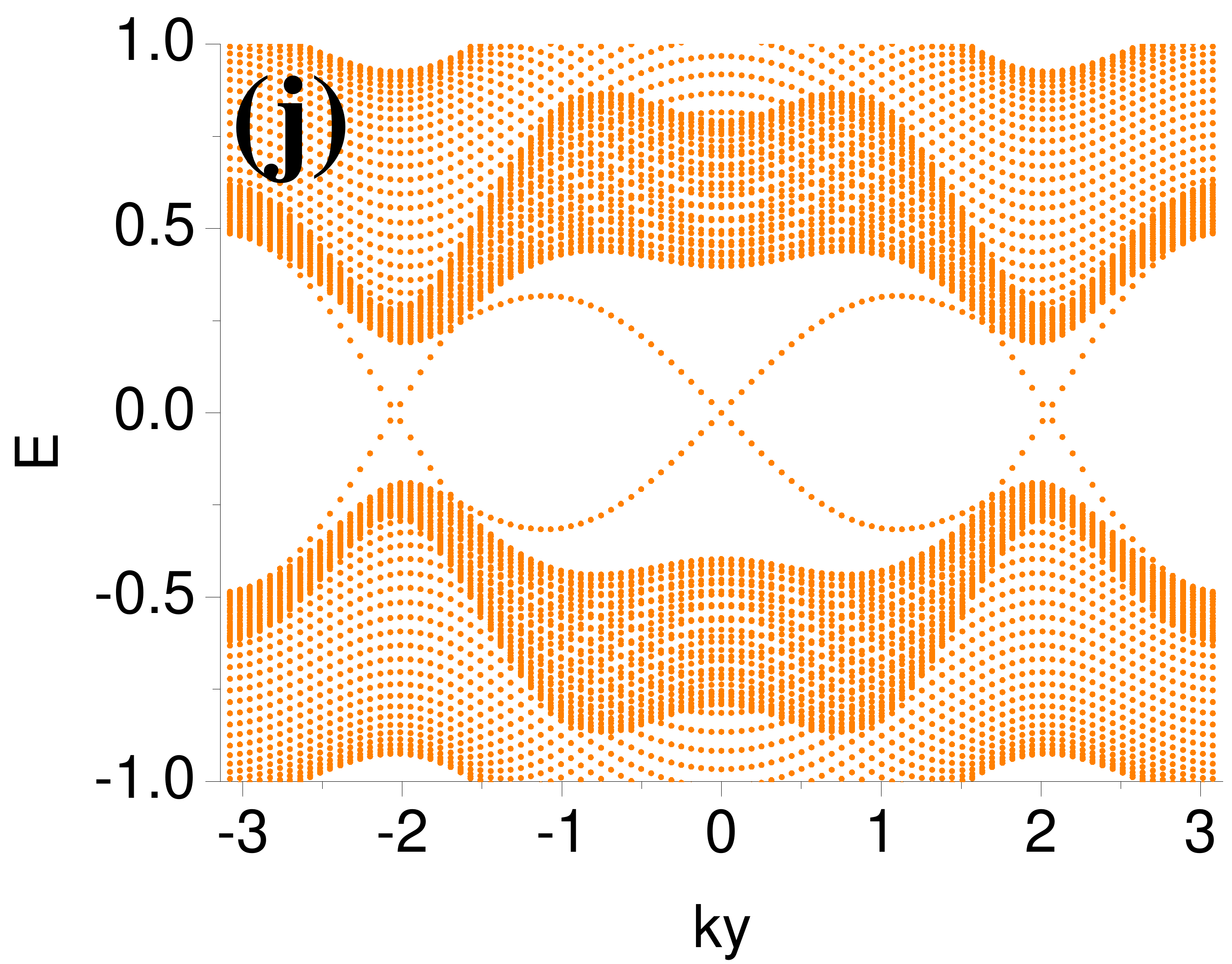} \\
\end{tabular}
\caption{(color online). (a)-(j) are the edge spectra of the
$d_{x^2-y^2}+id_{xy}+s$-wave superconductor with Rashba spin-orbit
coupling in case (a) of Tab. (\ref{symmandti}). For the phase
diagram of Fig. (\ref{phd}b), the edge spectra are demonstrated in
(a), (b) and (c). The parameters are $t=2$, $\alpha=1$,
$\Delta_{s_{1}}=1$, $\Delta_{s_{2}}=0$, $\Delta_{d_{1}}=2$,
$\Delta_{d_{2}}=1$ and (a) $\mu=0, V_{z}=4$, (b) $\mu=-2.5,
V_{z}=6$, (c) $\mu=-4, V_{z}=2\sqrt{5}$, which correspond to regions
I, II and III in Fig. (\ref{phd}b), respectively. For the phase
diagram of Fig. (\ref{phd}c), the edge spectra are demonstrated in
(d), (e), (f) and (g). The parameters are $t=1$, $\alpha=1$,
$\Delta_{s_{1}}=1$, $\Delta_{s_{2}}=0$, $\Delta_{d_{1}}=2$,
$\Delta_{d_{2}}=1$ and (d) $\mu=0, V_{z}=2\sqrt{3}$, (e) $\mu=0,
V_{z}=2\sqrt{5}$, (f) $\mu=-1.8, V_{z}=\sqrt{30}$, (g) $\mu=-4.5,
V_{z}=5$, which correspond to regions I, II, III and IV in Fig.
(\ref{phd}c), respectively. For the phase diagram of Fig.
(\ref{phd}d), the edge spectra are demonstrated in (h), (i) and (j).
The parameters are $t=0.5$, $\alpha=1$, $\Delta_{s_{1}}=1$,
$\Delta_{s_{2}}=0$, $\Delta_{d_{1}}=2$, $\Delta_{d_{2}}=1$ and (h)
$\mu=0, V_{z}=4$, (i) $\mu=-7, V_{z}=9$, (j) $\mu=-3, V_{z}=3$,
which correspond to regions I, II and III in Fig. (\ref{phd}d),
respectively.}\label{casea}
\end{figure*}

\begin{figure*}
\begin{tabular}{ccccc}
\includegraphics[width=3cm]{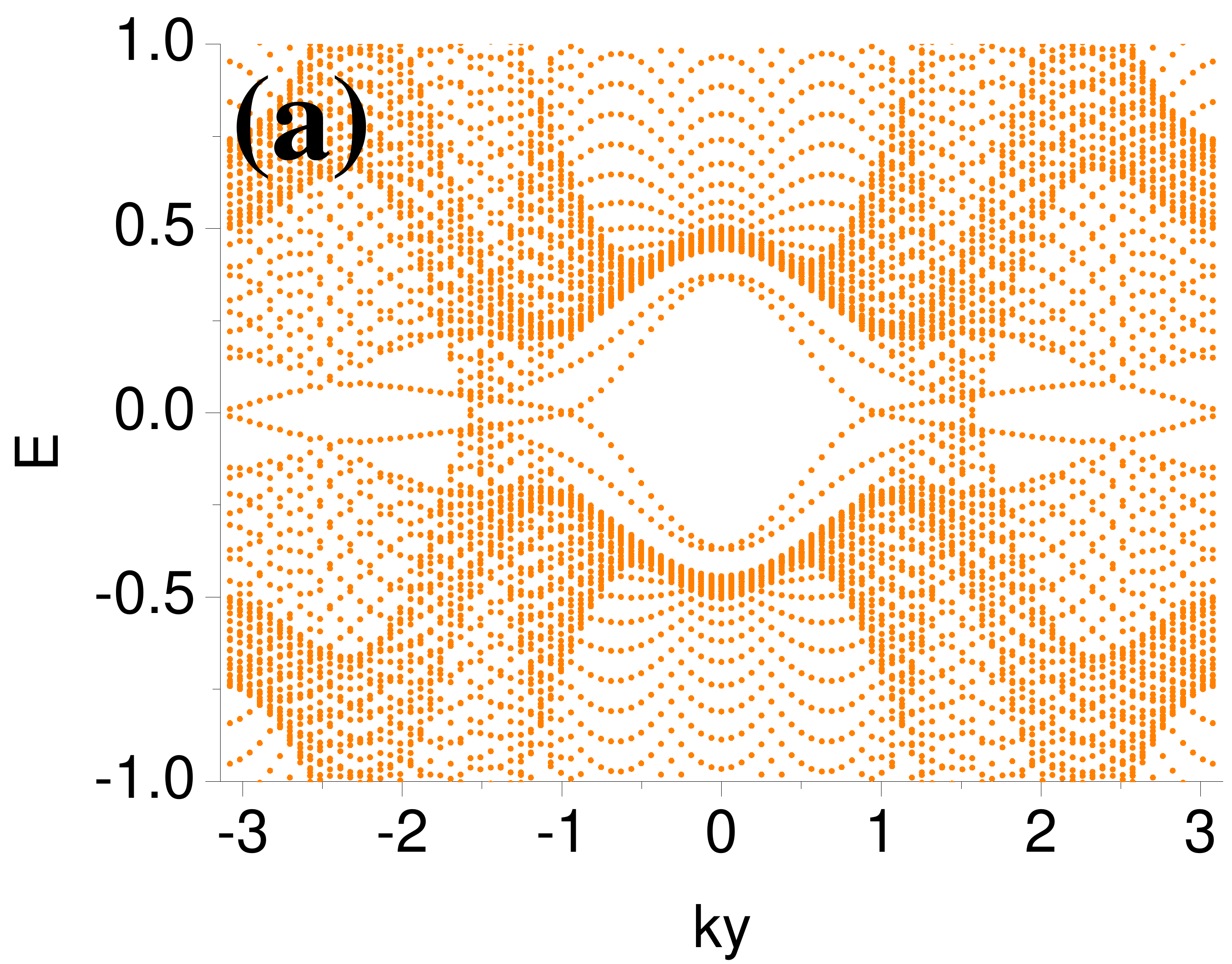} &
\includegraphics[width=3cm]{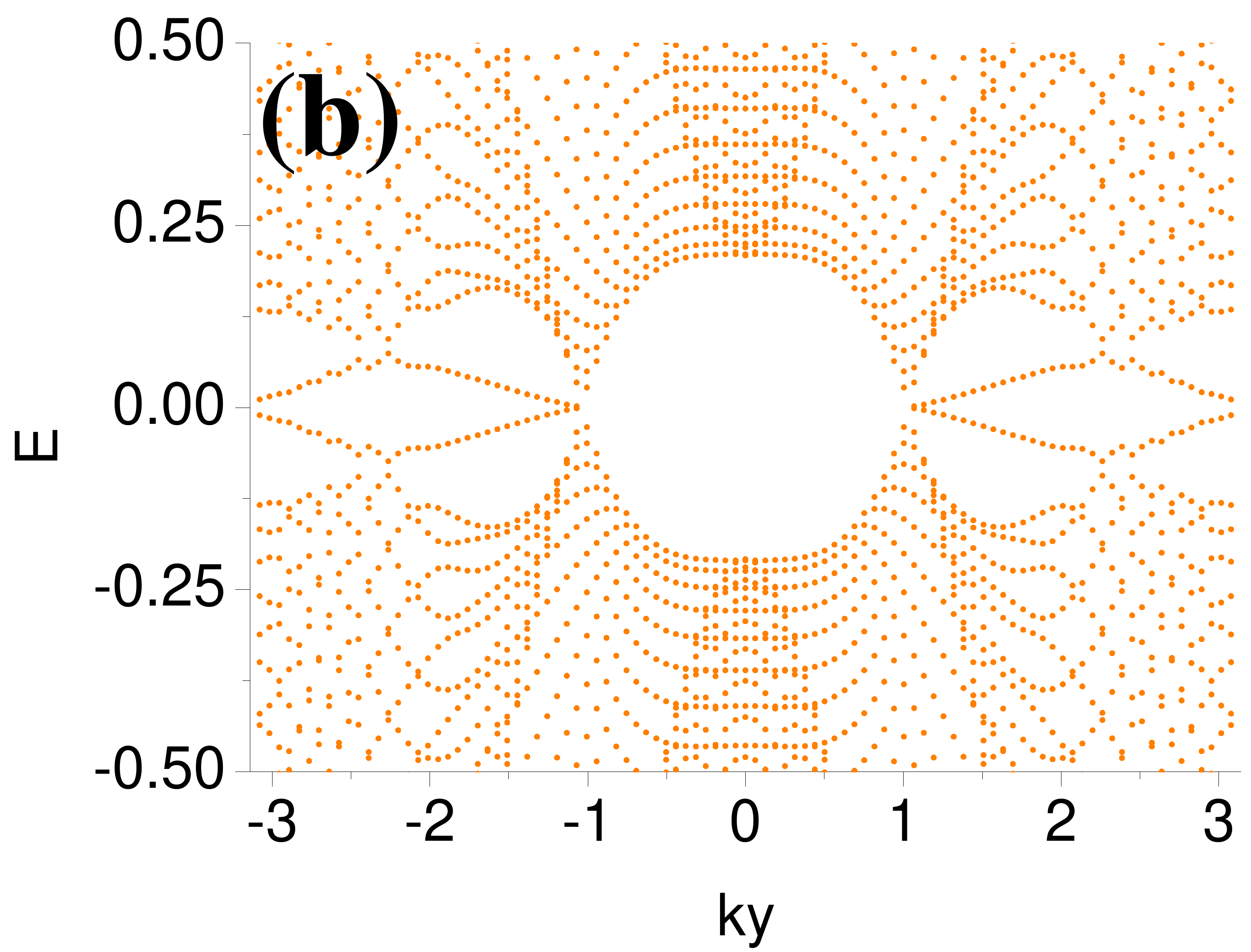} &
\includegraphics[width=3cm]{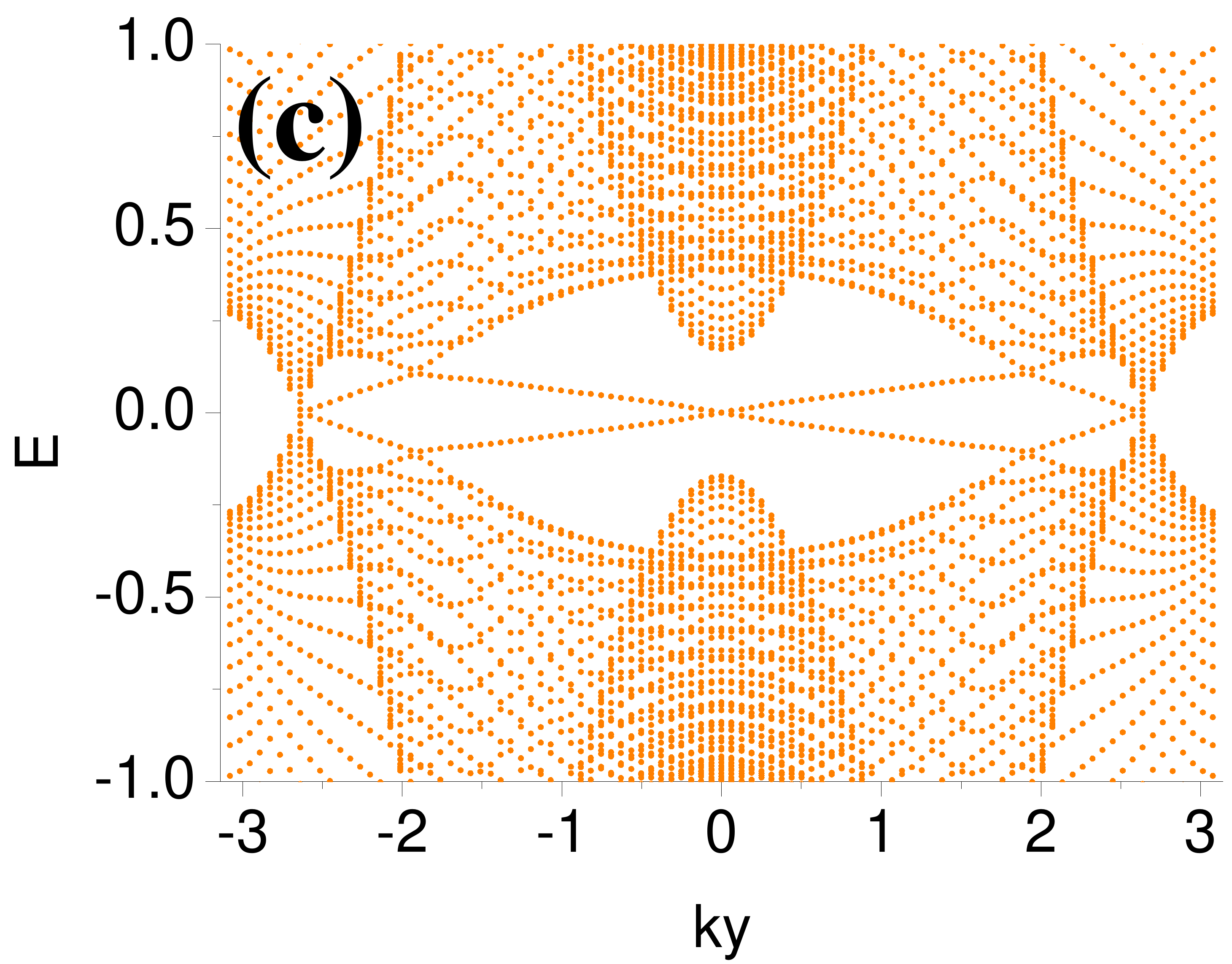} &
\includegraphics[width=3cm]{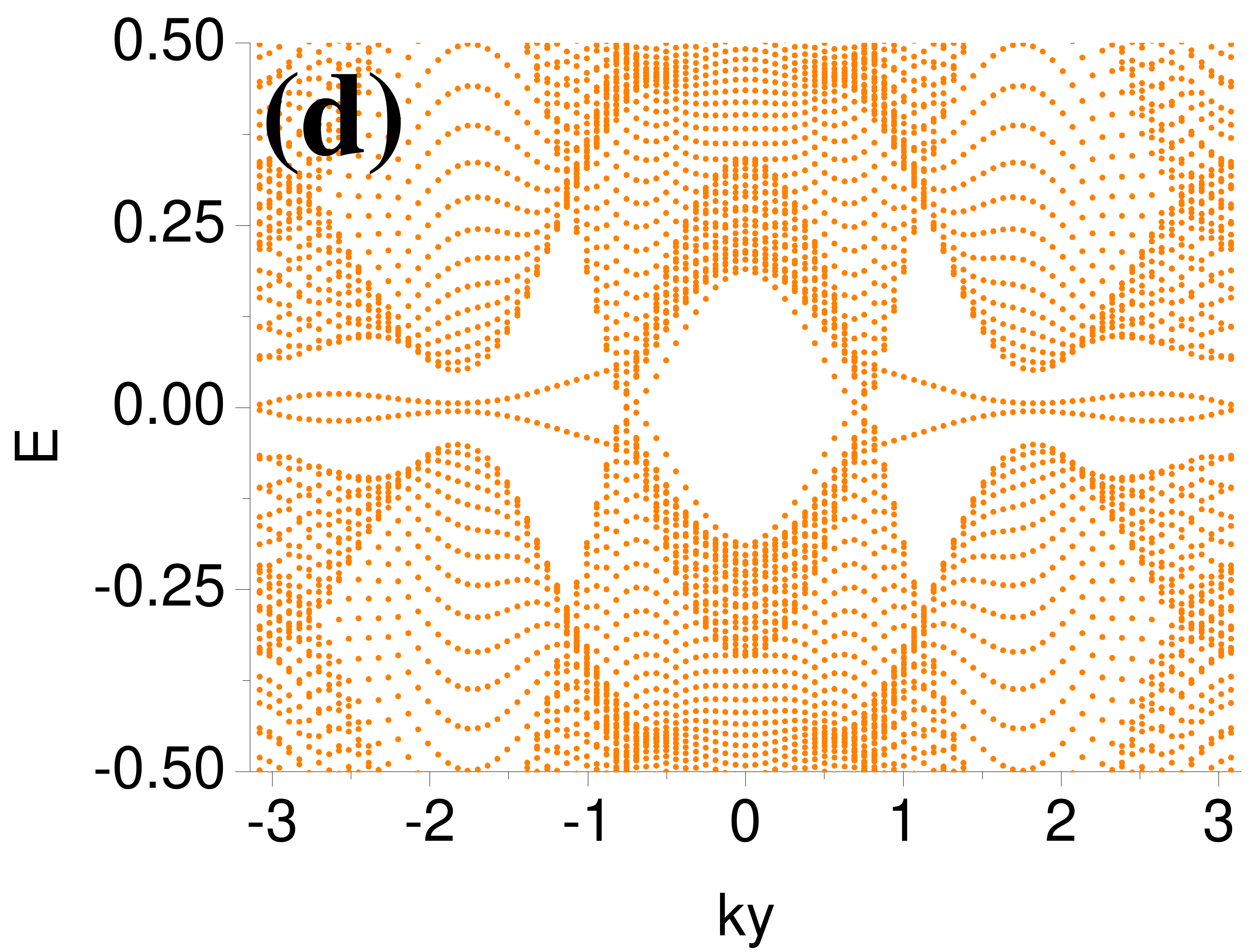} &
\includegraphics[width=3cm]{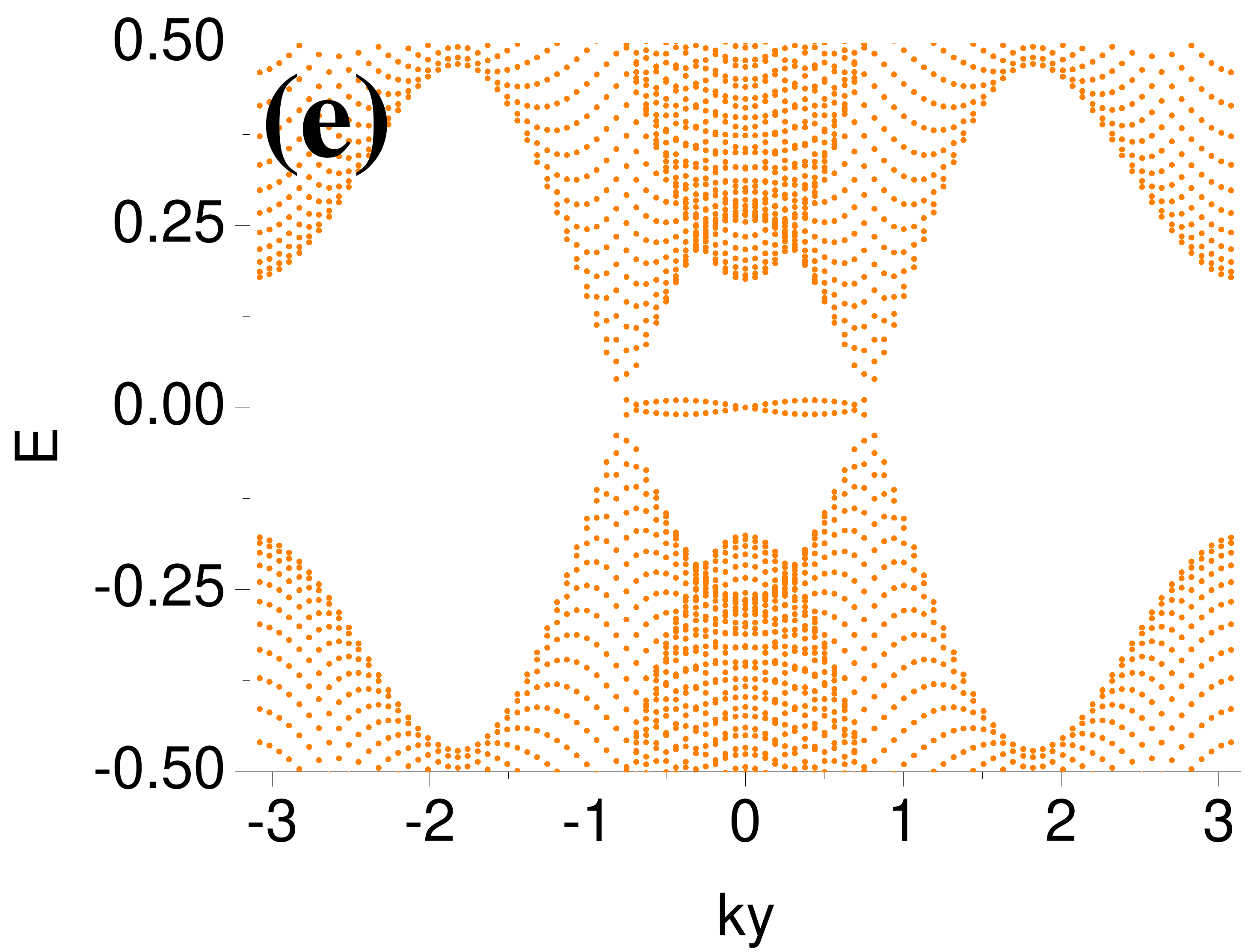} \\
\includegraphics[width=3cm]{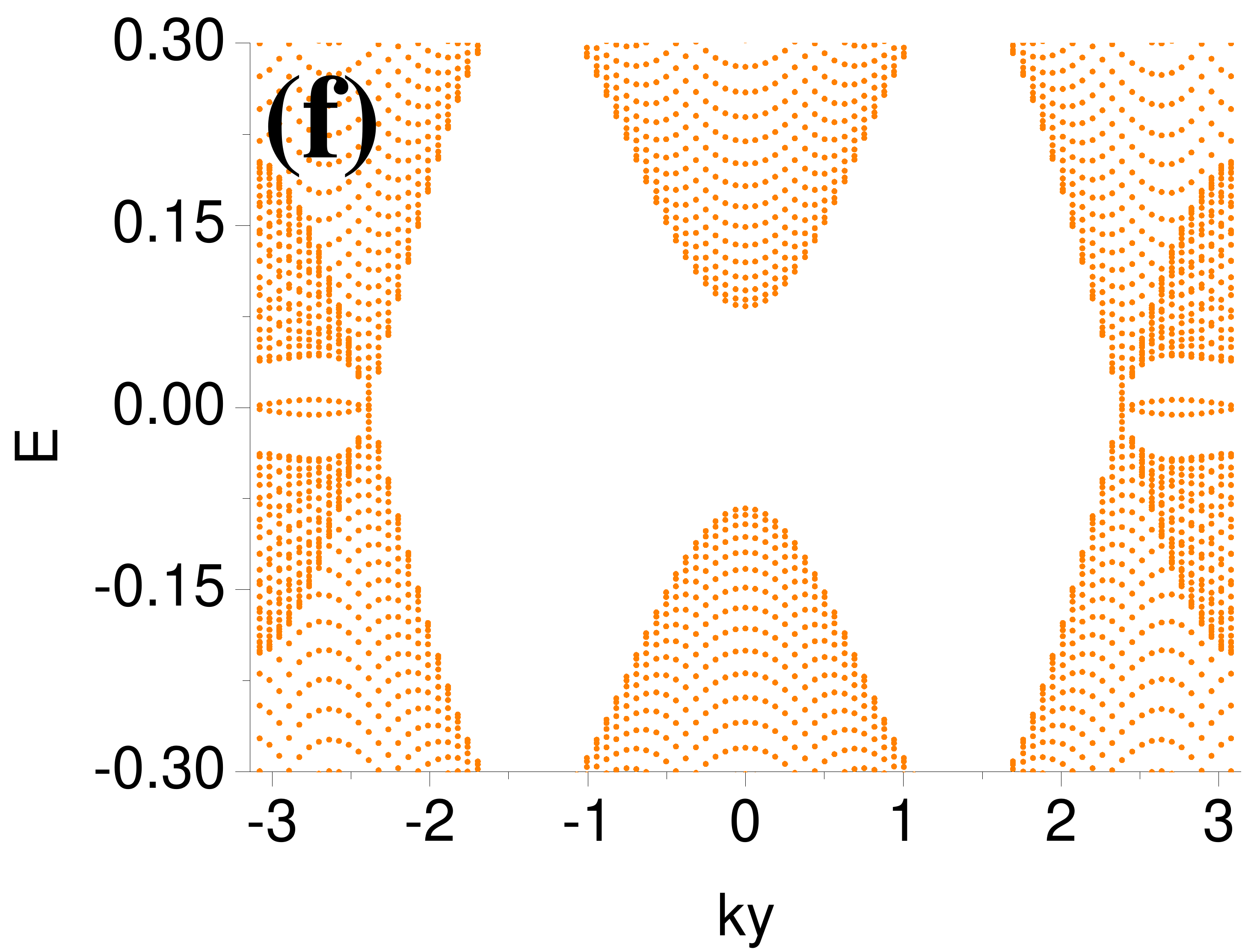} &
\includegraphics[width=3cm]{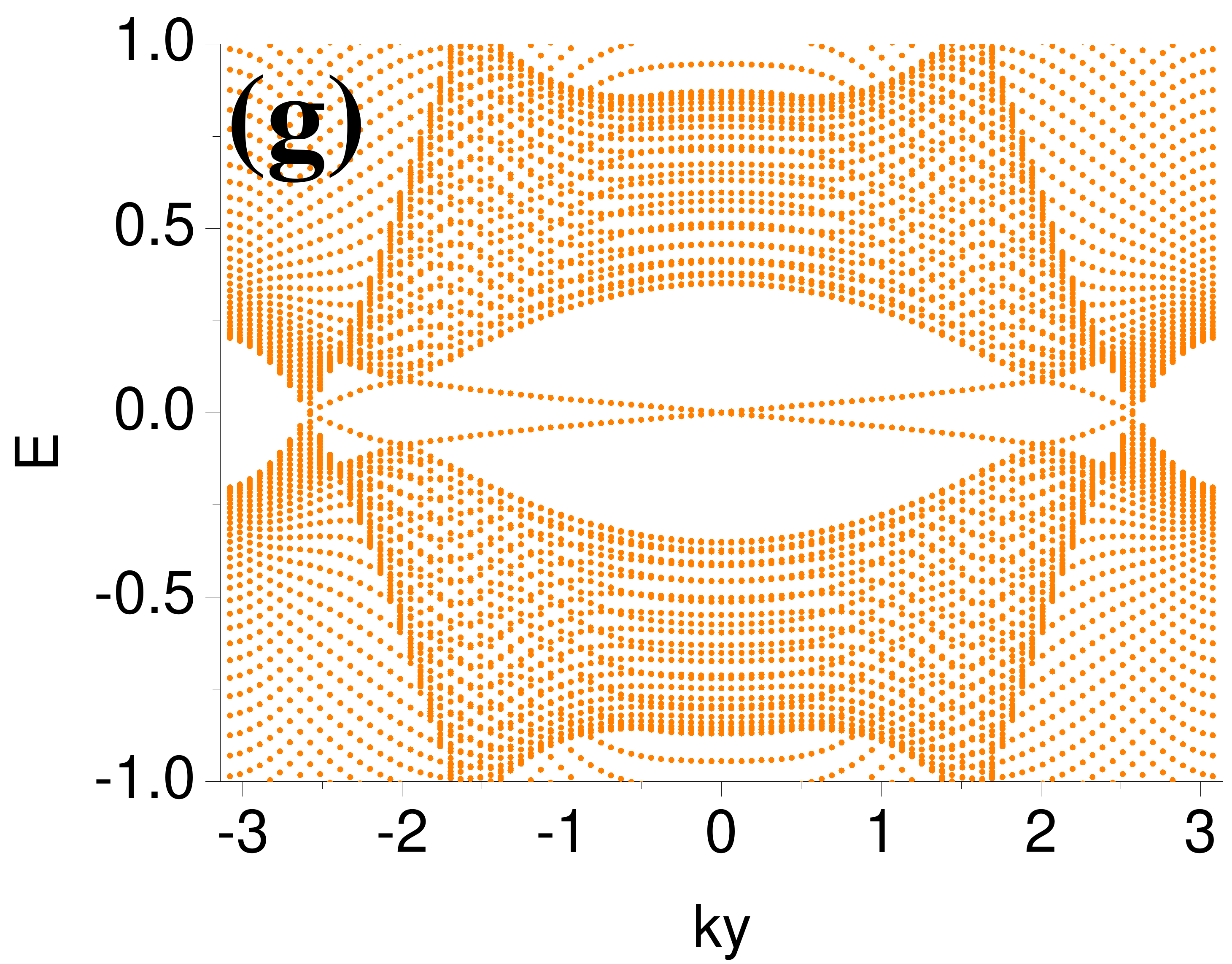} &
\includegraphics[width=3cm]{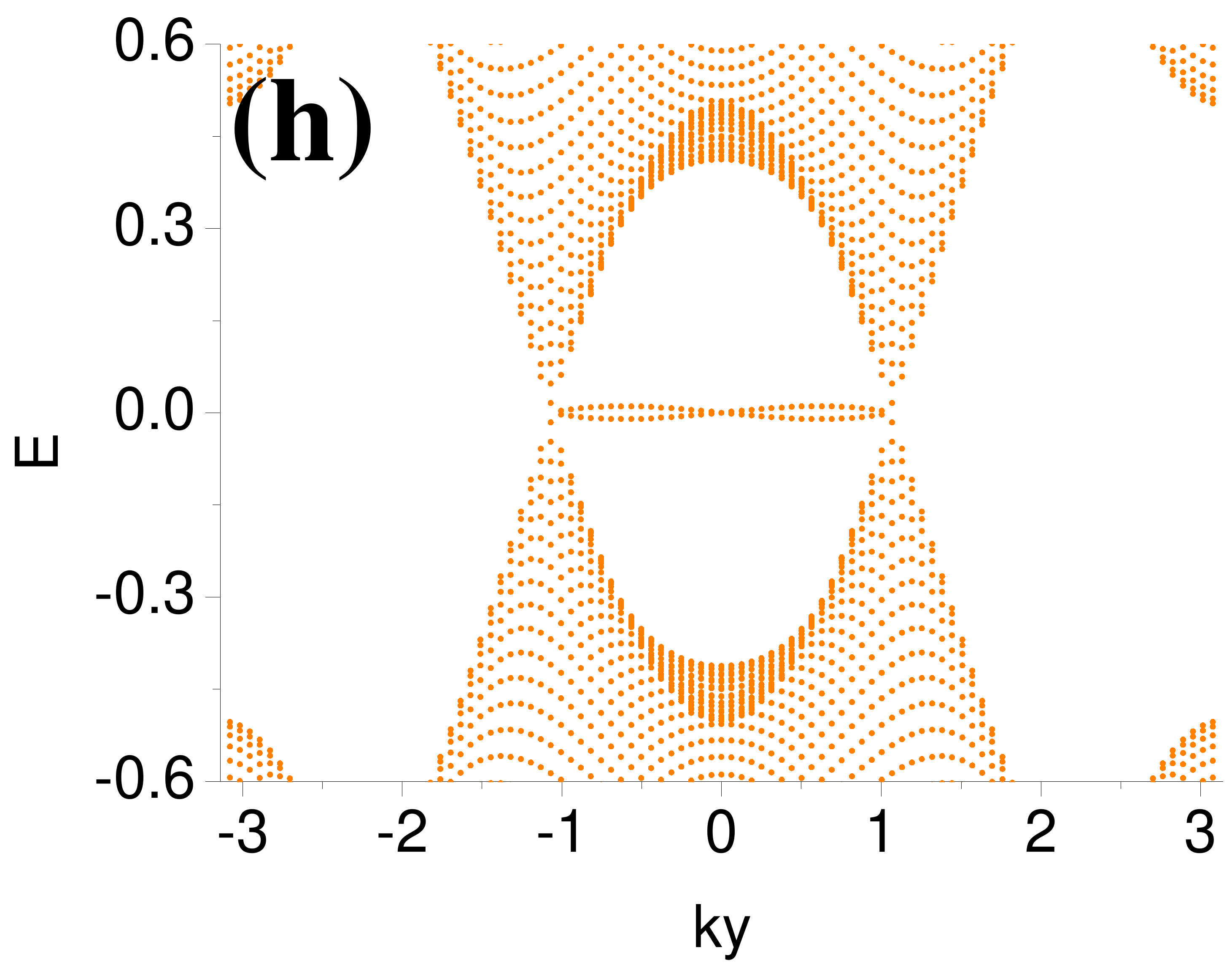} &
\includegraphics[width=3cm]{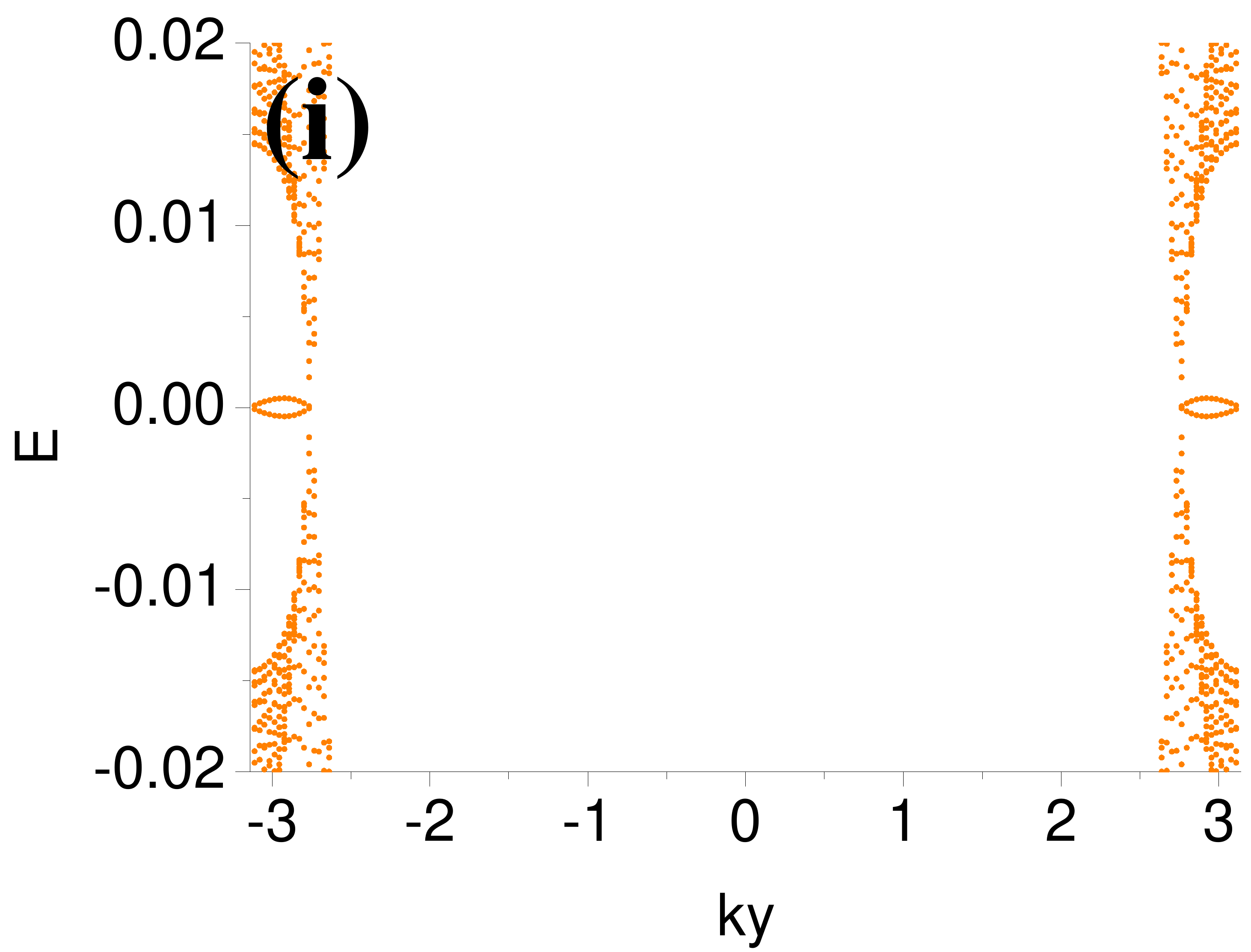} &
\includegraphics[width=3cm]{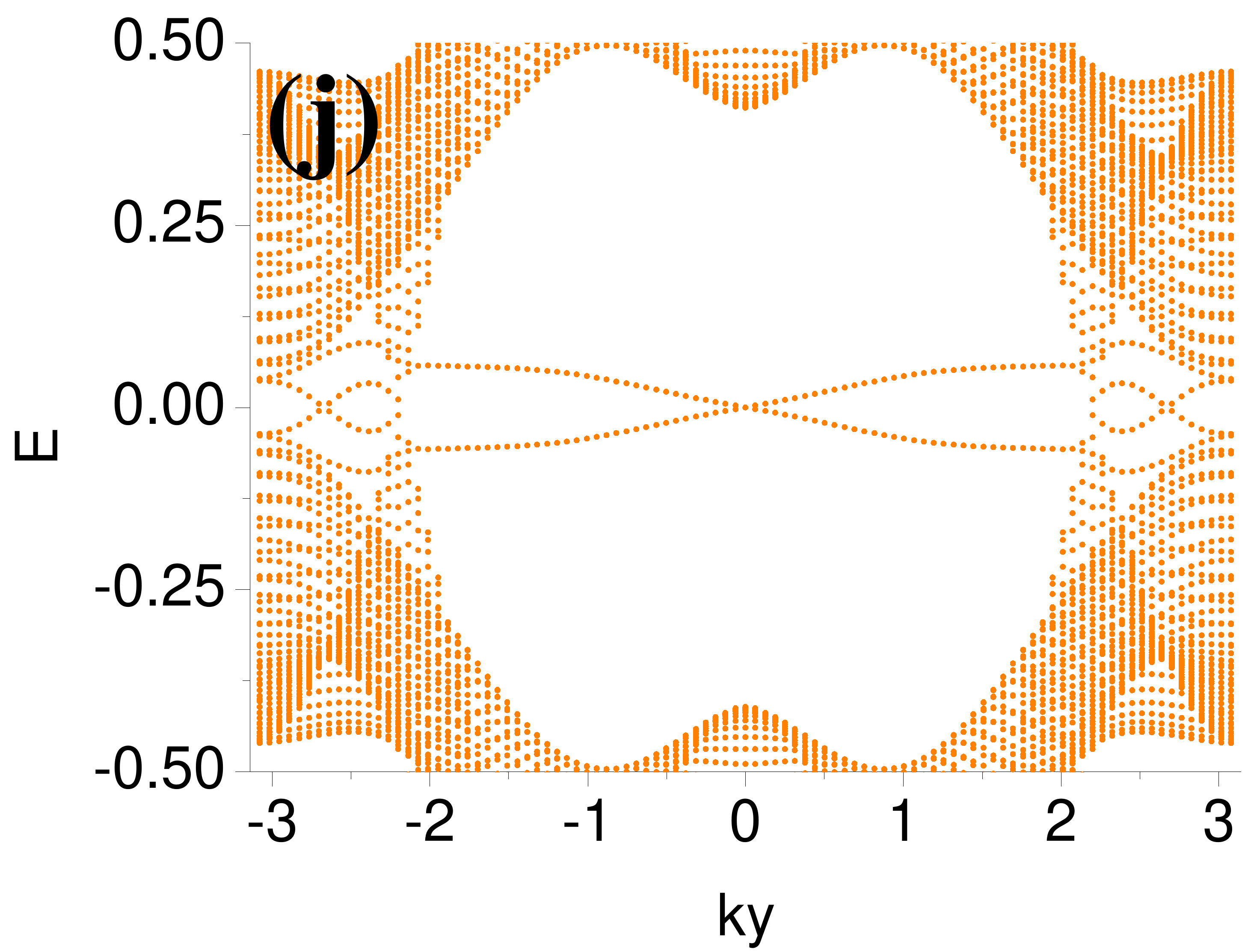} \\
\end{tabular}
\caption{(color online). (a)-(j) are the edge spectra of the
$d_{x^2-y^2}+id_{xy}+s$-wave superconductor with Dresselhaus (110)
spin-orbit coupling in case (g) of Tab. (\ref{symmandti}). For the
phase diagram of Fig. (\ref{phd}b), the edge spectra are
demonstrated in (a), (b) and (c). The parameters are $t=2$,
$\beta=1$, $\Delta_{s_{1}}=1$, $\Delta_{s_{2}}=0$,
$\Delta_{d_{1}}=2$, $\Delta_{d_{2}}=1$ and (a) $\mu=0, V^{2}=16$,
(b) $\mu=-2.5, V^{2}=36$, (c) $\mu=-4, V^{2}=20$, which correspond
to regions I, II and III in Fig. (\ref{phd}b), respectively. For the
phase diagram of Fig. (\ref{phd}c), the edge spectra are
demonstrated in (d), (e), (f) and (g). The parameters are $t=1$,
$\beta=1$, $\Delta_{s_{1}}=1$, $\Delta_{s_{2}}=0$,
$\Delta_{d_{1}}=2$, $\Delta_{d_{2}}=1$ and (d) $\mu=0, V^{2}=12$,
(e) $\mu=0, V^{2}=20$, (f) $\mu=-1.8, V^{2}=30$, (g) $\mu=-4.5,
V^{2}=25$, which correspond to regions I, II, III and IV in Fig.
(\ref{phd}c), respectively. For the phase diagram of Fig.
(\ref{phd}d), the edge spectra are demonstrated in (h), (i) and (j).
The parameters are $t=0.5$, $\beta=1$, $\Delta_{s_{1}}=1$,
$\Delta_{s_{2}}=0$, $\Delta_{d_{1}}=2$, $\Delta_{d_{2}}=1$ and (h)
$\mu=0, V^{2}=16$, (i) $\mu=-7, V^{2}=81$, (j) $\mu=-1, V^{2}=5$,
which correspond to regions I, II and III in Fig. (\ref{phd}d),
respectively.}\label{caseg}
\end{figure*}

The superconductors with mixed $s$-wave and $d$-wave pairing
symmetries are in case (a), (f) and (g). For each case, there are
three different kinds of phase diagrams depending on the hopping
amplitude $t$ as demonstrated in Fig. (\ref{phd}b)-(\ref{phd}d).
Although the edge spectrum becomes more complicated, there are no
qualitative differences in the edge spectrum between the $d+s$-wave
superconductor and the pure $d$-wave or $s$-wave superconductor. The
edge spectra for the $d_{x^2-y^2}+s$-wave superconductor with
Dresselhaus (110) spin-orbit coupling are shown in Fig.
(\ref{casef}). Due to the partial particle-hole symmetry and the
chiral symmetry, the edge spectra of this kind of superconductor are
the Majorana flat bands protected by the Pfaffian invariant
$\mathcal{P}(k_{y})$ and the winding number $\mathcal{W}(k_{y})$.
Notice that in Fig. (\ref{casef}f) and Fig. (\ref{casef}t), the
winding number $\mathcal{W}(k_{y})$ in some range of $k_{y}$ is 2,
however, its parity namely the Pfaffian invariant
$\mathcal{P}(k_{y})$ is 1. Therefore, the phase is trivial in this
range of $k_{y}$. The edge spectra for the
$d_{x^2-y^2}+id_{xy}+s$-wave superconductor with Rashba spin-orbit
coupling and Dresselhaus (110) spin-orbit coupling are shown in Fig.
(\ref{casea}) and Fig. (\ref{caseg}), respectively. Without the
protection of the partial particle-hole symmetry or the chiral
symmetry, the Majorana flat bands disappear and become Dirac cones.
In the nontrivial topological phase, we find that the edge spectra
have an odd number of Dirac cones.

\begin{figure}
\includegraphics[width=8cm]{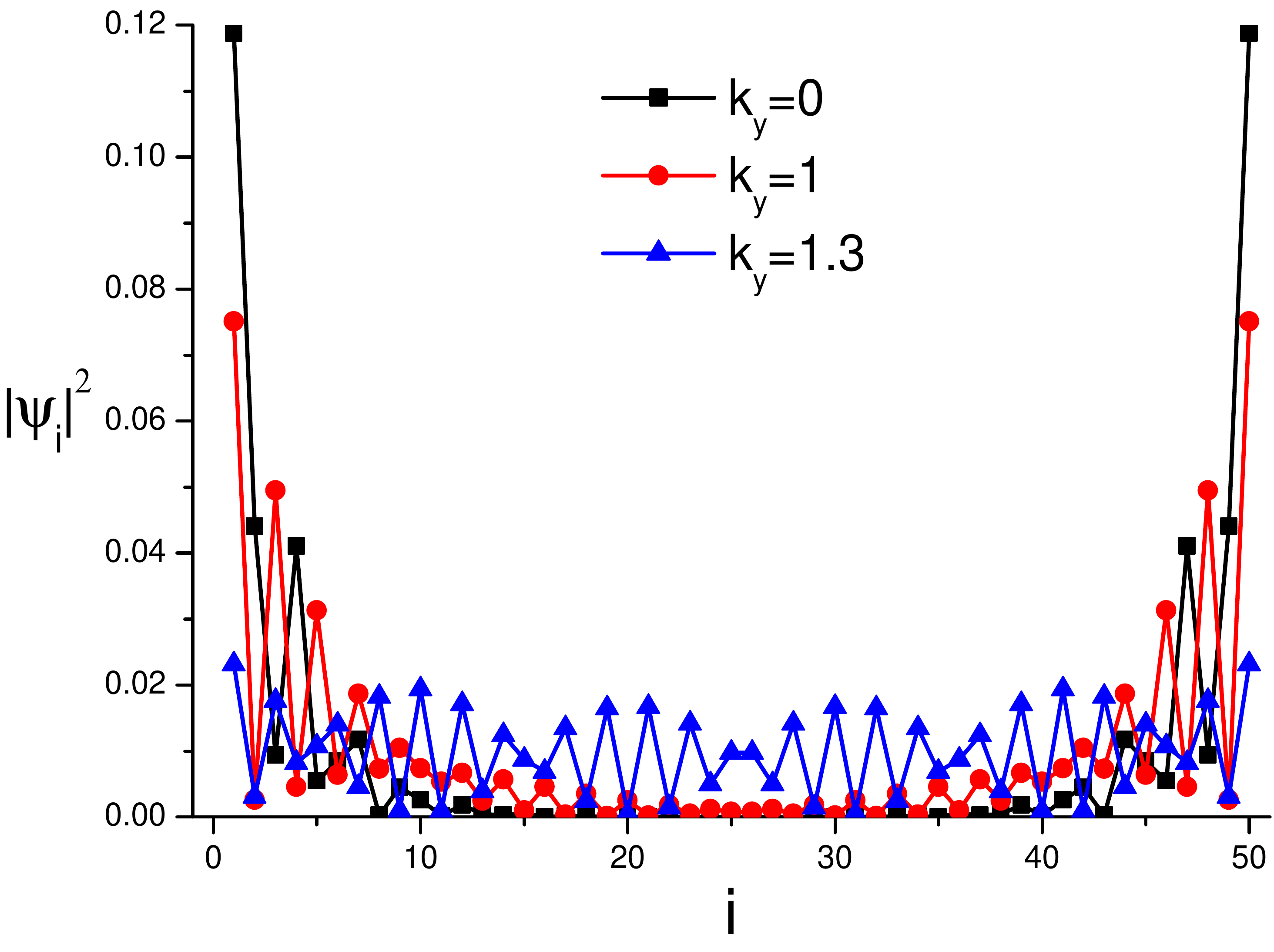}
\caption{(color online). The probability distributions of the
quasiparticle in the $d_{x^2-y^2}$-wave superconductor with
Dresselhaus (110) spin-orbit coupling in the edge Brillouin zone of
$k_{y}=0, 1, 1.3$. $i$ is the lattice site. $|\psi_{i}|^{2}$ is the
probability of quasiparticle at site $i$.}\label{MBS}
\end{figure}

We can define the $k_{y}$-dependent Pfaffian invariant
$\mathcal{P}(k_{y})$ or winding number $\mathcal{W}(k_{y})$ for the
partial particle-hole symmetry or the chiral symmetry. At fixed
$k_{y}$, there exist zero-energy states when
$\mathcal{P}(k_{y})=(-1)^{\mathcal{W}(k_{y})}=-1$. Therefore, there
are zero-energy Majorana flat bands in some range of $k_{y}$ in the
edge Brillouin zone when the Hamiltonian has the partial
particle-hole symmetry or the chiral symmetry. Notice that the
Majorana flat band does not always situate at the edge of the
system. At a fixed $k_{y}$, the bigger the gap of bulk state is, the
more localized the Majorana bound state is. Let us take the edge
spectra of the $d_{x^2-y^2}$-wave superconductor with the
Dresselhaus (110) spin-orbit coupling in Fig. (\ref{cased}a) as an
example. The probability distributions of the quasiparticle at
$k_{y}=0, 1, 1.3$ are shown in Fig. (\ref{MBS}). From Fig.
(\ref{cased}a), we see that the gap of the bulk state decreases as
$k_{y}$ increases from $0$ to $1.3$. At the same time, the
probability distribution of the quasiparticle becomes more and more
delocalized and extends into the bulk. Therefore, only the big-gap
Majorana bound states in the flat bands are well-defined Majorana
particles.

\section{summary}\label{summary}

In summary, we have investigated the topological phase and the
Majorana bound state in the spin-singlet superconductor with the
Rashba and Dresselhaus (110) spin-orbit couplings. We find that
apart from the particle-hole symmetry, the BdG Hamiltonian can also
exhibit the chiral symmetry, partial particle-hole symmetry and
partial chiral symmetry as listed in Tab. (\ref{symmandti}). The
topological invariants corresponding to these symmetries have been
discussed in Sec. (\ref{tiofBdG}) and are also listed in Tab.
(\ref{symmandti}). We have demonstrated all the possible phase
diagrams of the spin-singlet superconductor with the Rashba and
Dresselhaus (110) spin-orbit couplings in Sec. (\ref{pdofBdG}) by
the Pfaffian invariant $\mathcal{P}$ of the particle-hole symmetric
Hamiltonian. We find that only when the Hamiltonian has partial
particle-hole symmetry or chiral symmetry, the edge spectrum is flat
band protected by the one dimensional Pfaffian invariant
$\mathcal{P}(k_{y})$ or the winding number $\mathcal{W}(k_{y})$;
otherwise the edge spectrum is Dirac cone. The zero-energy state of
the Dirac cones and the zero-energy flat bands are the Majorana type
which is a precious source for topological quantum computing. The
edge spectra of the cases listed in Tab. (\ref{symmandti}) are shown
in Sec. (\ref{MBSofBdG}). We find that the Pfaffian invariant
$\mathcal{P}(k_{y})$ and the winding number $\mathcal{W}(k_{y})$ can
be used in determining the location of the zero-energy flat bands.

\begin{acknowledgments}
This work is partly supported by the National Research Foundation
and Ministry of Education, Singapore (Grant No. WBS:
R-710-000-008-271).
\end{acknowledgments}



%

\end{document}